\documentclass[aps,twocolumn,english,showpacs,superscriptaddress,footinbib,bibnotes,floatfix
]{revtex4-1}

\usepackage{graphicx}
\usepackage{grffile}            
\usepackage[usenames,dvipsnames]{color}
\usepackage{dcolumn}
\usepackage{bbm}
\usepackage{bm}
\usepackage{amsmath}
\usepackage{amssymb}
\usepackage{times}
\usepackage{hyperref}
\hypersetup{colorlinks=false}

\renewcommand{\Re}{\operatorname{Re}}
\renewcommand{\Im}{\operatorname{Im}}

\newcommand{\Id}[0]{\mathbbm{1}}

\newcommand\at[2]{\left.#1\right|_{#2}}

\begin{document}

\title{Quantum critical properties of a metallic spin density wave transition}

\author{Max H. Gerlach}
\thanks{These authors have contributed equally to this work.}
\affiliation{Institute for Theoretical Physics, University of Cologne, 50937 Cologne, Germany}

\author{Yoni Schattner}
\thanks{These authors have contributed equally to this work.}
\affiliation{Department of Condensed Matter Physics, The Weizmann Institute of Science, Rehovot, 76100, Israel}

\author{Erez Berg}
\affiliation{Department of Condensed Matter Physics, The Weizmann Institute of Science, Rehovot, 76100, Israel}

\author{Simon Trebst}
\affiliation{Institute for Theoretical Physics, University of Cologne, 50937 Cologne, Germany}

\date{\today}

\begin{abstract}
We report on numerically exact determinantal quantum Monte Carlo simulations of the onset of spin-density
wave (SDW) order in itinerant electron systems captured by a sign-problem-free two-dimensional lattice model.
Extensive measurements of the SDW correlations in the vicinity of the phase transition reveal that the critical dynamics
of the bosonic order parameter are well described by a dynamical critical exponent $z=2$, consistent
with Hertz-Millis theory, but are found to follow a finite-temperature dependence that does not fit the predicted behavior
of the same theory. The presence of critical SDW fluctuations is found to have a strong impact on the
fermionic quasiparticles, giving rise to a dome-shaped superconducting phase near the quantum
critical point. In the superconducting state we find a gap function that has an opposite sign between the two bands of
the model and is nearly constant along the Fermi surface of each band.
Above the superconducting $T_c$ our numerical simulations reveal a nearly temperature and frequency independent
self energy causing a strong suppression of the low-energy quasiparticle spectral weight in the vicinity of the hot spots
on the Fermi surface. This indicates a clear breakdown of Fermi liquid theory around these points. \\[3mm]
\end{abstract}


\maketitle

\section{Introduction}
Metallic spin-density wave (SDW) transitions are ubiquitous to strongly correlated materials such as the electron-doped cuprates~\cite{Armitage2010}, the Fe-based superconductors~\cite{Paglione2010}, heavy fermion systems~\cite{Gegenwart2008}, and organic superconductors~\cite{Brown2015}. In all these materials, unconventional superconductivity is found to emerge
near the onset of SDW order, with the maximum of the superconducting T$_c$ occuring either at or near the underlying SDW quantum phase transition (QPT).
In addition, the vicinity of the SDW transition is often characterized by strong deviations from Fermi liquid theory - both in thermodynamic and in single electron properties.

More broadly, understanding the properties of quantum critical points (QCPs) in itinerant fermion systems has attracted much interest over the past several decades~\cite{Hertz1976, Millis1993, Abanov2000, Abanov2003, Abanov2004,Rech2006,Lohneysen2007,Lee2009,Mross2010,Metlitski2010,Metlitski2010a,Hartnoll2011,Berg2012,Dalidovich2013,Efetov2013,Lee2013,Fitzpatrick2014,Holder2015,Raghu2015,Varma2015a, Varma2015b,Maier2016,Schattner2015a,Lee2016}. Unlike thermal critical phenomena and QCPs in insulating systems, here the critical order parameter fluctuations are strongly interacting with low-energy fermionic quasiparticles near the Fermi surface. In the traditional approach to this problem due to Hertz~\cite{Hertz1976}, later refined by Millis~\cite{Millis1993}, the fermions are integrated out from the outset, leading to an effective \emph{bosonic} action for the order parameter fluctuations. The dynamics of the order parameter is found to be overdamped due to the coupling to the fermions, that act as a bath. The effective bosonic action is then treated using conventional renormalization group (RG) techniques. While physically appealing, this approach has the drawback that integrating out low-energy modes is dangerous, since it generates non-analytic terms in momentum and frequency that are difficult to treat within the RG scheme. An alternative popular approach has been to use a $1/N$ expansion \cite{Abanov2003}, where $N$ is the number of fermion flavors. However, in the important case of two spatial dimensions, this approach turns out to be uncontrolled, as well~\cite{Lee2009,Metlitski2010a}. Alternative expansion parameters have been proposed~\cite{Mross2010,Dalidovich2013,Lee2016}, but a fully controlled analytical treatment of QCPs in itinerant electron systems has remained one of the grand challenges in strongly correlated electron physics.

In addition to the bosonic critical fluctuation dynamics, an important open question regards the behavior of the \emph{fermionic} quasiparticles in the vicinity of the transition. The exchange of SDW critical fluctuations can mediate a superconducting instability; however, the same fluctuations also strongly scatter the quasiparticles, causing them to lose their coherence and leading to the formation of a non-Fermi liquid metal. It is not clear which of these effects dominates; i.e., is there a well-defined non-Fermi liquid regime that precedes superconductivity, or does pairing always preempt the formation of a non-Fermi liquid~\cite{Metlitski2015}?

In this work, we perform extensive numerically exact determinantal quantum Monte Carlo (QMC) simulations~\cite{Blankenbecler1981,White1989,LohJr.1992,Assaad2008,Gubernatis2016} of a metallic system in the vicinity of an SDW transition. We use the approach of Refs.~\cite{Berg2012,Schattner2015a}, that introduced a two-dimensional multi-band lattice model that captures the generic structure of the ``hot spots'' - points on the Fermi surface where quasiparticles can scatter off critical SDW fluctuations resonantly. The universal properties of metallic SDW transitions are believed to depend only on the vicinity of the hot spots. At the same time, the model is amenable to QMC simulations without a sign problem \cite{Berg2012}. Our goal here is both to understand the generic properties of the transition, and to provide a controlled benchmark to analytic theories. We present detailed information about the bosonic and fermionic correlations and the interplay with unconventional superconductivity in the vicinity of the QCP.

Previously, the finite-temperature phase diagram of the model has been characterized, and a dome-shaped superconducting phase peaked near the SDW transition was found~\cite{Schattner2015a, Wang2015}. Here, we measure the SDW correlations in the vicinity of the transition, above the superconducting $T_c$. We find that, over a broad range of parameters, the SDW susceptibility is well described by the following form:
\begin{multline}
\chi_0(\mathbf q , \omega_n, r, T) =\\ \frac{1}{a_q (\mathbf{q}-\mathbf{Q})^2 + a_\omega |\omega_n| + a_r(r-r_{c0}) + f(r, T)}.
\label{eq:hertz}
\end{multline}
Here $\mathbf Q$ is the ordering wavevector [chosen to be $(\pi,\pi)$
in our model], $\omega_n = 2\pi n T$ is a Matsubara frequency, and $r$
is a parameter used for tuning through the SDW QCP, while $a_q$, $a_\omega$,
$a_r$, and $r_{c0}$ are nonuniversal constants.  The function $f(r,T)$
extrapolates to $0$ as $T\rightarrow 0$. Importantly,
Eq.~(\ref{eq:hertz}) captures the behavior of both the bosonic SDW
correlations and the susceptibility of a fermion bilinear operator
with the same symmetry, establishing the consistency of our analysis.

Interestingly, $\chi_0(\mathbf q , \omega_n, r, T \rightarrow 0)$ has precisely the form predicted by Hertz and Millis; in particular, the bosonic critical dynamics are characterized by a dynamic critical exponent $z=2$. The function $f(r,T)$ does not follow the predicted behavior, however. In a window of temperatures above T$_c$, we find a power-law dependence $f(r\approx r_{c0},T) \propto T^\alpha$ with $\alpha\simeq 2$, in contrast to the linear behavior predicted by Millis~\cite{Millis1993, SubirBook}.

The single-fermion properties above $T_c$ are found to depend strongly on the distance from the hot spots. Away from the hot spots, a behavior consistent with Fermi liquid theory is observed. At the hot spots, a substantial loss of spectral weight is seen upon approaching the QCP. In a temperature window above T$_c$, the fermionic self energy is only weakly frequency and temperature dependent, 
corresponding to a nearly-constant lifetime of quasiparticles at the hot spots. This behavior indicates a strong breakdown of Fermi liquid theory at these points. It is not clear, however, whether it represents the asymptotic behavior at the putative underlying SDW QCP, since superconductivity intervenes before the QCP is reached.

Finally, in order to probe the interplay between magnetic quantum criticality and superconductivity, we measure the superconducting gap, $\Delta_{\mathbf{k}}$, and the momentum-resolved superconducting susceptibility, $P_{\mathbf{k},\mathbf{k}'}$, across the phase diagram. No strong feature is found in $\Delta_{\mathbf{k}}$ at the hot spots. Rather, $\Delta_{\mathbf{k}}$ and $P_{\mathbf{k},\mathbf{k}'}$ vary smoothly on the Fermi surface. While the pairing interaction may be strongly peaked at wavevector $\mathbf{Q}$, the resulting gap function does not reflect
such a strong momentum dependence.

This paper is organized as follows. In Sec.~\ref{sec:model} we describe the model and review its phase diagram. Sec.~\ref{sec:SDW_corr} presents a detailed analysis of the SDW susceptibility across the phase diagram. 
In Sec.~\ref{sec:fermion_corr} we study the single-fermion properties, providing evidence for the breakdown of Fermi liquid theory in the vicinity of the hot spots. Sec.~\ref{sec:SC} analyzes the gap structure in the superconducting state. The cumulative results are put into perspective in Sec.~\ref{sec:discussion}. Supplementary data sets and some technical details are presented in the appendices.

\section{The Model and the phase diagram}
\label{sec:model}

\begin{figure}[t]
  \centering
  \includegraphics[width=.55\columnwidth]{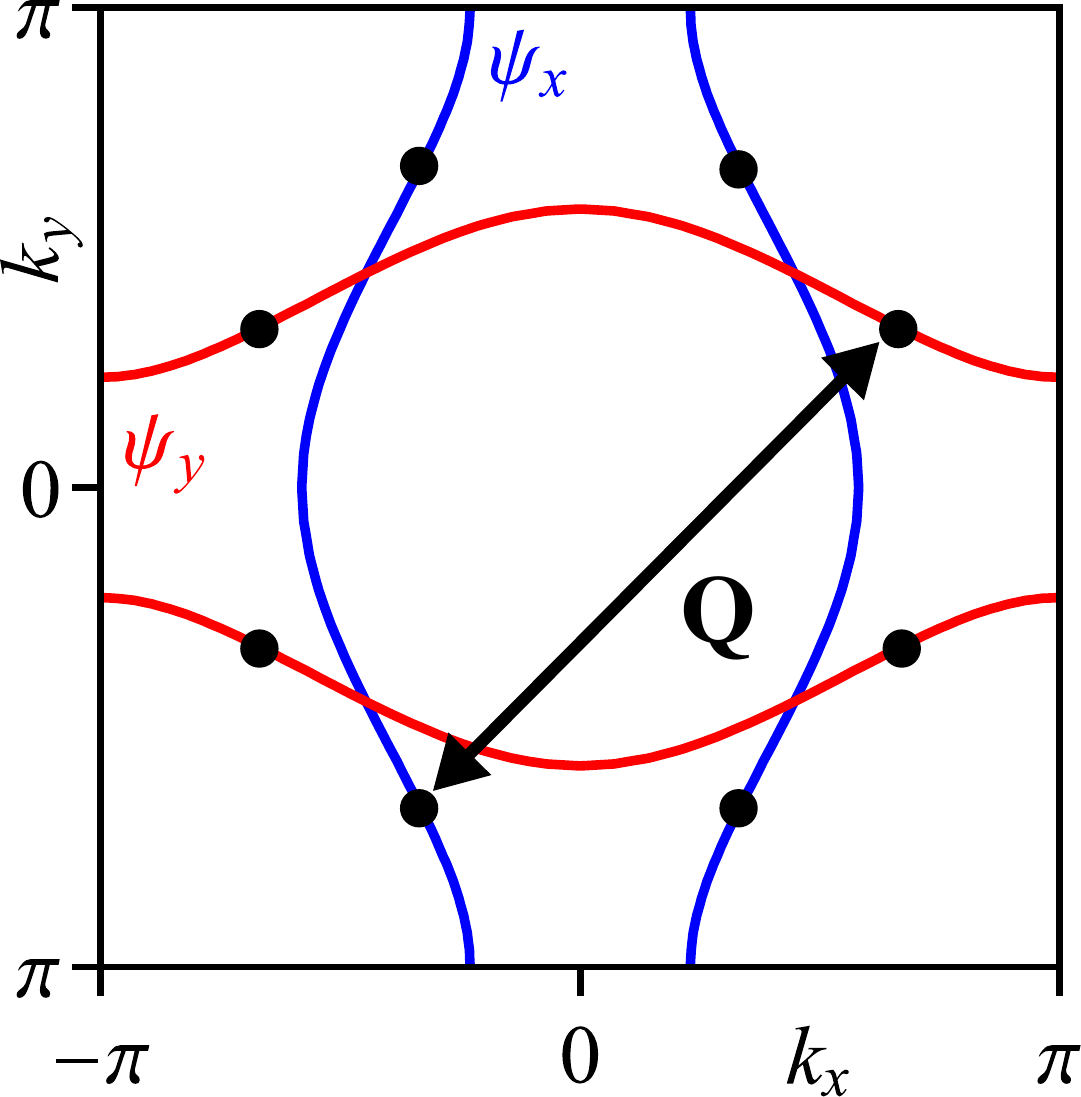}
  \caption{Noninteracting Fermi surfaces of the $\psi_x$ and $\psi_y$
    fermions.  hot spots (indicated by the black dots) are linked by
    the vector $\mathbf{Q} = (\pi,\pi)$ with points on the other band.
    The $\psi_x$ and $\psi_y$ fermions have stronger dispersion in
    direction $k_x$ and $k_y$, respectively. }
  \label{fig:fermisurfaces}
\end{figure}

\begin{figure*}[th!]
  \centering
  {
    \newlength{\figheight}
    \setlength{\figheight}{0.235\linewidth}
    \includegraphics[height=\figheight]{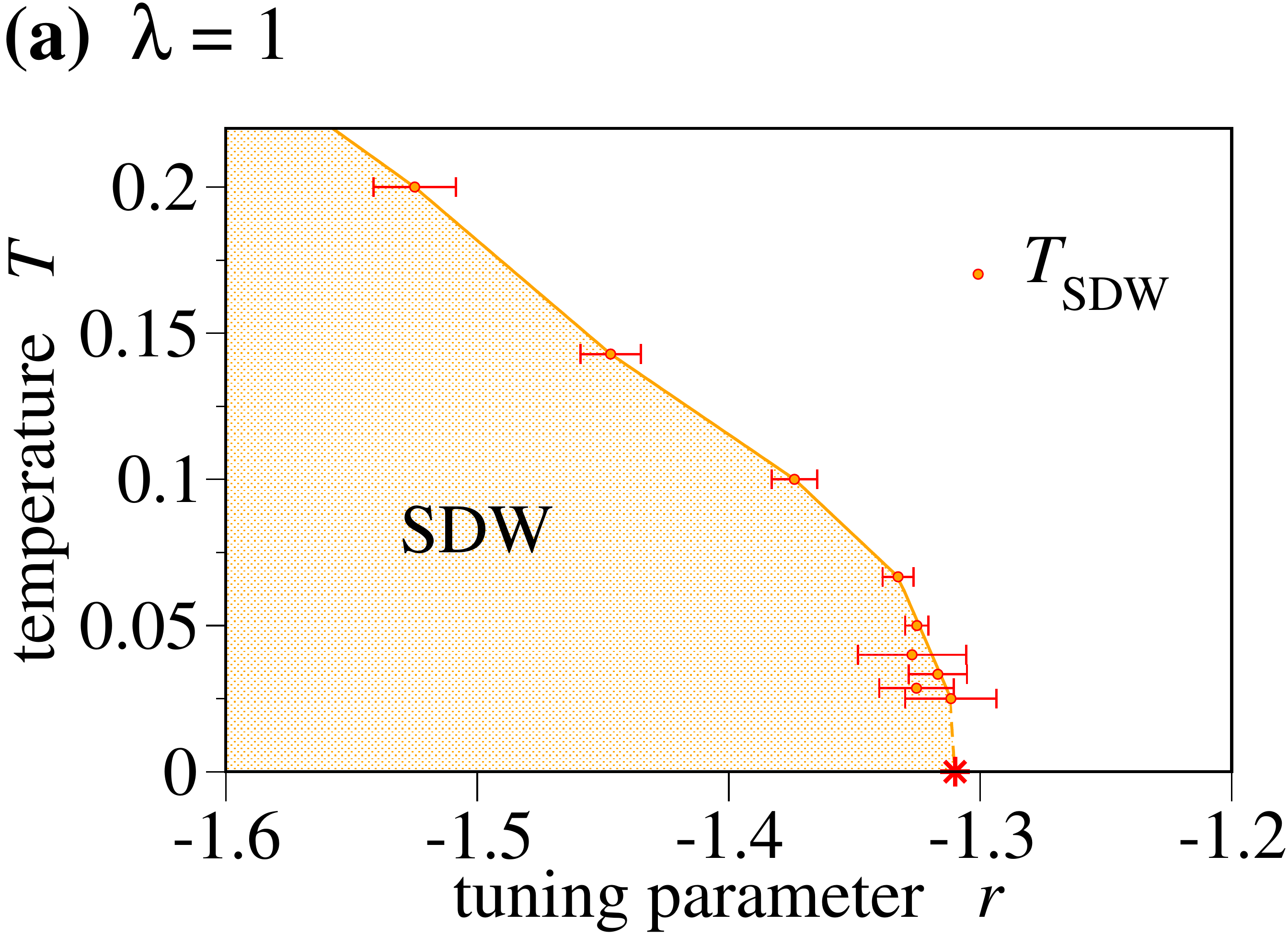} \hspace{0.01\linewidth}
    \includegraphics[height=\figheight]{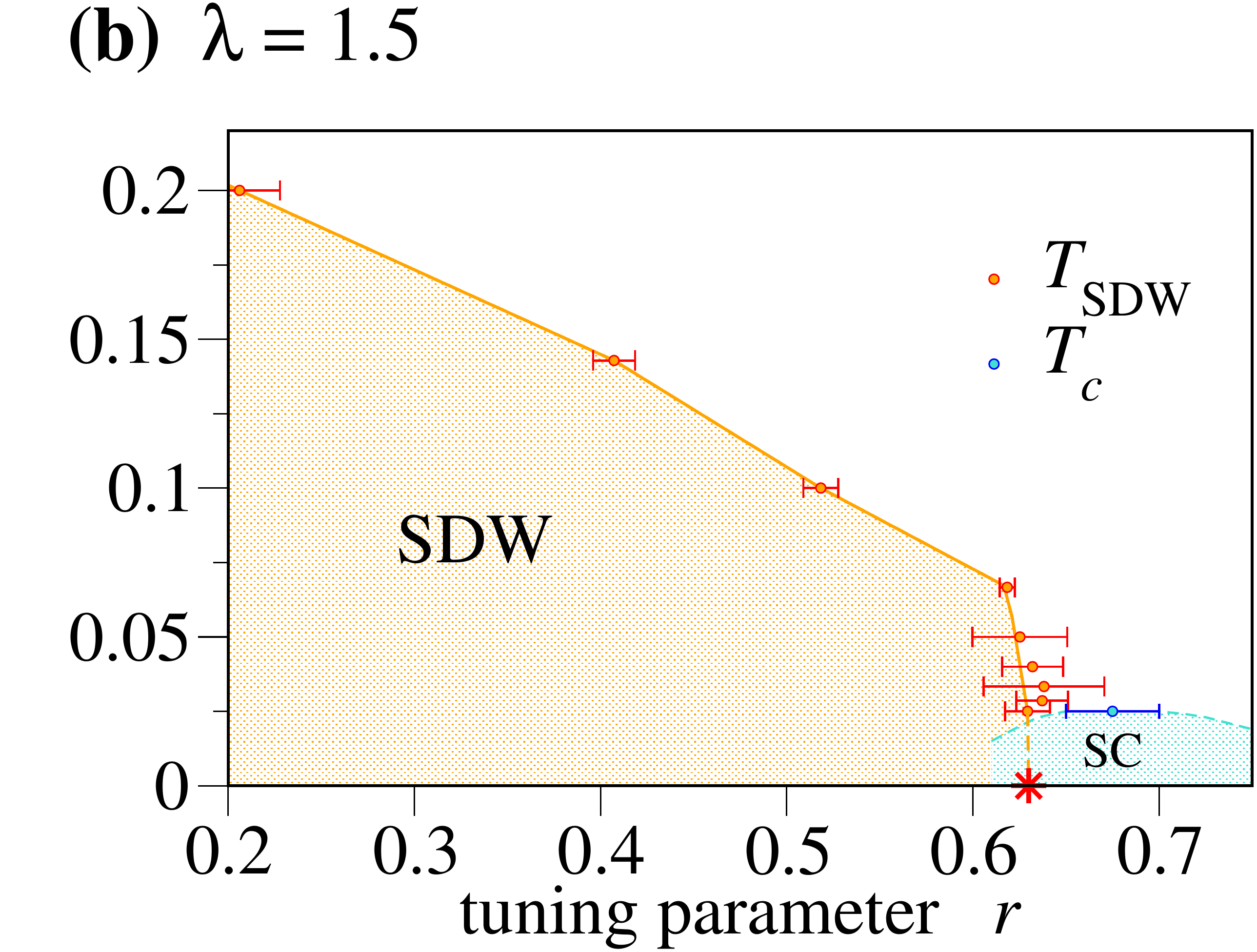} \hspace{0.02\linewidth}
    \includegraphics[height=\figheight]{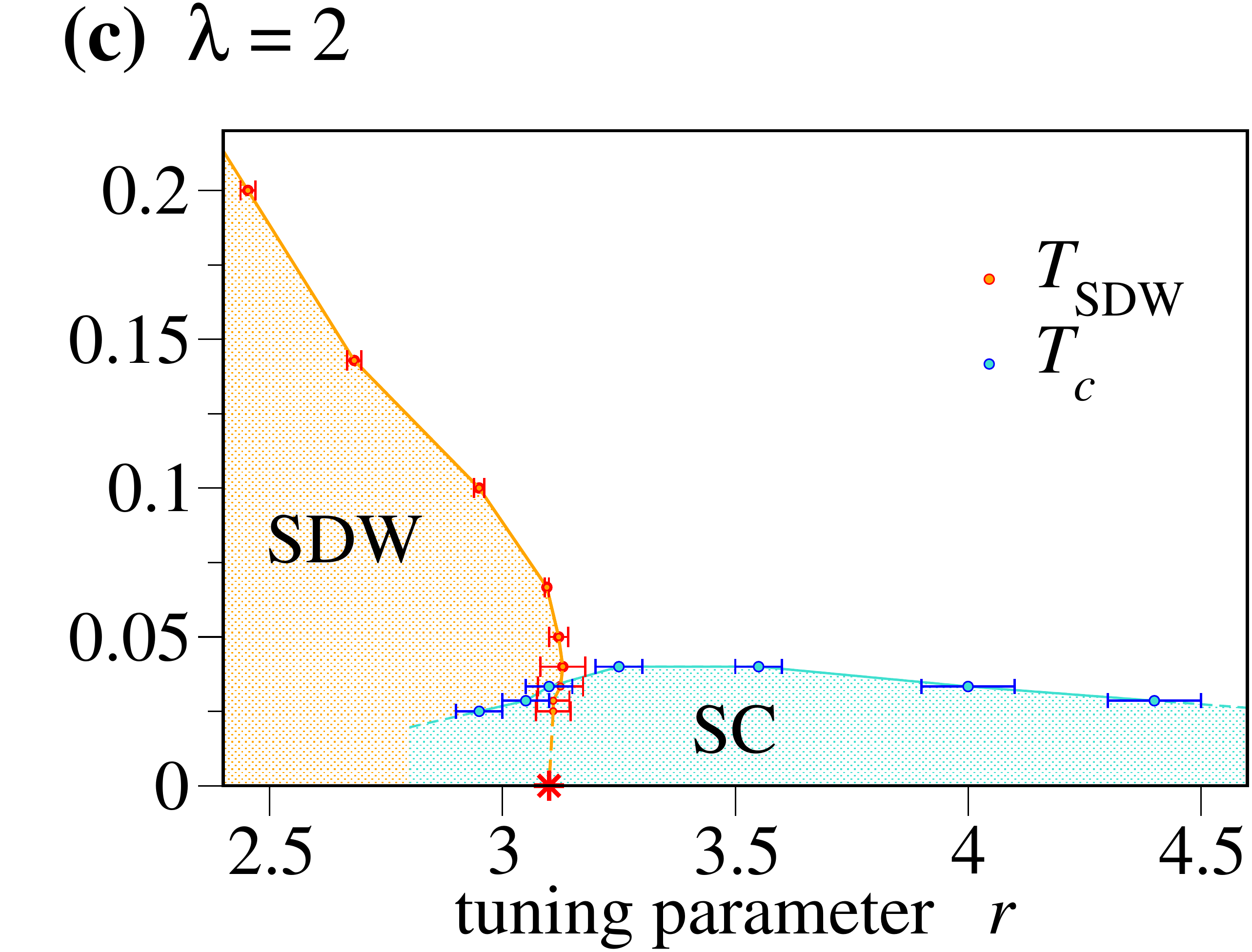}
  }
  \caption{Finite-temperature phase diagram of model \eqref{eq:action}
    for three different values of the Yukawa coupling $\lambda$ and
    bare bosonic velocity $c=3$.  Shown are the transition temperature
    $T_{\rm SDW}$ to magnetic spin density wave (SDW) order as well as
    the estimated location of the zero-temperature phase transition
    point $r_{c}$ (indicated by the red star).  The superconducting (SC)
    transition temperature $T_{\rm c}$ is shown where applicable.  In
    the same units the Fermi energy is $E_F = 2.5$.  Dashed lines are
    a guide to the eye.  }
  \label{fig:phasediagrams}
\end{figure*}

Our lattice model is composed of two flavors of spin--$\frac{1}{2}$
fermions, $\psi_x$ and $\psi_y$,
that are coupled to a real bosonic vector field $\vec{\varphi}$, which represents fluctuations of a
commensurate SDW order parameter at wavevector
$\mathbf{Q} = (\pi, \pi)$.
The two types of fermions exhibit quasi one-dimensional dispersions along momenta $k_x$ and $k_y$, respectively,
which in the absence of intreractions give rise to the Fermi surfaces illustrated in Fig.~\ref{fig:fermisurfaces}.
It is precisely this two-flavor structure that fundamentally allows us to set
up an action completely devoid of the fermion sign problem in QMC  simulations~\cite{Berg2012}.  Yet,
the Fermi surfaces of this model capture the generic structure of the hot spots, which is generally believed to
determine the universal physics near the QCP.

As in our previous work on this model~\cite{Schattner2015a},
we assume an $O(2)$ symmetric SDW order parameter, i.e. $\vec{\varphi}$ is restricted to the $XY$ plane.
In contrast to the case of an $O(3)$ order parameter (studied in Refs.~\cite{Berg2012,Wang2015}), the easy-plane order parameter
implies the existence of a {\it finite-temperature} SDW phase transition of Berezinskii-Kosterlitz-Thouless (BKT)
character, which we can track in our numerics. In addition, the reduced dimensionality of the order parameter
brings a welcome computational benefit as it enables a reduction of the
dimensions of all single-fermion matrices by half, greatly improving the efficiency of the
numerical linear algebra. 

Our lattice model is given by the action $S=S_F + S_{\varphi} =
\int_0^{\beta}d\tau(L_F+L_{\varphi})$ with
\begin{align}
  \label{eq:action}
  L_F ={}& \sum_{\substack{i,j,s\\ \alpha = x,y}}
  \psi^{\dagger}_{i \alpha s} \left[ (\partial_{\tau}
    -\mu)\delta_{ij} - t_{\alpha ij} \right]
  \psi_{j \alpha s} \notag\\
  {}& + \lambda \sum_{\substack{
      i, s,s'}}
  [\vec{s} \cdot
  \vec{\varphi}_i ]_{ss'} \psi^{\dagger}_{ixs}
  \psi_{iys'}^{\vphantom{\dagger}} + \mathrm{h.c.}, \notag\\
  L_{\varphi} ={}& \frac{1}{2} \sum\limits_{i}
  \frac{1}{c^2}
  \left( \frac{d\vec{\varphi}_i}{d\tau} \right)^2
  + \frac{1}{2} \sum\limits_{\left\langle i,j
    \right\rangle} \left(
    e^{i \mathbf{Q}\cdot \mathbf{r}_i}\vec{\varphi}_i -
    e^{i \mathbf{Q}\cdot \mathbf{r}_j} \vec{\varphi}_j\right)^2
  \notag\\
  &+ \sum\limits_{i} \left[ \frac{r}{2}\vec{\varphi}_i^2 +
    \frac{u}{4} (\vec{\varphi}_i^2)^2 \right] .
\end{align}
This action is defined on a square lattice with sites labeled by
$i,j = 1, \dotsc, N_s$, where $\langle i,j \rangle$ are nearest
neighbors.  The two fermion flavors are indexed by $\alpha=x,y$, while
$s,s'={\uparrow},{\downarrow}$ index the spin polarizations and
$\vec{s}$ are the Pauli matrices.  Imaginary time is denoted by $\tau$
and $\beta = 1/T$ is the inverse temperature.  The fermionic
dispersions are implemented by setting different
hopping amplitudes along the horizontal and vertical lattice
directions.  For the $\psi_x$-fermions they are given by $t_{x,h}=1$
and $t_{x,v}=0.5$, respectively, and for the $\psi_y$-fermions by
$t_{y,h}=0.5$ and $t_{y, v}=1$. Note that our model is fully $C_{4}$-symmetric
with a $\pi/2$ rotation mapping the $\psi_x$-band to the $\psi_y$-band
and vice versa.  The tuning parameter $r$ allows to tune the system to
the vicinity of an SDW instability.  In an experimental system, $r$
could be proportional to a physical tuning parameter such as pressure
or doping. We set the chemical potential to $\mu=-0.5$, such that the
Fermi energy, measured relative to the band bottom, is $E_F=2.5$. The quartic
coupling is set to $u=1$.  In the following we mostly focus on the case of a
bare bosonic velocity of $c = 3$ and a Yukawa coupling between
fermions and bosons of $\lambda = 1.5$. Occasionally, we also consider
other values of $c$ and $\lambda$.

Our numerical analysis of model \eqref{eq:action} is based on
extensive finite-temperature
DQMC~\cite{Blankenbecler1981,White1989,LohJr.1992,Assaad2008,Gubernatis2016}
simulations.  For the general setup and technical details on the
implementation of our DQMC simulations we refer to our earlier
paper~\cite{Schattner2015a} and its detailed supplemental online
material. Here we want to single out a few conceptual aspects of our
setup, which have allowed to push our simulations down to temperatures
of $T = 1/40$ for system sizes up to $16 \times 16$ sites.  First,
using a replica-exchange scheme~\cite{HukushimaNemoto,Katzgraber2006}
in combination with a global update procedure \cite{Schattner2015a}
thermal equilibration of our simulations is decidedly improved.  Most
of our simulations were performed in the presence of a weak fictitious
perpendicular ``magnetic" field (designed not to break time-reversal
symmetry), which serves to greatly speed up convergence to the
thermodynamic limit for metallic systems~\cite{Assaad2002a,
  Schattner2015,Schattner2015a}.  Since this technique breaks
translational invariance of the fermionic Green's function, we cannot
make use of it to study $\mathbf{k}$-resolved fermionic observables.
For this reason we have carried out additional simulations without the
magnetic flux, but with twisted boundary conditions, which allows to
increase the momentum space resolution.  We give details on these
procedures in Appendix \ref{sec:DQMCAppendix}

To set the stage for our discussion in the following sections, we
summarize the finite-temperature phase diagram of model
\eqref{eq:action} for $c=3$ and three different values of the Yukawa
coupling $\lambda = 1$, $1.5$, and $2$ in
Fig.~\ref{fig:phasediagrams}.  Besides a paramagnetic regime for
sufficiently high temperature, the dominant feature of these phase
diagrams is a quasi-long-range ordered SDW phase whose transition
temperature $T_{\mathrm{SDW}}$ is suppressed with increasing tuning
parameter $r$. Extrapolating the SDW transition down to the zero
temperature provides an estimate of the location of the quantum phase
transition at $r=r_c$ (indicated by the red star).
At low temperatures, the SDW transition may become weakly first
order~\cite{Schattner2015a}. However, in the temperature range
considered here, the transition is either continuous or
(possibly) very weakly first order.
While for $\lambda=1$ this
SDW phase is the only ordered phase down to temperatures of
$T=1/40$, 
there is an additional superconducting phase emerging in the vicinity
of the QPT for the two larger values of the Yukawa coupling.  For
$\lambda = 1.5$ we barely observe the tip of this quasi-long-range
ordered phase with a maximum critical temperature of
$T^{\mathrm{max}}_c \approx 1/40$, which is our numerical temperature
limit. For $\lambda = 2$, we can clearly map out a superconducting
dome with the maximum of the critical temperature
$T_c^{\mathrm{max}} \approx 1/20$.

At finite temperatures, both the SDW and the SC finite-temperature transitions are expected to be of
BKT type.  The SDW susceptibility
$\chi = \int d\tau \sum_i e^{i\mathbf{Q} \cdot \mathbf{r}_i}\left\langle\vec{\varphi}_i(\tau)
  \vec{\varphi}_0(0) \right\rangle$
is found to  follow a scaling law
$\chi \propto L^{2-\eta}$ with a continuously changing exponent $\eta$.
We identify the transition temperature $T_{\mathrm{SDW}}$ with the
point where the exponent takes the universal value $\eta = 1/4$.
The SC transition temperature can be both determined via a similar $\eta$-fit
or by the point where the superfluid density $\rho_s$ obtains the
universal value of $2 T_c / \pi$.  Both estimates are found to agree.
The  error bars in Fig.~\ref{fig:phasediagrams} are mostly due to finite-size effects.
For further details on the procedures employed to identify the
different phase transitions see Ref.~\cite{Schattner2015a} and its
accompanying supplementary online material.

A common feature to all three phase diagrams is a change of slope
of the SDW phase boundary curve at low temperatures ($T\approx 0.07$).
This apparent ``bending" is more pronounced at larger Yukawa coupling
$\lambda$. Such behavior is generally expected to occur at the onset of superconductivity~\cite{Moon2010,Schattner2015a}. Here, however, the bending does not visibly track the SC transition temperature.
The curvature of the $T_{\mathrm{SDW}}$ line
is also reflected in the SDW susceptibility in the paramagnetic region of
the phase diagram, see Fig.~\ref{fig:chi-inv-color-l1} for $\lambda=1$
and Fig.~\ref{fig:chi-inv-color-l1.5-l2} in
Appendix~\ref{sec:magn-corr-at} for $\lambda=1.5, 2$.

\begin{figure}[t]
  \centering
  \includegraphics[width=\linewidth]{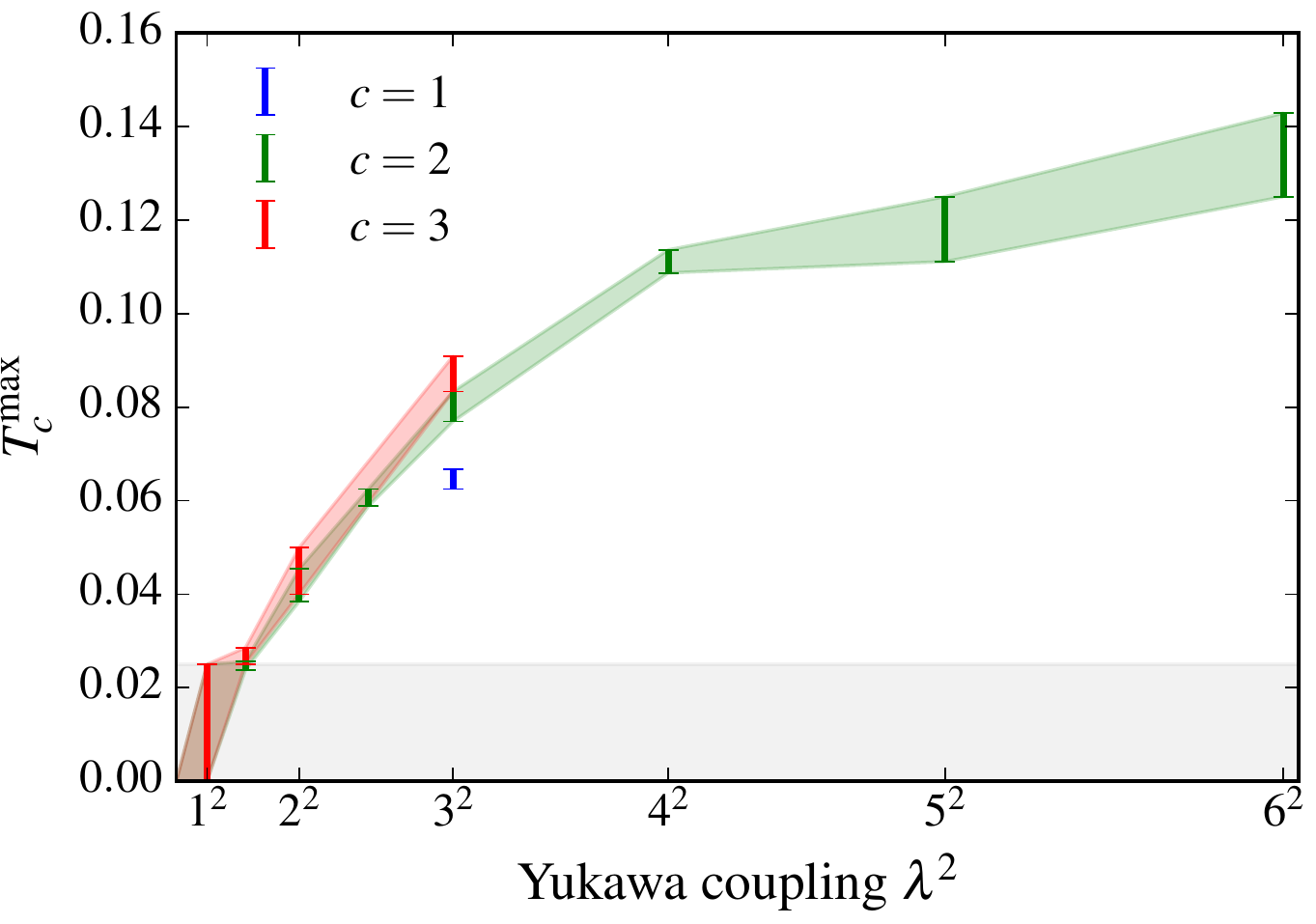}
  \caption{The maximal superconducting transition temperature $T_c^{\mathrm{max}}$ for different
   values of the Yukawa coupling $\lambda$ and (bare) boson velocity $c$. The hatched gray region
   indicates temperatures $T<0.025$, which are beyond our numerically accessible temperature range.}
  \label{fig:Tc_vs_lambda}
\end{figure}

Over a broader range of parameters, the maximum SC transition temperature $T_c^{\mathrm{max}}$
grows monotonically with increasing of either the Yukawa coupling or the boson velocity $c$, as illustrated in Fig.~\ref{fig:Tc_vs_lambda}.
Up to an intermediate coupling strength $\lambda\approx 3$, $T_c^{\mathrm{max}}$ rapidly grows as $T_c^{\mathrm{max}}\propto \lambda^2$, eventually
saturating at stronger coupling. Qualitatively, these trends are in agreement with results from Eliashberg theory \cite{Wang2016}.

\begin{figure}
  \newlength{\fw}
  \setlength{\fw}{0.9\linewidth}
  \begin{minipage}{\fw}
    \raggedright
    \includegraphics[width=\linewidth]{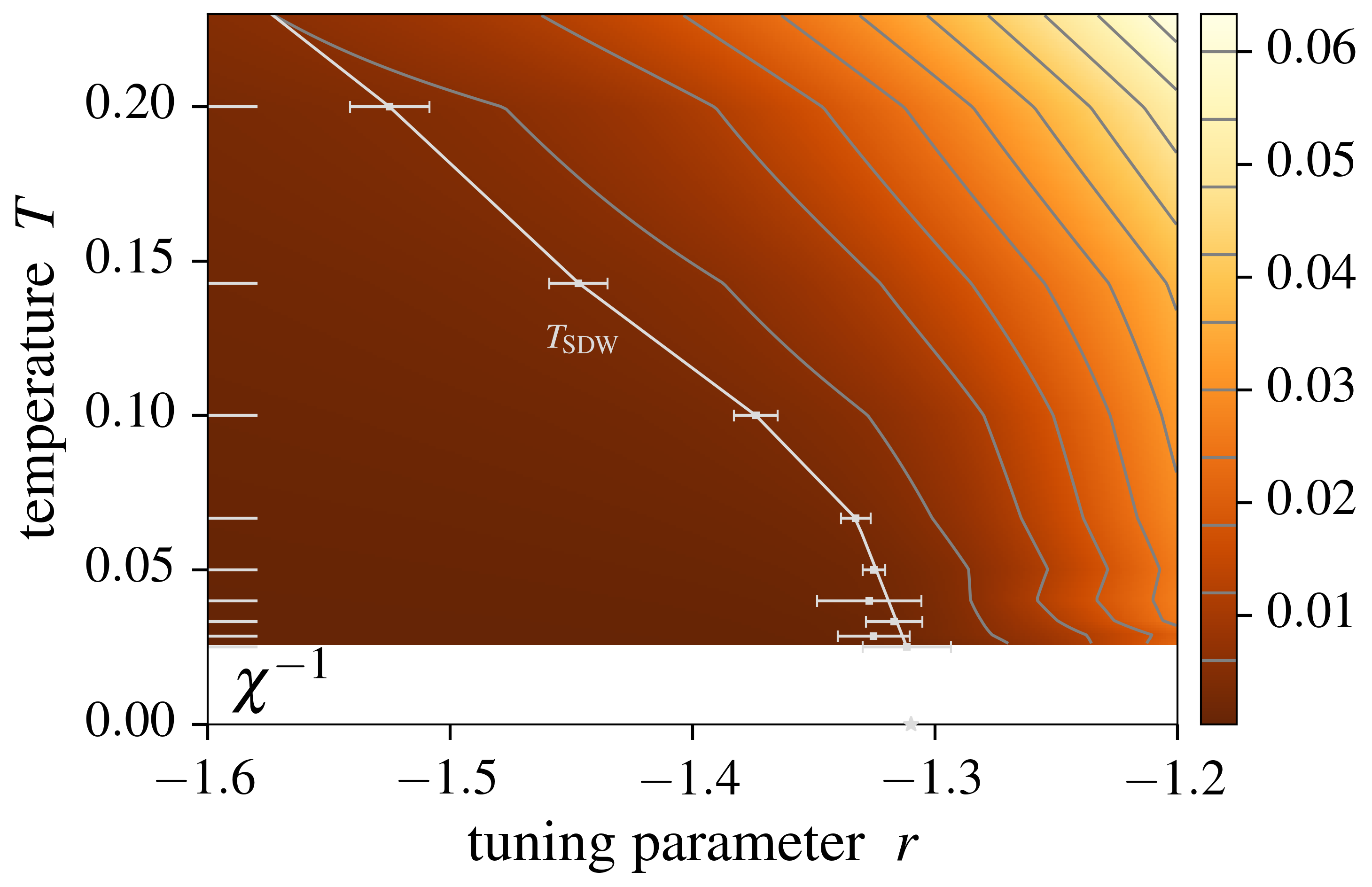}
  \end{minipage}
  \caption{Inverse SDW susceptibility $\chi^{-1}$ across the phase
    diagram for $\lambda=1$.  We show numerical data obtained at
    $L=14$ at the temperatures that are indicated by the ticks on the
    inside of the plot.  Intermediate temperatures are interpolated
    linearly, while the high resolution in the tuning parameter $r$ is
    achieved by reweighting.  Contour lines of $\chi^{-1}$ are marked
    gray. }
  \label{fig:chi-inv-color-l1}
\end{figure}

\section{Magnetic correlations}
\label{sec:SDW_corr}

We start our discussion of the quantum critical behavior of model
\eqref{eq:action} with an examination of magnetic fluctuations across
its entire finite-temperature phase diagram.  We probe the formation of
magnetic correlations both through the susceptibility of the bosonic order
parameter, $\vec{\varphi}$, and through a fermionic bilinear of the same
symmetry, which we evaluate independently in the same numerical
simulations of the action \eqref{eq:action}.
As we show below, both susceptibilities
exhibit the same behavior, supporting the robustness of our
conclusions to be presented.

\subsection{Bosonic SDW susceptibility}
\label{sec:boson-sdw-susc}

We first consider the bosonic susceptibility calculated from the
SDW order parameter $\vec{\varphi}$ in action \eqref{eq:action}
\begin{align}
  \label{eq:chi}
  \chi(\mathbf{q}, i\omega_n, r, T) &= \sum_i \int_0^{\beta}\!\! d\tau e^{i
    \omega_n \tau - i \mathbf{q} \cdot \mathbf{r}_i} \langle
  \vec{\varphi}_i(\tau) \cdot \vec{\varphi}_0(0) \rangle
\end{align}
for a given momentum $\mathbf{q}$ and Matsubara frequency
$\omega_n = 2\pi n T$.  The expectation values are estimated in a DQMC
simulation run at finite temperature $T$ and for a specific value of
the tuning parameter $r$, indicated here as explicit parameters. At
low temperature, we use the following form to fit the data, inspired
by Hertz theory \cite{Hertz1976}:
\begin{multline}
  \label{eq:chi_functional}
  \chi_0^{-1}(\mathbf{q}, i\omega_n, r, T\to0) = \\
  a_q(\mathbf{q}-\mathbf{Q})^2 + a_\omega |\omega_n| + a_r(r-r_{c0}) ,
\end{multline}
where $a_q,a_\omega$ and $a_r$ are non-universal fitting parameters that
describe the momentum dependence in the vicinity of the ordering
wavevector $\mathbf Q=(\pi,\pi)$, Landau damping, and the dependence on the tuning parameter $r$,
respectively. The fitting parameter $r_{c0}$ indicates the location of
the divergence of $\chi_0$. Due to the appearance of a
superconducting phase at low temperatures, $r_{c0}$ may differ from the actual
location of the QPT at $r=r_c$. However, within our numerical resolution,
we find $r_{c0}\approx r_c$, where $r_c$ is obtained by extrapolating the
finite-temperature transition line $T_{\mathrm{SDW}} \to 0$, as shown in the
phase diagrams of Fig.~\ref{fig:phasediagrams}.

\begin{figure}
  \centering
  \includegraphics[width=\linewidth]{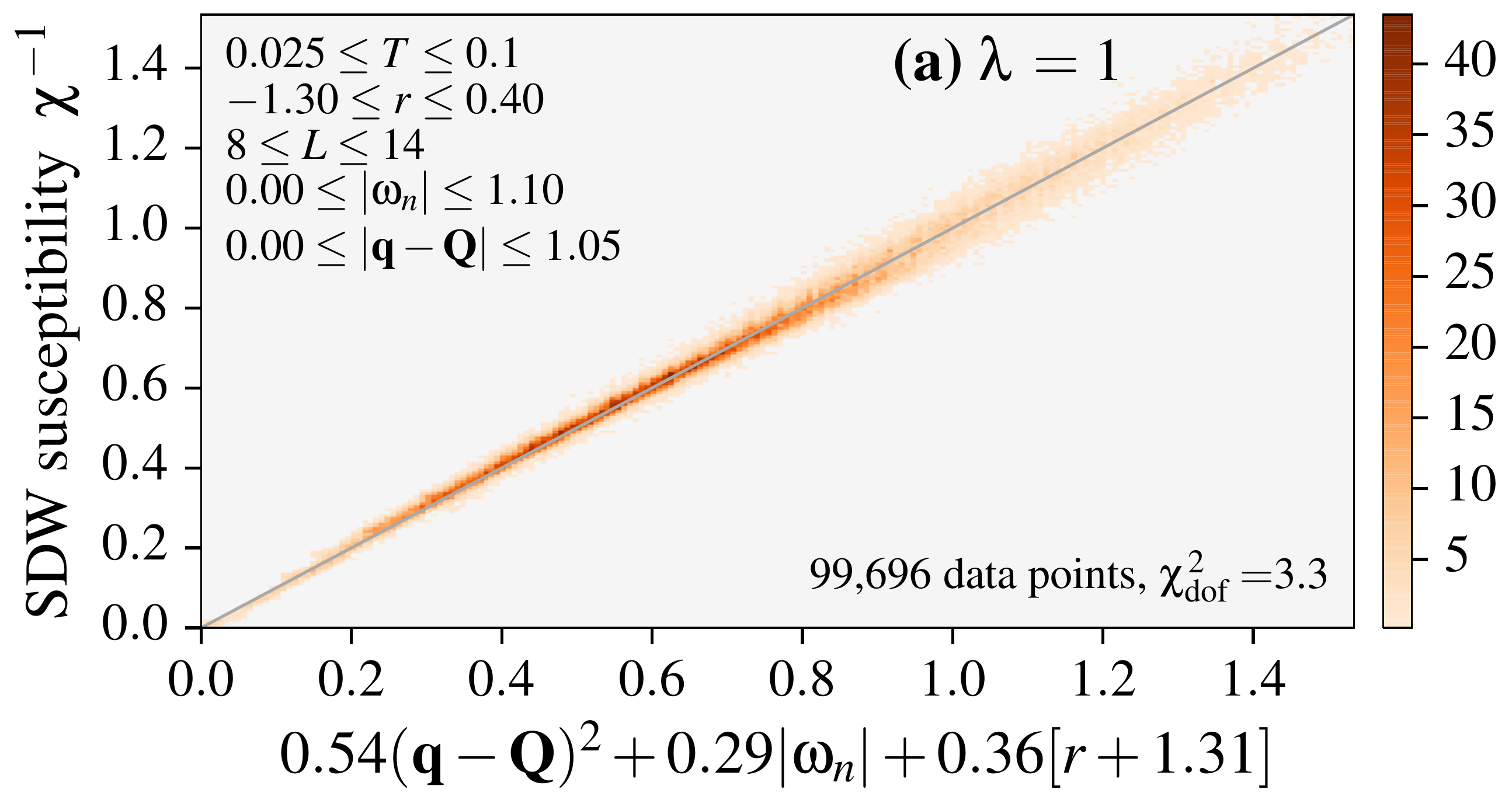}\\
  \includegraphics[width=\linewidth]{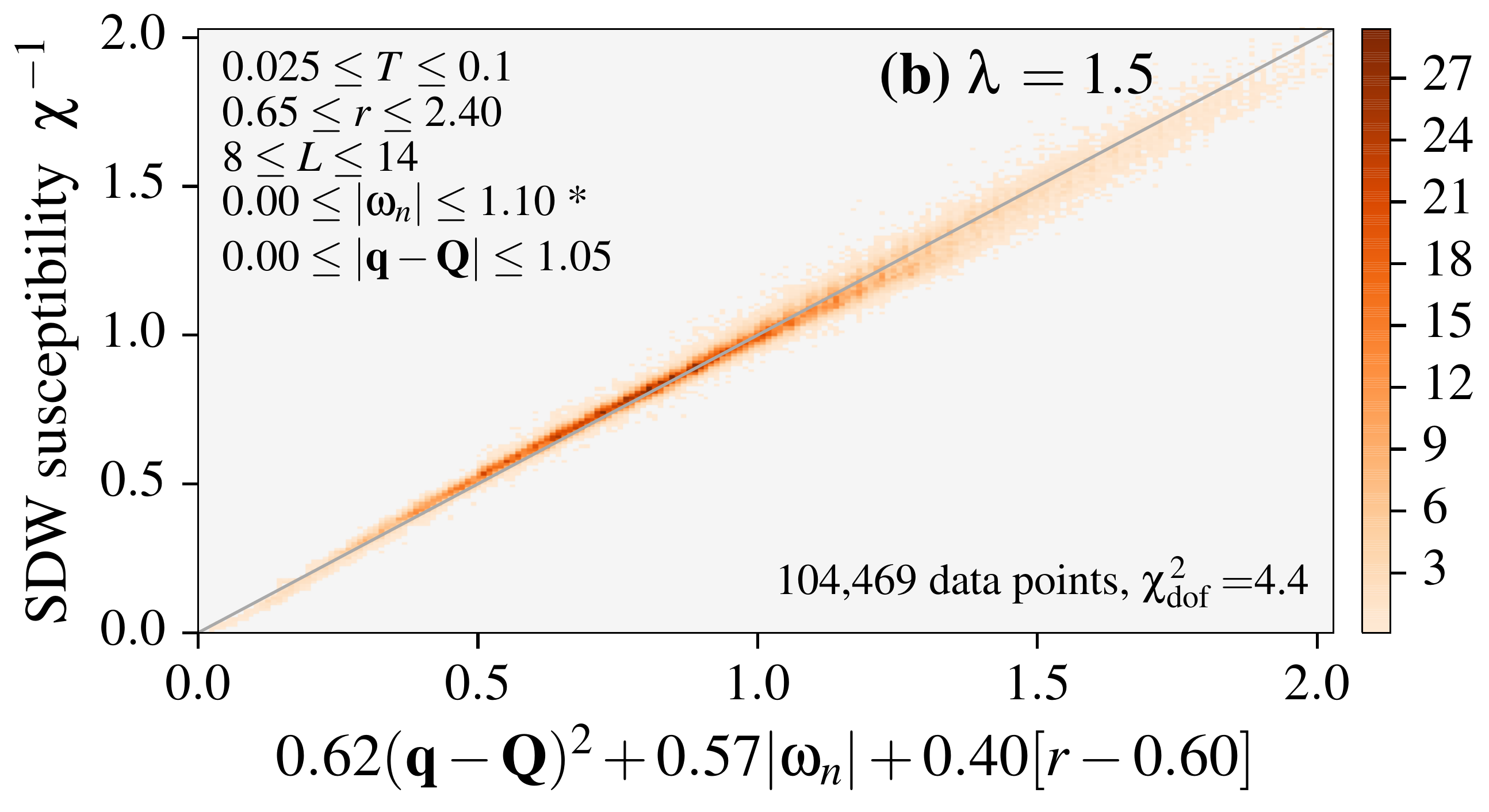}\\
  \includegraphics[width=\linewidth]{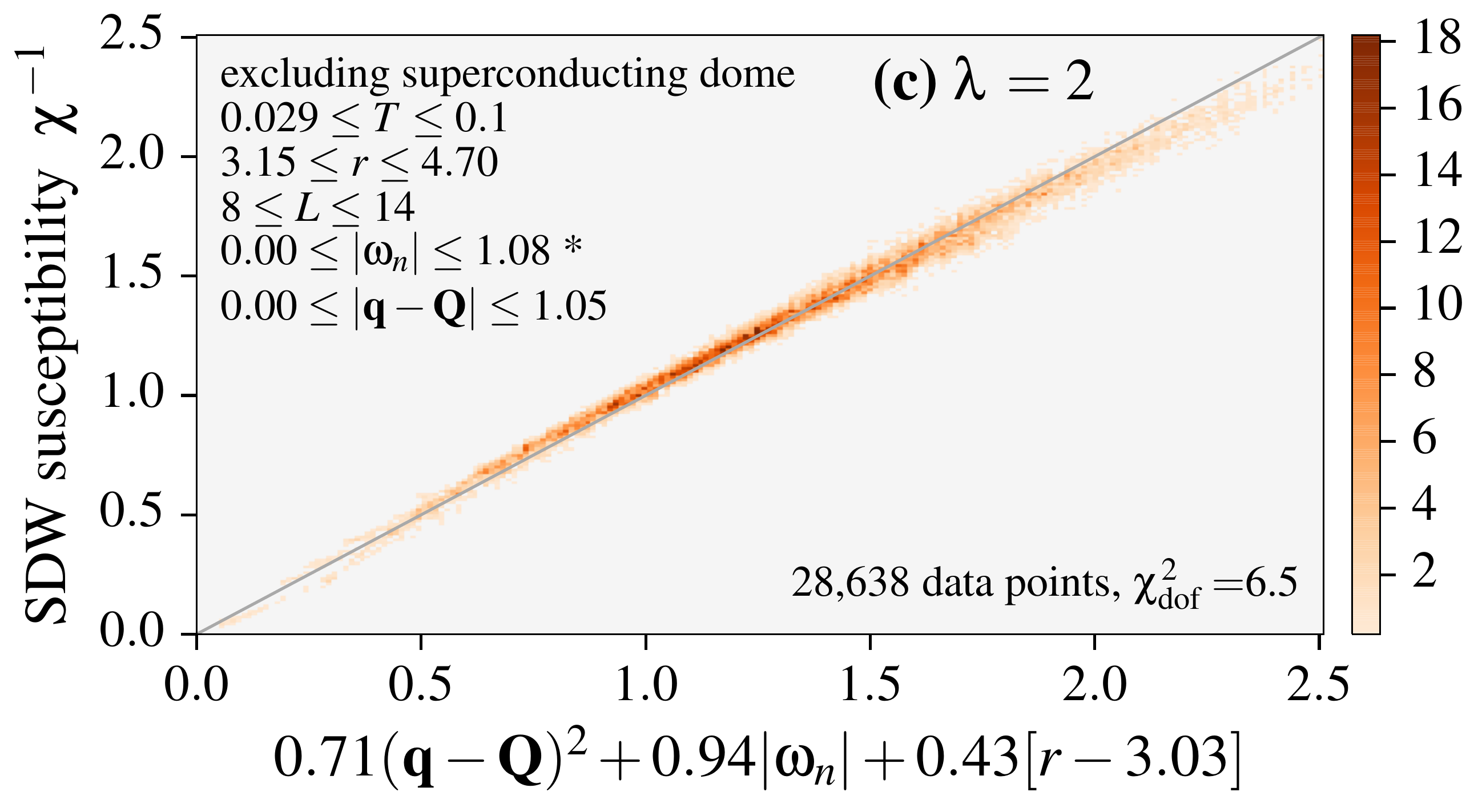}
  \caption{Comparison between the inverse SDW susceptiblity
    $\chi^{-1}$ and the functional form
    $\chi_0^{-1} = a_q (\mathbf{q} - \mathbf{Q})^2 + a_\omega|\omega_n| + a_r(r -
    r_{c0})$, which has been fitted for small frequencies $\omega_{n}$
    and momenta $\mathbf{q}-\mathbf{Q}$ at low temperatures $T$
    and tuning parameters $r > r_{c0}$ in the magnetically disordered phase,
     for (a)
    $\lambda=1$, (b) $\lambda=1.5$, and (c) $\lambda=2$.  Data inside
    the superconducting phase has been excluded from the fit.  For
    temperatures $T \leq 2 T_c^{\mathrm{max}}$  we restrict the fit
    to finite frequencies $|\omega_n| > 0$. The
    correspondence of $\chi^{-1}$ with the fitted form is shown in the
    form of 2D histograms over all data points, which are normalized
    over the total area.  In each fit we have minimized
    $\chi^2_{\mathrm{dof}} =\frac{1}{N_{\mathrm{dof}}} \sum \Big[
    \frac{\chi^{-1} - \chi^{-1}_0}{\varepsilon} \Big]^2$, 
    where $N_{\mathrm{dof}}$ is the number of degrees of freedom of
    the fit and $\varepsilon$ is the statistical error of the data.
  }
\label{fig:chi_collapses}
\end{figure}

Running extensive DQMC simulations for system sizes
$L=8,10, \ldots 14$, we have evaluated $\chi$ across the three
principal phase diagrams of Fig.~\ref{fig:phasediagrams} for different
values of the Yukawa coupling $\lambda = 1$, $1.5$, $2$ and bare
bosonic velocity $c=3$.  Restricting our analysis to the magnetically
disordered side for each coupling and to temperature scales above the
superconducting phase, we find that our calculated susceptibilities
are in good agreement with the functional form of
Eq.~\eqref{eq:hertz}. The consistency with
Eq.~\eqref{eq:chi_functional} is illustrated in the panels of
Fig.~\ref{fig:chi_collapses}, which show data collapses 
for a range of small momenta $\mathbf{q}-\mathbf{Q}$,
small Matsubara frequencies, low temperatures $T \leq 0.1$ and tuning
parameters $r \ge r_{c0}$.
Finite-size effects are rather small. Considering the
variation of the Yukawa coupling $\lambda$, we find that the fit to
the functional form \eqref{eq:chi_functional} is slightly worse for
stronger coupling $\lambda$, which is also indicated by the larger
spread of the data points. This decreasing fit quality may be a
consequence of the smaller temperature window available above the superconducting
$T_c$, as well as the associated regime of superconducting fluctuations at $T\gtrsim T_c$~\cite{Schattner2015a}, which
increases with Yukawa coupling (see also Fig.~\ref{fig:Tc_vs_lambda}).

\begin{figure}
  \centering
  \includegraphics[width=\linewidth]{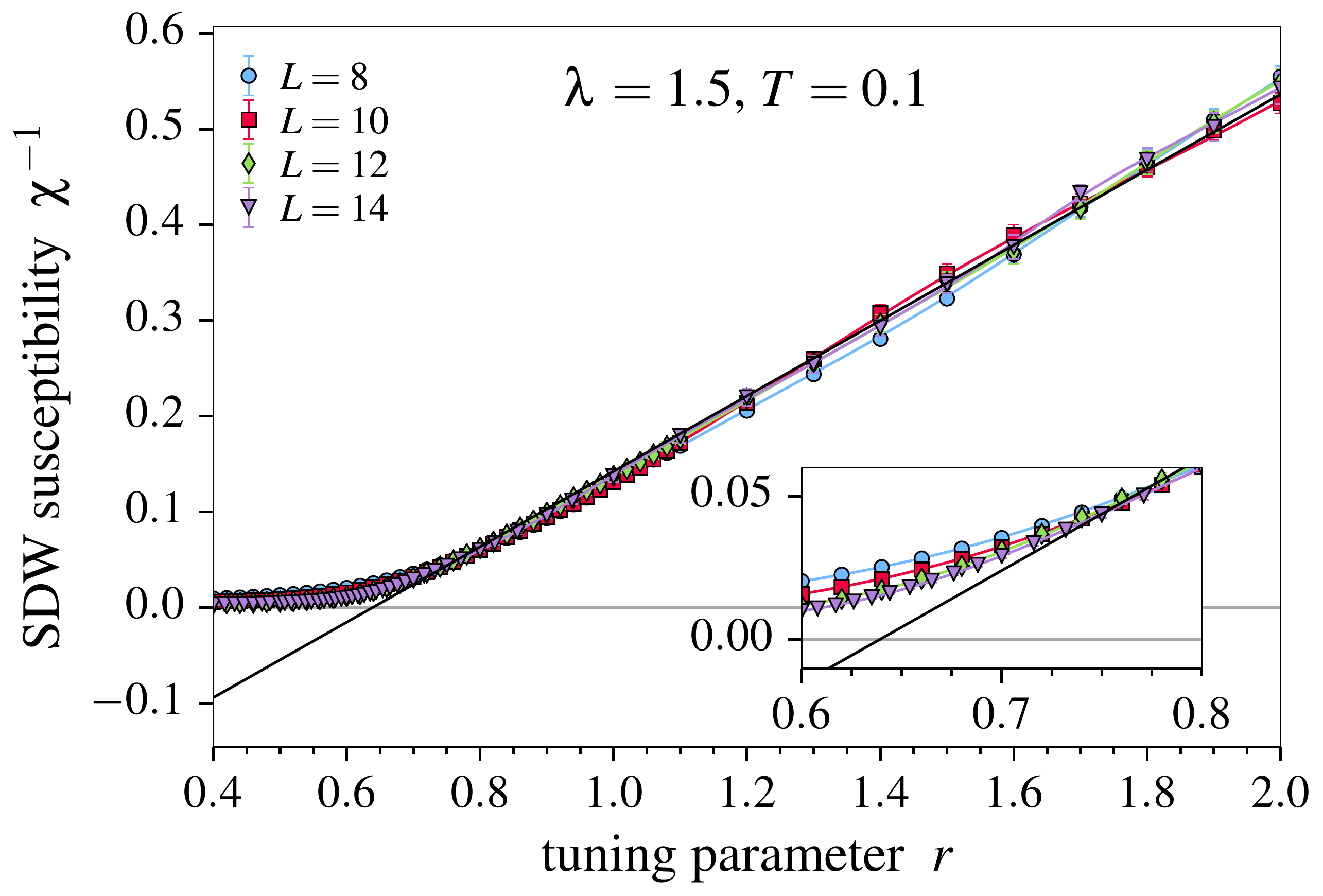}
  \caption{Bosonic SDW susceptibility
    $\chi^{-1}(\mathbf{q}=\mathbf{Q}, i\omega_n=0)$ as a function of the tuning
    parameter $r$ for $\lambda = 1.5$ at $T=0.1$.  The black line is a
    linear fit for $r > 0.7$ and $L=14$.  Continuous colored lines
    through data points have been obtained by a reweighting analysis.}
  \label{fig:chi_r_l1.5}
\end{figure}

With the data collapse of Fig.~\ref{fig:chi_collapses} 
asserting the general validity of the functional
form~(\ref{eq:chi_functional}), we now take a closer look at its
individual dependence on tuning parameter, frequency and momentum.
First, the dependence on the tuning parameter $r$ is
illustrated for the inverse susceptibility
$\chi^{-1}(\mathbf{q} = \mathbf{Q}, i \omega_n = 0)$ in
Fig.~\ref{fig:chi_r_l1.5} (for $\lambda = 1.5$ and $T = 0.1$).
For tuning parameters $r \gtrsim r_{c0} = 0.6$ we find that the
data for different system sizes follows a linear dependence.
The moderate deviation from a perfect kink-like behavior
at $r_{c0}$ is likely a combination of finite-size and finite-temperature
effects (see also the finite-size trend shown in the inset of
Fig.~\ref{fig:chi_r_l1.5}).  A very similar picture emerges for the
two other coupling parameters $\lambda = 1$ and $\lambda = 2$, for
which we show analogous plots in Fig.~\ref{fig:chi_r_l1_l2} of
Appendix~\ref{sec:magn-corr-at}.

\begin{figure}
  \centering
  \includegraphics[width=\linewidth]{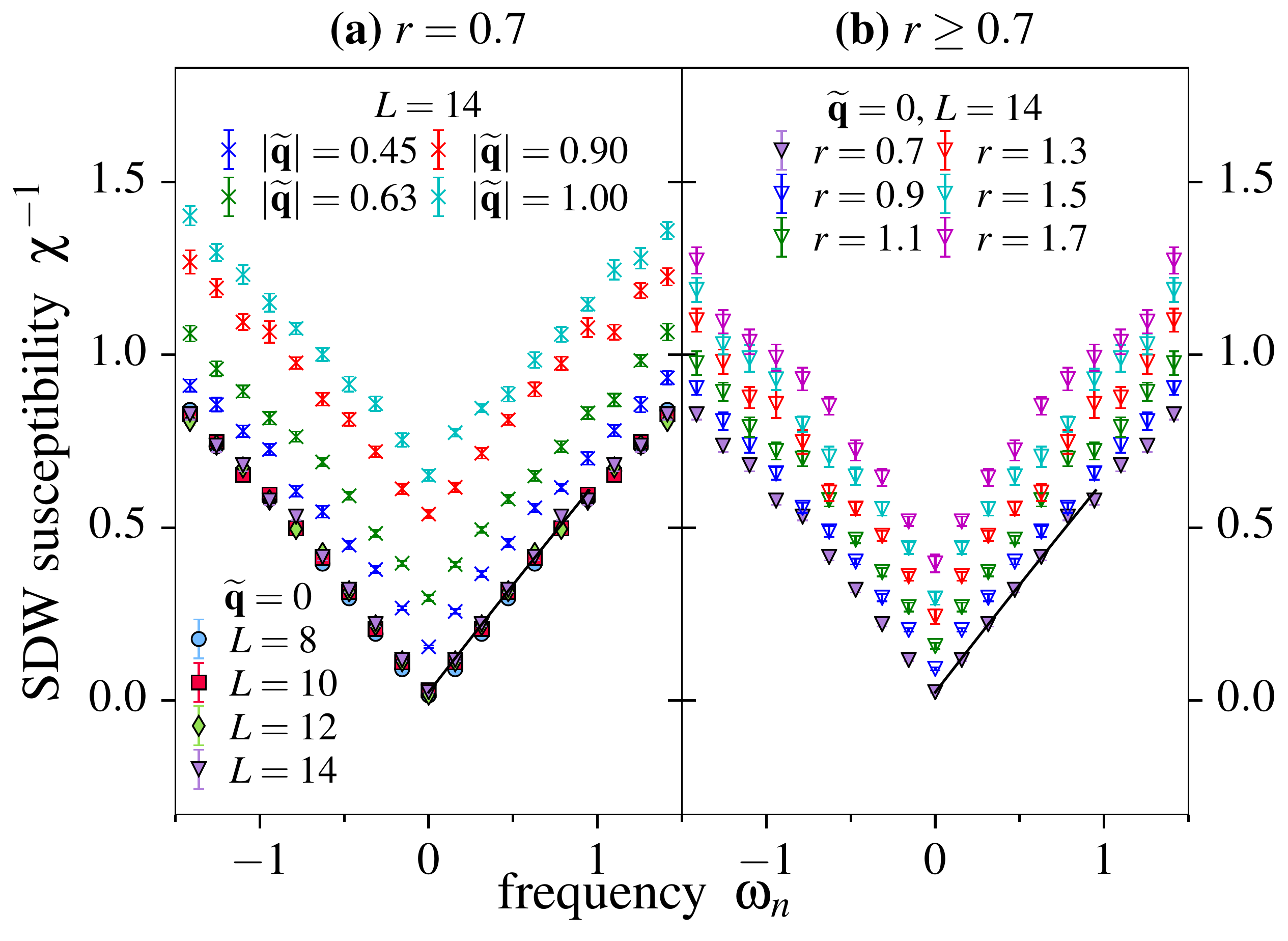}\\
  \caption{Frequency dependence of the inverse bosonic SDW
    susceptibility $\chi^{-1}$ for $\lambda = 1.5$ at $T=1/40$ (a)
    shown at $r \approx r_{c0}$ for various momenta
    $\mathbf{q} = \mathbf{Q}+\mathbf{\widetilde{q}}$ and (b) shown at
    various values $r > r_{c0}$ for $\mathbf{q} = \mathbf{Q}$.  The
    black line is the best fit of a second degree polynomial
    $b_0 + b_1 |\omega_n| + b_2 \omega_n^2$ to the
    $\mathbf{q}=\mathbf{Q}$, $L=14$ low-frequency data, yielding a
    basically straight line.}
  \label{fig:chi_freq_multir_l1.5}
\end{figure}

Turning to the frequency dependence of
$\chi^{-1}(\mathbf{q}, i \omega_n)$ next, we find that for
a range of values $r \ge r_{c0}$ the
frequency dependence is linear for small Matsubara frequencies
$\omega_n$ with an apparent cusp at $\omega_n=0$, signaling overdamped dynamics of the order parameter field.  This holds both for
$\mathbf{q}=\mathbf{Q}$ and for small finite momentum differences
$\mathbf{q} - \mathbf{Q}$.  See Fig.~\ref{fig:chi_freq_multir_l1.5}
for an illustration at $\lambda = 1.5$ and
Appendix~\ref{sec:magn-corr-at} with
Fig.~\ref{fig:chi_freq_multir_l1_l2} for $\lambda = 1$ and
$\lambda = 2$. At finite Matsubara frequencies $\omega_n$,
finite-size effects are negligibly small, as evident in the data
collapse of $\chi^{-1}$ for different system sizes in the left panel in
Fig.~\ref{fig:chi_freq_multir_l1.5}.

To establish the presence of a $|\omega_n|$ term in $\chi^{-1}$, we fit it at low frequencies to the form $b_0+b_1|\omega_n|+b_2 \omega_n^2$. The fits are shown in Fig.~\ref{fig:chi_freq_multir_l1.5}. The $|\omega_n|$ contribution is clearly dominant in this frequency range.
Inside the superconducting phase, the $|\omega_n|$ term is suppressed (see Fig.~\ref{fig:chi_freq_l3_c2} in Appendix~\ref{sec:magn-corr-at}). This is presumably due to gapping out of the fermions. 

\begin{figure}
  \centering
  \includegraphics[width=\linewidth]{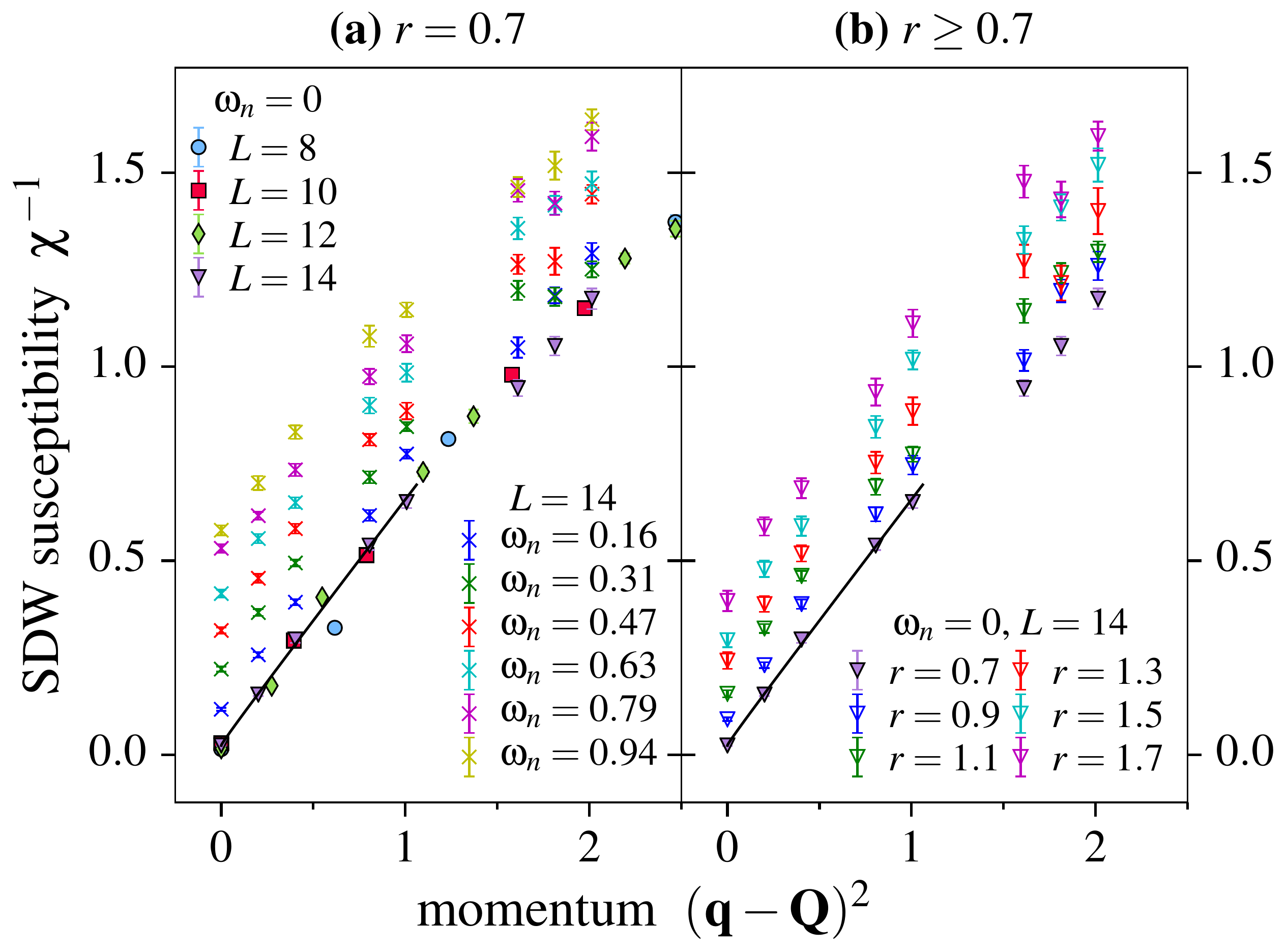}
  \caption{Inverse bosonic SDW susceptibility $\chi^{-1}$ as a
    function of momentum
    $\mathbf{q} = \mathbf{Q}+\mathbf{\widetilde{q}}$ for $\lambda=1.5$
    at $T=1/40$ (a) shown at $r \approx r_{c0}$ for various frequencies
    $\omega_n$ and (b) shown at various values $r \gtrsim r_{c0}$ for
    $\omega_n = 0$.  The black line is the best fit of
    $a_0 + a_2 \mathbf{\widetilde{q}}^2$ to the $\omega_n = 0$, $L=14$
    small-momentum data.}
  \label{fig:chi_mom_multir_l1.5}
\end{figure}

Third, for the same range of $r$ the momentum dependence
of $\chi^{-1}(\mathbf{q}, i \omega_n)$ is consistent with a quadratic form in
$\mathbf{q} - \mathbf{Q}$, which holds both for $\omega_n=0$ and small finite
frequencies $\omega_n$.  See Fig.~\ref{fig:chi_mom_multir_l1.5} for
$\lambda = 1.5$ and appendix~\ref{sec:magn-corr-at} with
Fig.~\ref{fig:chi_mom_multir_l1_l2} for $\lambda = 1$ and
$\lambda = 2$.  Note that due to the discretization of the Brillouin
zone finite-size effects are more pronounced here than for the
frequency dependence.

\subsection{Fermion bilinear SDW susceptibility}
\label{sec:ferm-sdw-susc}

\begin{figure}
  \centering  \includegraphics[width=\linewidth]{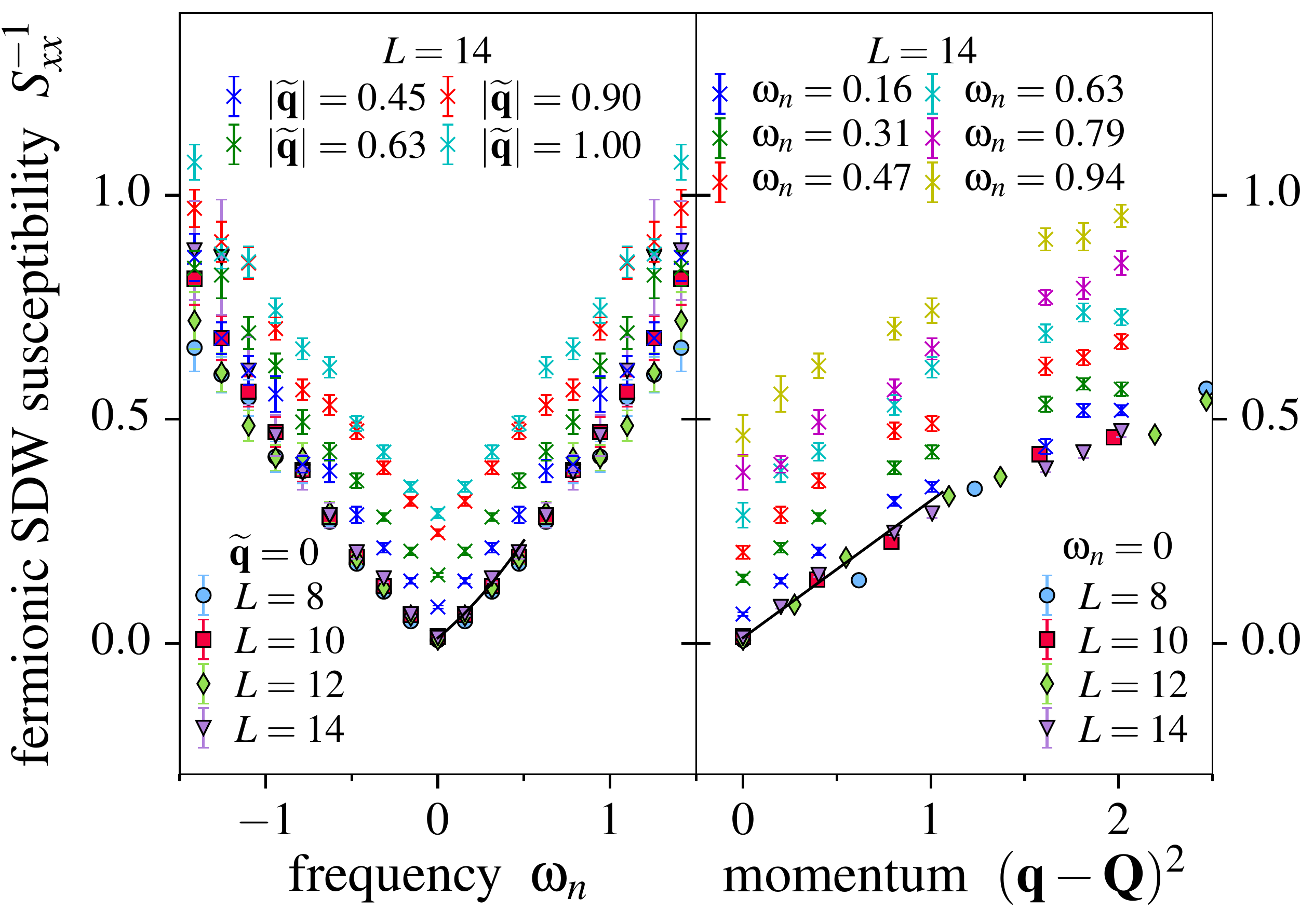}
  \caption{Inverse fermionic SDW susceptibility $S_{xx}^{-1}$ for
    $\lambda=1.5$ at $T=1/40$ and $r = 0.7 \approx r_{c0}$.  (Left hand side)
    Frequency dependence for various momenta
    $\mathbf{q} = \mathbf{Q}+\mathbf{\widetilde{q}}$.  The black line
    is a fit of the second degree polynomial
    $b_0 + b_1 |\omega_n| + b_2 \omega_n^2$ to the
    $\mathbf{q}=\mathbf{Q}$, $L=14$ low-frequency data, yielding a
    basically straight line.  (Right hand side) Momentum dependence
    for various frequencies $\omega_n$.  The black line is a fit of
    $a_0 + a_2 \mathbf{\widetilde{q}}^2$ to the $\omega_n = 0$, $L=14$
    small-momentum data.}
  \label{fig:chi_freq_mom_f_l1.5}
\end{figure}

An important independent confirmation that the form \eqref{eq:chi_functional}
is generic to the quantum critical regime is to affirm that it also holds for
other SDW order parameters that have the same symmetry.
We have examined the correlations of a {\it fermion} bilinear order parameter:
\begin{multline}
  \label{eq:sxx}
  S_{xx}(\mathbf{q}, i\omega_n, r, T) = \sum_i \int_0^{\beta}\!\!
  d\tau e^{i \omega_n \tau - i \mathbf{q} \cdot \mathbf{r}_i}
  \langle S_i^x(\tau) S_0^x(0) \rangle .
\end{multline}
In the estimation of $S_{xx}$ we make use of spin rotational symmetry
around the $z$ axis,
$\langle S_i^x(\tau) S_0^x(0) \rangle = \langle S_i^y(\tau) S_0^y(0)
\rangle$.  Here $S_i^x$ and $S_i^y$ are inter-flavor fermion spin
operators, which are given by
\begin{align}
  \label{eq:vecs}
  \vec{S}_i = (S_i^x, S_i^y, S_i^z)
  = \sum_{s,s'} \vec{s}_{ss'} \psi^{\dagger}_{xis} \psi_{yis'} + \text{ h.c.}
\end{align}
Indeed, we find that at small frequencies
and momenta, the fermion bilinear SDW susceptibility $S_{xx}$
follows the same functional form \eqref{eq:chi_functional}
as the bosonic SDW susceptibility $\chi$ discussed above.
The momenta and frequency dependences of the fermionic bilinear susceptibility at $\lambda=1.5$ are shown in
Fig.~\ref{fig:chi_freq_mom_f_l1.5}, with the respective dependences
of the bosonic susceptibility appearing in Figs.~\ref{fig:chi_freq_multir_l1.5} and
\ref{fig:chi_mom_multir_l1.5}.
Additional data for the
fermionic SDW susceptibility at $\lambda = 1$ and $\lambda = 2$ is
given in Fig.~\ref{fig:chi_freq_mom_f_l1_l2} of
Appendix~\ref{sec:magn-corr-at}.

In summary, the dependence of both the bosonic and fermionic SDW susceptibilities
on the tuning parameter, frequency, and momentum stand in
good agreement with the form \eqref{eq:chi_functional}.

\subsection{Temperature dependence}
\label{sec:temp-depend}

\begin{figure}
  \centering
  \includegraphics[width=\linewidth]{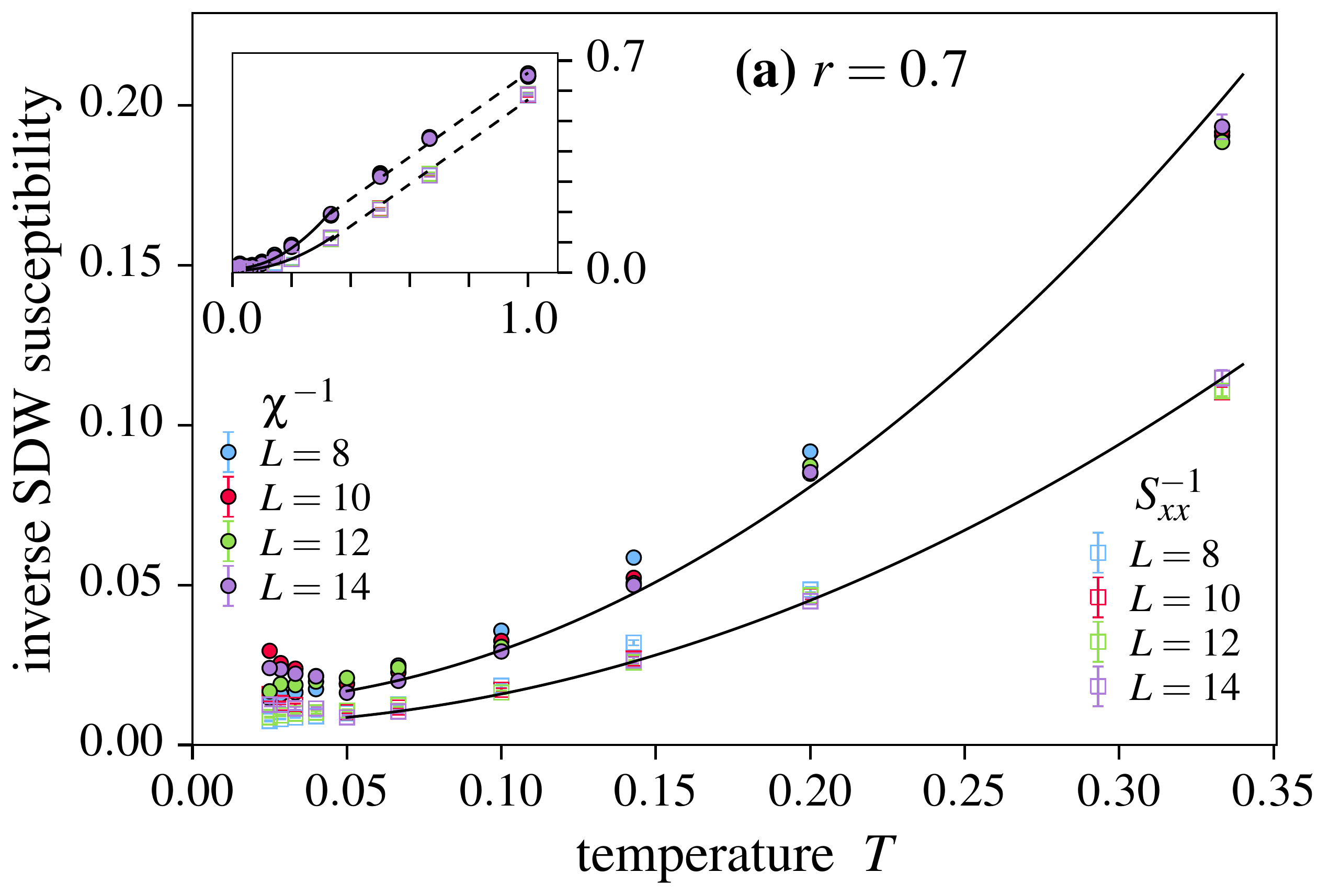}\\
  \includegraphics[width=\linewidth]{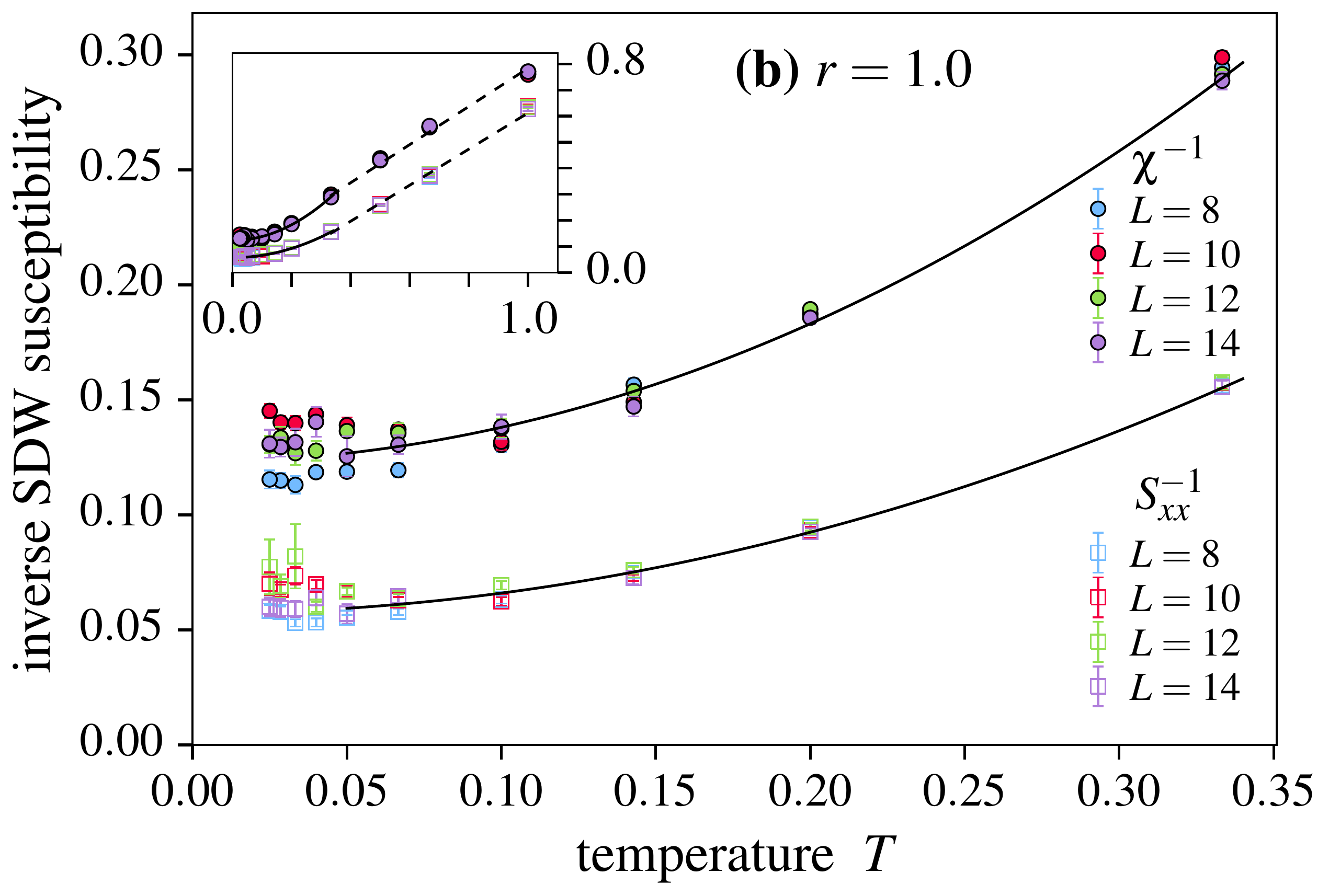}
  \caption{Inverse bosonic SDW susceptibility
    $\chi^{-1}(\mathbf{q}=\mathbf{Q}, i\omega_n=0)$ and inverse
    fermionic SDW susceptibility
    $S_{xx}^{-1}(\mathbf{q}=\mathbf{Q}, i\omega_n=0)$ as a function of
    temperature for $\lambda = 1.5$ at (a) $r = 0.7 \approx r_{c0}$ and
    at (b) $r = 1.0 > r_{c0}$.  Solid lines indicate fits of
    $a_0 + a_2 T^2$ to the $L=14$ data at intermediate temperatures.
    Dashed lines are linear fits to the high-temperature data.  In
    each figure the inset shows the same data as the main plot over a
    more extended temperature range. }
  \label{fig:chi_temp_l1.5}
\end{figure}

We now turn to the temperature dependence of the numerically computed
bosonic and fermionic SDW susceptibilities $\chi^{-1}$ and
$S_{xx}^{-1}$. Our
numerical data for the temperature dependence of $\chi^{-1}$ and
$S_{xx}^{-1}$ is shown in Fig.~\ref{fig:chi_temp_l1.5} for fixed Yukawa coupling
$\lambda=1.5$ and two different values of the tuning parameter on the
paramagnetic side of the QCP, i.e. for $r > r_{c0}$.  This data is
complemented with similar results for $\lambda=1$ and $\lambda = 2$ in
Fig.~\ref{fig:chi_temp_l1_l2} of Appendix~\ref{sec:magn-corr-at}.

Evidently, the data shows different scaling regimes with increasing
temperature.  At sufficiently high temperatures, $T \gtrsim 0.35$ the
susceptibilities $\chi^{-1}(\mathbf{q} = \mathbf{Q}, i\omega_n=0)$ and
$S^{-1}_{xx}(\mathbf{q} = \mathbf{Q}, i\omega_n=0)$ are approximately
linearly dependent on temperature,  as shown in the insets of
Fig.~\ref{fig:chi_temp_l1.5}.  In an intermediate temperature regime,
however, we observe a crossover to a different functional temperature
dependence as shown in the main panels of
Fig.~\ref{fig:chi_temp_l1.5}.  In this intermediate temperature window
$0.05 \lesssim T \lesssim 0.35 $ our numerical data is found to
reasonably fit functions of the {\it quadratic} form
$a_0 + a_2 T^{\alpha}$ with $\alpha \simeq 2 \pm 0.3$.
Unlike the leading
dependences on the tuning parameter, frequency and momentum discussed
in the previous section, this power-law dependence is not as
robust.
Note that the crossover temperature between the high-$T$ linear and
intermediate-$T$ quadratic behaviors does not depend strongly on the
tuning parameter $r$.  Notably, even for $r \approx r_{c0}$ this
intermediate regime does not disappear.

At still lower temperatures $T \lesssim 0.05 $ our data might indicate
a second crossover to yet different behavior. With the tuning
parameter $r$ tuned close to its critical value $r_{c0}$ both $\chi$
and $S_{xx}$ are found to be non-monotonic for the smallest
temperatures and largest system sizes accessed in this study. The
apparent upturn, whose precise location is hard to determine due to
finite-size effects (which are strongest for $r\approx r_{c0}$) and
the enhanced statistical uncertainty at low temperatures, is most
likely a precursor effect of superconductivity \cite{Schattner2015a},
which for $\lambda=1.5$ sets in just at the lowest temperature we have
accessed in this work, $T_c \approx 1/40$.  For larger Yukawa coupling
$\lambda = 2$, where $T_c$ is higher, this non-monotonic behavior is
indeed found to be more pronounced as shown in
Fig.~\ref{fig:chi_temp_l1_l2} of appendix~\ref{sec:magn-corr-at}.

Note that over the range of temperatures displayed in
Fig.~\ref{fig:chi-inv-color-l1} the leading temperature dependence of $\chi^{-1}$
is quadratic.  This is reflected in the contour lines of
$\chi^{-1}$ in the $r-T$ plane, which have a form
$T \sim \sqrt{\chi^{-1} - a_r(r - r_{c0})}$, approaching
infinite slope at low temperatures.

Since the data does not allow us to
identify a simple functional form for the temperature dependence of $\chi^{-1}$, we have opted
against taking into account any temperature dependence in the fits for
the data collapses shown in Fig.~\ref{fig:chi_collapses}.  Instead we
have constrained the included data to $T < 0.1$ where the overall
temperature dependence is rather weak.

\section{Single-Fermion correlations}
\label{sec:fermion_corr}

We now turn to examine the fermionic spectral properties in the
metallic state above the superconducting $T_c$.  As our DQMC
simulations are performed in imaginary time, there is an inherent
difficulty in probing real-time dynamics.
To partially circumvent this issue, we use the relation
\cite{Trivedi1995}
\begin{equation}
  G_{\mathbf{k}}(\tau)=\int_{-\infty}^\infty d\omega
  \frac{e^{-\omega(\tau-\beta/2)}}{2\cosh{\beta\omega/2}}A_{\mathbf k}(\omega) \,,
  \label{eq:G_tau}
\end{equation}
which connects the readily available imaginary-time ordered Green's
function
$G_{\mathbf{k}}(\tau) = \langle \psi_{\mathbf{k}}(\tau)
\psi_{\mathbf{k}}^{\dagger}(0) \rangle$, where $0 \le \tau \le \beta$, with the spectral
function $A_{\mathbf{k}}(\omega)$ of interest. Here and in the following, we
focus on a single flavor of fermions $\psi_y$, suppressing band and spin
indices.  Close inspection of \eqref{eq:G_tau} reveals that the
behavior of the Green's function $G_\mathbf{k}(\tau)$ at long times,
i.e for imaginary times close to $\tau=\beta/2$, provides information
about the spectral function integrated over a frequency window of
width $T$.

\begin{figure*}[th]
  \includegraphics[width=\textwidth]{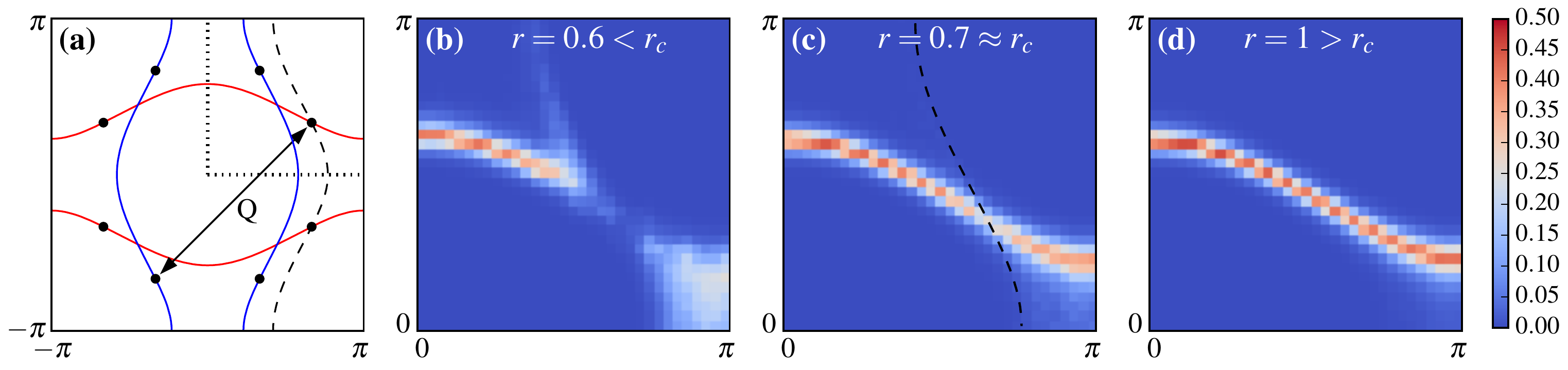}
  \caption{ (a) Noninteracting Fermi surfaces.  A pair of hot spots
    is connected by the magnetic ordering wavevector $\mathbf{Q}$.
    The dashed curve corresponds to the Fermi surface of the $\psi_x$
    band, shifted by $\mathbf{Q}$, with a hot spot now at the
    intersection with the $\psi_y$ band (b-d) Color-coded Green's
    function $G_{\mathbf{k}}(\tau=\beta/2)$ evaluated for the $\psi_y$
    fermions on a quadrant of the Brillouin zone, dotted in (a), for
    three values of the tuning parameter $r$.  The dashed curve in
    panel (c) corresponds to the shifted noninteracting $\psi_x$
    Fermi surface.  The parameters used here are $L=16$, $T=0.05$,
    $\lambda=1.5$, and $c=3$.  Results of simulations with different
    boundary conditions are combined for enhanced momentum resolution.
  }
  \label{fig:FS}
\end{figure*}
In Fig.~\ref{fig:FS} we present the evolution of the Fermi surface
across the phase diagram. For orientation, the Fermi surfaces of the
noninteracting system are shown in panel (a). In panels (b-d) we show
$G_{\mathbf{k}}(\tau=\beta/2)$ across a quadrant of the Brillouin zone
for a low temperature $T = 0.05 \approx 2 T_c^{\mathrm{max}}$.  Near the magnetic QCP (panel c),
there is a clear loss of spectral weight in the immediate vicinity of
the hot spots, as compared to the magnetically disordered phase (panel
d). Upon entering the magnetic phase (panel b), a gap opens around the
hot spots. In this section we focus on the parameter set $\lambda = 1.5$ and $c=3$.
The DQMC simulations are carried out with different sets of twisted boundary conditions (see Appendix
\ref{sec:DQMCAppendix} for details), providing a four-fold enhancement in $\mathbf{k}$-space resolution.

A Fermi liquid is usually characterized by the quasiparticle weight $Z_{\mathbf{k}_F}$ and the
Fermi velocity $\mathbf{v}_{\mathbf{k}_F}$. We note that these  quantities are only strictly
defined at zero temperature. Given that the zero-temperature ground state of our model is probably always superconducting,
our strategy is to consider finite-temperature proxies for $Z_{\mathbf{k}_F}$ and $\mathbf{v}_{\mathbf{k}_F}$,
and study their behavior over an intermediate temperature range $E_F> T > T_c$.
Such proxies, $Z^\tau_{\mathbf{k}_F}(T)$ and $\mathbf{v}^\tau_{\mathbf{k}_F}(T)$, can be extracted by considering the
imaginary time dependence of $G_\mathbf{k}(\tau)$ near $\tau=\frac{\beta}{2}$ and fitting it to the Fermi liquid form~\cite{Schattner2015}
\begin{equation}
G_\mathbf{k} (\tau\sim{\beta/2}) = Z^\tau_{\mathbf{k}}(T) \frac{e^{-\epsilon_{\mathbf{k}} \left(\tau-\frac{\beta}{2}\right)}}{2\cosh\left(\frac{\beta\epsilon_{\mathbf{k}}}{2}\right)},
\end{equation}
where $\epsilon_\mathbf{k}=\mathbf{v}^\tau_{\mathbf{k}_F}(T) \cdot (\mathbf{k}-\mathbf{k}_F)$.

In a complementary approach we consider the Matsubara frequency
dependence of the Green's function
$G_{\mathbf{k}}(\omega_n) = \int_0^{\beta}\!d\tau\, e^{i
  \omega_n \tau} G_{\mathbf{k}}(\tau)$. In a Fermi liquid at low
temperatures we have~\cite{FradkinBook}
\begin{equation}
  G_{\mathbf{k}}(\omega_n) \approx Z_{\mathbf k} \left[i\omega_n -\mathbf{v}_{\mathbf{k}_F} \cdot (\mathbf{k}-\mathbf{k}_F) \right]^{-1}
  \label{eq:G_FL}
\end{equation}
up to higher order terms in temperature, frequency or the distance from the Fermi surface.
It is then natural to define the finite-temperature quantities
\begin{equation}
  Z^\omega_{\mathbf{k}_F}(T)= \frac{\omega_1}{\Im G^{-1}_{\mathbf{k}_F}(\omega_1)}
  \label{eq:Z-proxy}
\end{equation}
and
\begin{equation}
  \mathbf{v}^\omega_{\mathbf{k}_F}(T) = \omega_1 \at{\frac{\partial}{\partial_{\mathbf{k}}} \frac{\Re G_{\mathbf{k}}(\omega_1)}{\Im G_{\mathbf{k}}(\omega_1)}}{\mathbf{k}=\mathbf{k}_F} \,,
  \label{eq:v-proxy}
\end{equation}
where $\omega_1 = \pi T$ is the first Matsubara frequency at
temperature $T$. In the zero temperature limit,
$ Z^\omega_{\mathbf{k}_F}(T \rightarrow 0) = Z^\tau_{\mathbf{k}_F}(T \rightarrow 0) = Z_{\mathbf{k}_F}$,
and similarly for $\mathbf{v}_{\mathbf{k}_F}$. We therefore use the finite-temperature observables \eqref{eq:Z-proxy} and \eqref{eq:v-proxy} as alternative proxies for the quasiparticle spectral weight and Fermi velocity, respectively.
\begin{figure}[t]
  \includegraphics[width=\columnwidth]{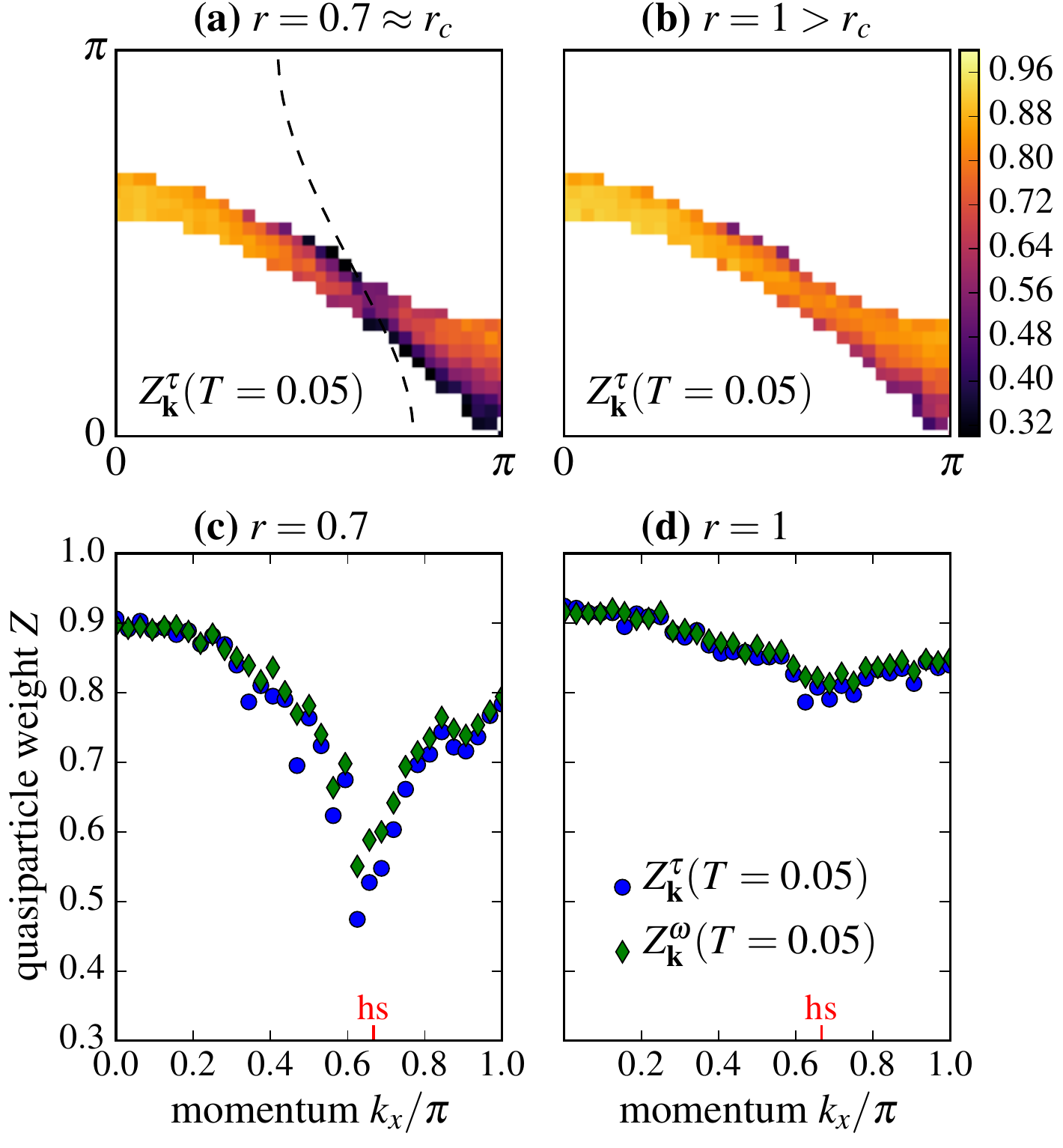}
  \caption{
    (a-b) The quasiparticle weight $Z^\tau_{\mathbf{k}} (T=0.05)$ in a quadrant of the Brillouin zone.
    The dashed line in panel (a) corresponds to the noninteracting Fermi surface of the $\psi_x$ fermions, shifted by $\mathbf{Q}$.
    (c-d) The quasiparticle weights $Z^\tau_{\mathbf{k}} (T=0.05)$ and $Z^\omega_{\mathbf{k}} (T=0.05)$ along the Fermi surface.
    The location of the hot spot is indicated by the red
    marker.  Here we show data obtained for $L=16$. }
  \label{fig:Z_vs_k}
\end{figure}

Figure \ref{fig:Z_vs_k} shows the momentum dependence of $Z^\tau_{\mathbf{k}}$
for temperature $T=1/20$.
With $r$ tuned close to the location of the QCP at $r_c$,
$Z^\tau_{\mathbf{k}}$ is suppressed in the vicinity of the hot spots,
as shown for one quadrant of the Brillouin zone in
Fig.~\ref{fig:Z_vs_k}(a) and along the Fermi surface in
Fig.~\ref{fig:Z_vs_k}(c).  This
stands in sharp contrast to the featureless behavior of
$Z^\tau_{\mathbf{k}}$ in the magnetically disordered
phase, as shown in Figs.~\ref{fig:Z_vs_k}(b,d).  We find qualitative
agreement between the two proxies $Z^\tau_{\mathbf{k}}$ and
$Z^\omega_{\mathbf{k}}$ throughout, as illustrated in panels (c) and
(d) of Fig.~\ref{fig:Z_vs_k}.
Here, we numerically identify and track the Fermi surface as the maxima of
$G_\mathbf{k} (\tau=\beta/2)$ at fixed $k_x$.

\begin{figure}[b]
  \includegraphics[width=\columnwidth]{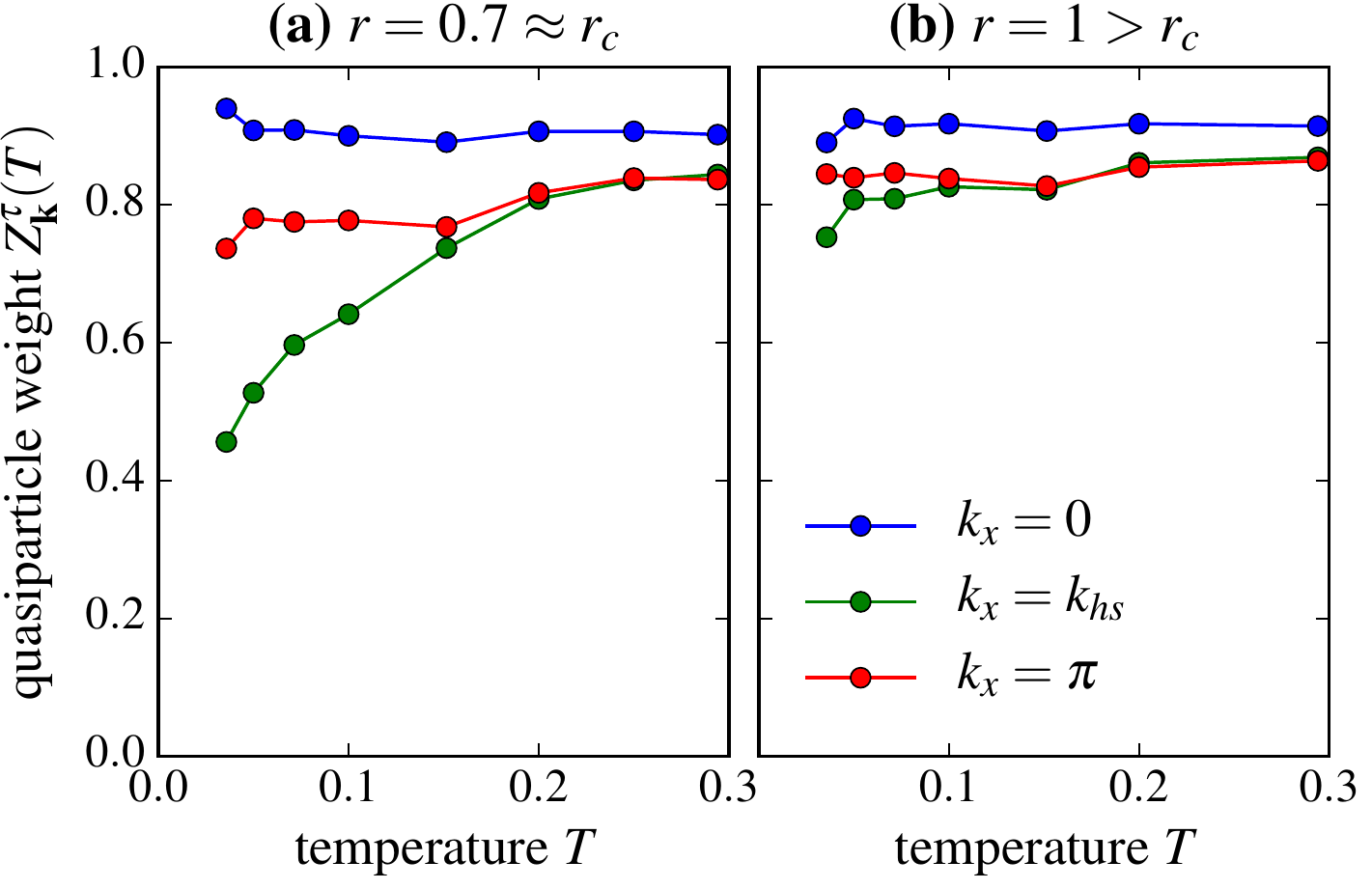}
  \caption{ Temperature dependence of the quasiparticle weight
    $Z^{\tau}_{\mathbf{k}}(T)$ for different momenta $k_x$ along the Fermi surface
    (a) in the vicinity of the QCP at $r_c$ and (b) in the disordered side.
     Here we show data obtained for $L=16$.}
  \label{fig:Z_vs_T}
\end{figure}

\begin{figure}[t!]
  \includegraphics[width=\columnwidth]{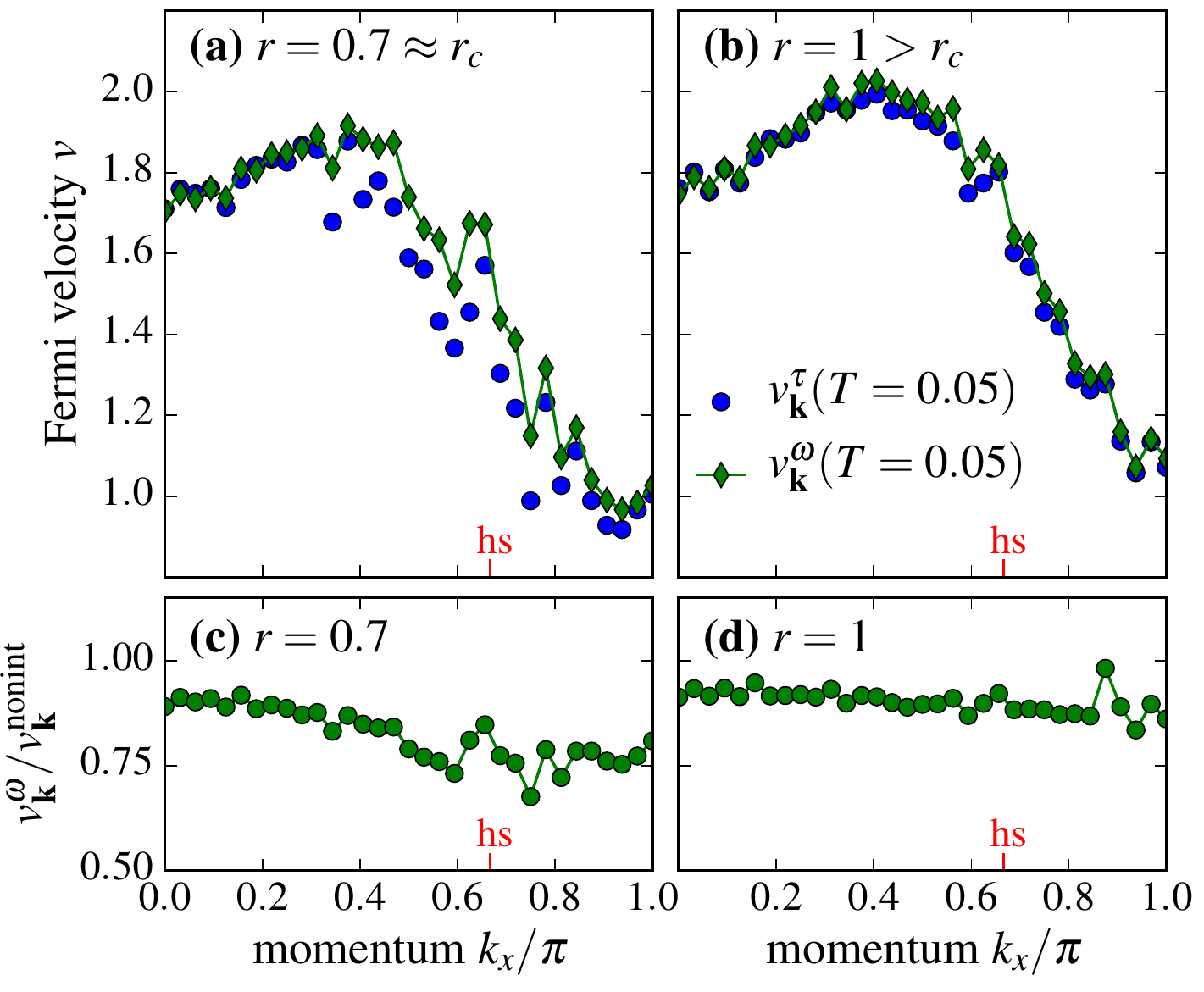}
  \caption{ (a)-(b) The finite-temperature proxies to the velocity
  	$v_{\mathbf{k}}$ along the Fermi surface.
    (c)-(d) The velocity renormalization
    $v^\omega_{\mathbf{k}}(T=0.05)/v^{\textrm{nonint}}_{\mathbf{k}}$.
    The location of the hot spot is indicated by the red marker.
    Here we show data obtained for $L=16$ at $T=0.05$.}
  \label{fig:vf_vs_k}
\end{figure}

The temperature dependence of $Z^\tau_{\mathbf{k}}(T)$ is shown in
Fig.~\ref{fig:Z_vs_T}.  For momenta away from the hot spots,
we find $Z^\tau_\mathbf{k}(T)$  to be nearly flat in temperature and to
approach a constant as $T\rightarrow 0$. A different picture emerges
at the hot spots, i.e. for $\mathbf{k} = \mathbf{k}_{hs}$.  Here
$Z^\tau_{\mathbf{k}=\mathbf{k}_{hs}}(T)$ remains flat only in the
magnetically disordered phase $r>r_c$, whereas the quasiparticle weight
$Z^\tau_{\mathbf{k}=\mathbf{k}_{hs}}(T)$ decreases substantially at
the critical coupling $r_c\approx 0.7$ as the temperature is lowered
towards the QCP, see Fig.~\ref{fig:Z_vs_T}(a). While our numerical
data does not allow for a simple extrapolation towards $T=0$, the
results are not inconsistent with a vanishing of
$Z_{\mathbf{k}=\mathbf{k}_{hs}}$,
indicating a breakdown of Fermi-liquid behavior at this point.

\begin{figure}[th!]
  \includegraphics[width=\columnwidth]{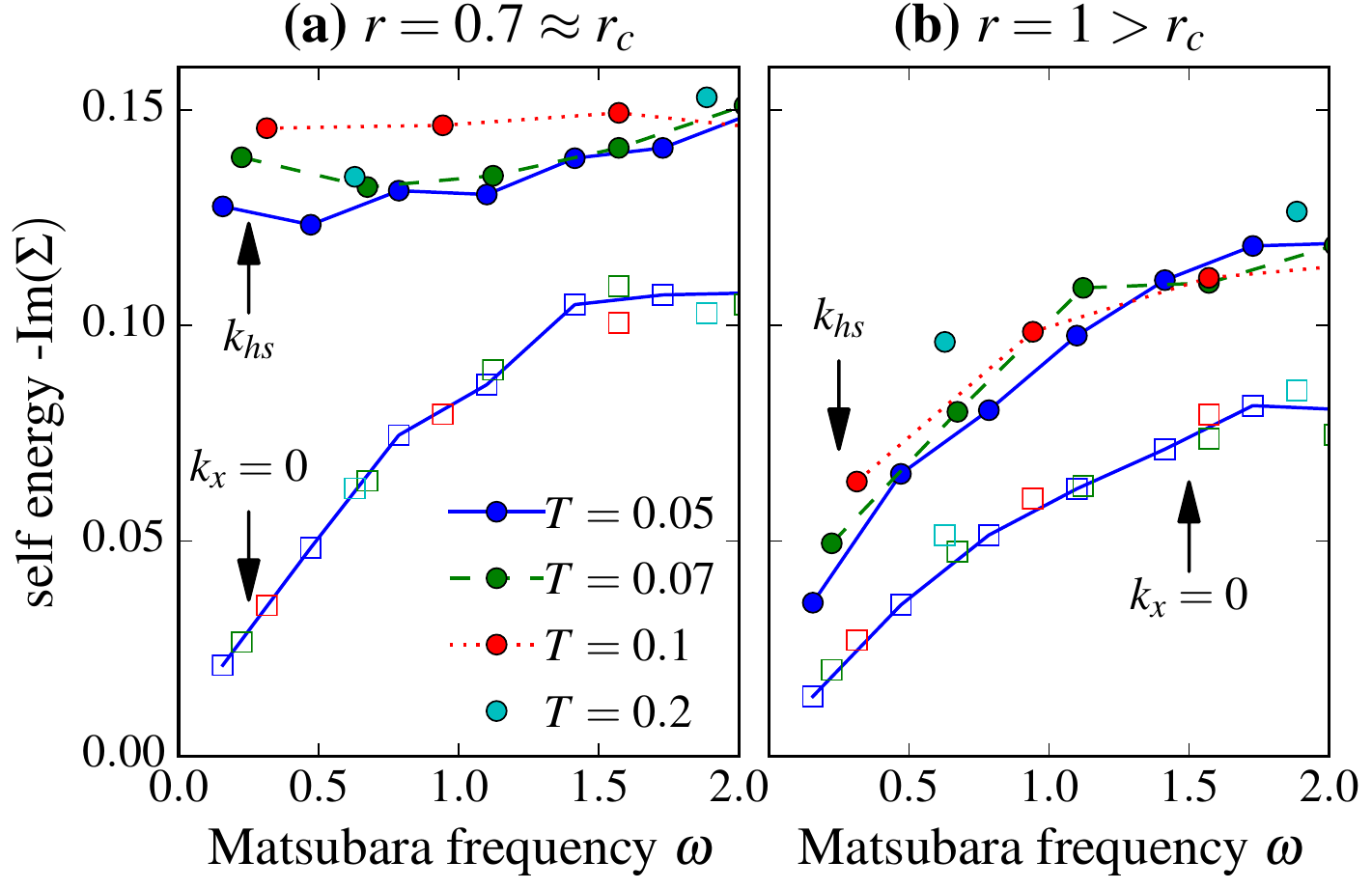}
  \caption{ The imaginary part of the Matsubara self energy $\Im \Sigma$
  for different temperatures and two momenta along the Fermi surface,
    (a) in the vicinity of the QCP at $r_c$ and (b)
    on the disordered side.  Here we show data obtained for
    $L=14$. The data for $\mathbf{k}=\mathbf{k}_{hs}$ is indicated by full circles,
    the momentum away from the hotspot is indicated by empty squares.}
  \label{fig:self_energy}
\end{figure}

Figure \ref{fig:vf_vs_k} shows the velocity $v_{\mathbf{k}_F}(T=0.05)$ along the Fermi surface.
The qualitative behavior of the velocity does not differ substantially between $r=0.7\approx r_c$
[Fig.~\ref{fig:vf_vs_k}(a)]
and $r=1>r_c$ [Fig.~\ref{fig:vf_vs_k}(b)]. The insets show the ratio $\mathbf{v}_{\mathbf{k}_F}(T=0.05)/v^{\textrm{nonint}}_{\mathbf{k}}$, where $v^{\textrm{nonint}}_{\mathbf{k}}$ is the Fermi velocity of the noninteracting system. A small feature might be visible in the vicinity of the
hot spot, but there is certainly no evidence of a substantial suppression of $v_{\mathbf{k}_{hs}}$ close to $r_c$.

Having found a substantial suppression of the quasiparticle weight
tuned close to the QCP, we now directly examine the
frequency dependence of the self energy, $\Sigma_\mathbf{k}(\omega_n)$, defined via
$G_{\mathbf k}(\omega_n)=\left(i\omega-\epsilon_{\mathbf k} -\Sigma_{\mathbf k}(\omega_n)\right)^{-1}$.
The imaginary part of the self energy is shown in Fig.~\ref{fig:self_energy}.
With $r=0.7\approx r_c$ tuned close to the QCP, see
Fig.~\ref{fig:self_energy}(a), the self energy close to the hot spots
is found to be nearly frequency independent, consistent with a \emph{constant}, yet small,
scattering rate $\gamma =-\Im \Sigma_{\mathbf{k}_{hs}}(\omega_n\rightarrow 0^+) \approx 0.13$ which is only
weakly dependent on temperature.
Away from the hot spot the self energy is linear with frequency, with a substantially
smaller intercept.
Moving away from the critical point [Fig.\ref{fig:self_energy}(b)], the self energy
at all momenta decreases rapidly as the frequency is lowered.

\section{Superconducting state}
\label{sec:SC}

After concentrating our discussion on the manifestation of quantum
critical behavior in the normal state, we now consider the effect of the QCP on the superconducting state that emerges in
its vicinity.

\begin{figure}[t!]
  \centering
  \includegraphics[width=\columnwidth]{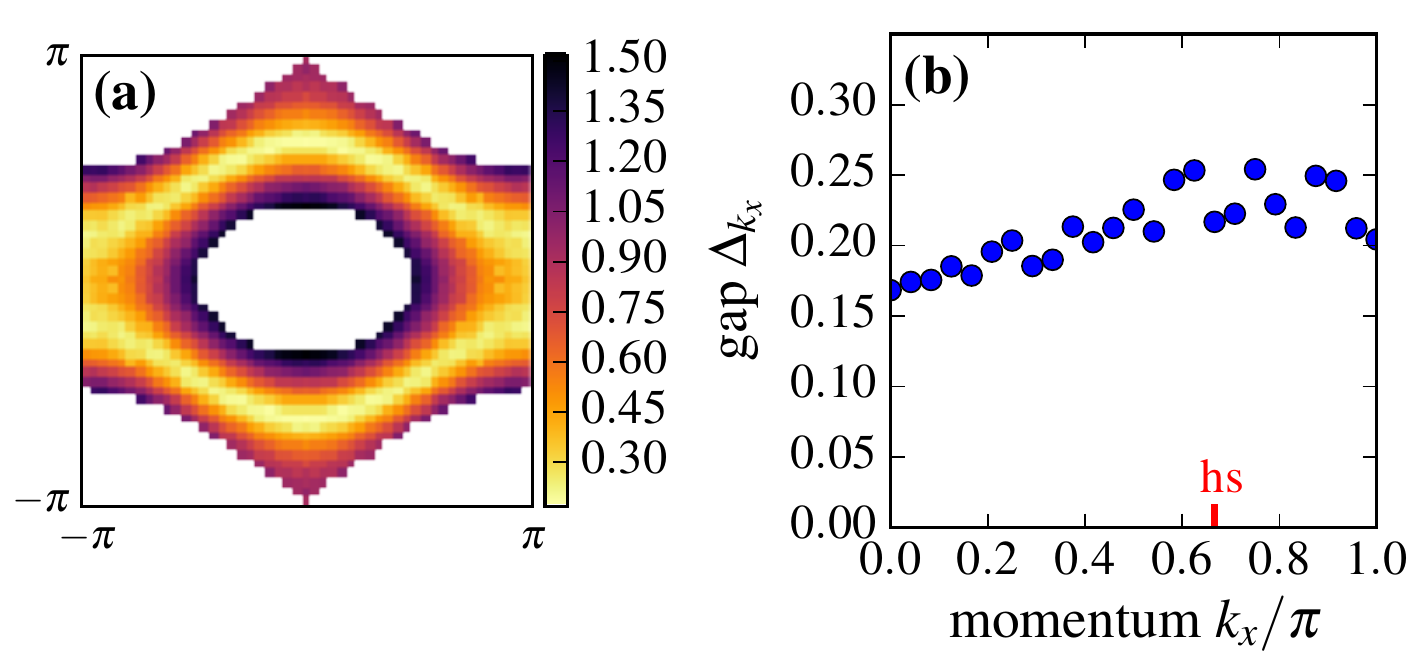}
  \caption{ (a) Single-particle excitation energy $E_{\mathbf{k}}$ of
    the $\psi_y$ fermions, as extracted from the imaginary-time
    evolution of the Green's function $G_{\mathbf{k}}(\tau)$ across
    the Brillouin zone, cf. Fig.~\ref{fig:Ek_from_Gtau_fits} of the
    appendix.  (b) Single-particle gap $\Delta_{k_x}$.
    For both panels data is for parameters
    $\lambda=3$, $c=2$, $r=10.2$ and
    $T=0.025\approx 0.3~T_c$~\cite{Schattner2015a} and a system size
    of $L=12$.  Several twisted boundary conditions were combined for
    a four-fold enhancement of the resolution in $\mathbf{k}$-space,
    see appendix \ref{sec:DQMCAppendix}.  }
  \label{fig:Ek_from_Gtau}
\end{figure}

We begin by considering the fermionic Green's function for temperatures $T\ll T_c$. In this regime, the single-particle excitation energy $E_{\mathbf{k}}$ can be extracted, as demonstrated in appendix \ref{sec:Delta}, from the decay of the single-particle Green's function at intermediate
times $\tau_0 < \tau < \beta/2$ (where $\tau_0$ is a microscopic scale).
The so-extracted excitation energy $E_{\mathbf{k}}$ is plotted in Fig.~\ref{fig:Ek_from_Gtau}, which shows that, across the Brillouin zone, $E_{\mathbf{k}}$ has a broad minimum in the vicinity of the noninteracting Fermi surface.
From these momentum-resolved energy bands we extract the superconducting gap $\Delta_{k_x}$
as the minimum of $E_{\mathbf{k}}$ with respect to $k_y$.
As seen in Fig.~\ref{fig:Ek_from_Gtau}~(b),  the superconducting gap $\Delta_{k_x}$ varies smoothly across momentum space, without any significant features at the hot spots.
In this section we choose parameters $\lambda = 3$ and $c = 2$, as in
Ref.~\cite{Schattner2015a}. The maximal $T_c$ for this value of
$\lambda$ is high enough to allow us to explore properties of the
superconducting state significantly below $T_c$.

\begin{figure}[t!]
  \centering
  \includegraphics[width=\columnwidth]{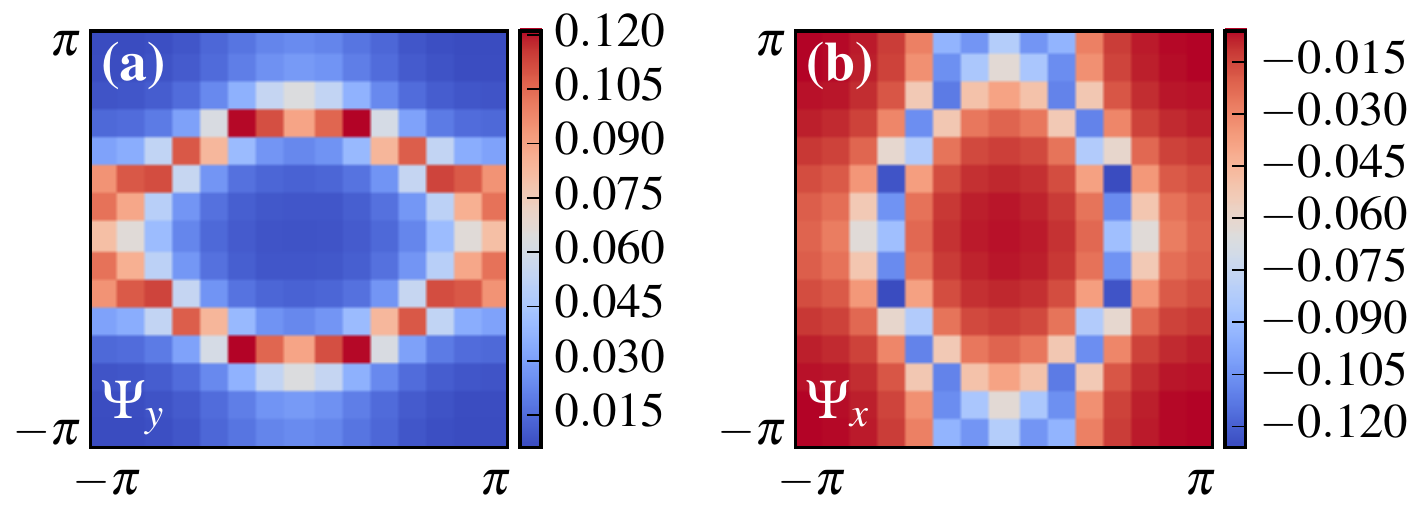}
  \caption{Optimal pair amplitude  $\Psi^{\textrm{opt}}_{\mathbf{k}\alpha}$
  		with the band $\alpha=y$ shown in panel (a) and the band $\alpha=x$ shown in (b).
		Data is calculated for parameters $\lambda=3$, $c=2$, $r=10.2$ and $T=0.1\approx 1.2~T_c$ and system size $L=14$.\
		}
  \label{fig:pairing_P}
\end{figure}

At higher temperatures, close to $T_c$, additional information can be obtained by considering the momentum-resolved superconducting susceptibility
\begin{equation}
  P_{\mathbf{k}\alpha; \mathbf{k'}\alpha'} = \int_0^\beta\! d\tau\, \langle\Psi_{\mathbf{k}\alpha}(\tau) \Psi^\dagger_{\mathbf{k'}\alpha'}(0)\rangle,
  \label{eq:pairing_susceptibility}
\end{equation}
where $\Psi_{\mathbf{k}\alpha} = \frac{1}{2} \left(\psi_{\mathbf{k} \alpha \uparrow} \psi_{\mathbf{-k} \alpha \downarrow} - \psi_{\mathbf{k}\alpha\downarrow} \psi_{\mathbf{-k}\alpha\uparrow}\right)$ is the singlet superconducting pair amplitude on the band $\alpha=x,y$.
Here we focus on the intraband, spin-singlet channel since it is the leading instability \cite{Berg2012,Schattner2015a}.
Figure \ref{fig:pairing_P} shows the optimal pair amplitude $\Psi^{\textrm{opt}}_{\mathbf{k}\alpha}$, corresponding to the maximal eigenvalue of the matrix  $P_{\mathbf{k}\alpha; \mathbf{k'}\alpha'}$ at a temperature slightly above $T_c$.
The pair amplitude of the band $\alpha=y$, shown in Fig.~\ref{fig:pairing_P}~(a), is of the opposite sign to the amplitude on the band $\alpha=x$,
shown in Fig.~\ref{fig:pairing_P}~(b). In fact, the two amplitudes are related precisely by a $\pi/2$ rotation,
highlighting the $d$-wave symmetry of the superconducting order parameter.
The optimal pair amplitude is found to be maximal around the (noninteracting) Fermi surface.
The variation of $\Psi^{\textrm{opt}}_{\mathbf{k}\alpha}$ along the Fermi surface is weak, again showing no strong features at the hot spots.

\section{Discussion}
\label{sec:discussion}
In this work, we have explored the properties of a metal on the verge of an SDW transition. We focused on the critical regime upon approaching the transition, characterized by a rapid growth of the SDW correlations, but still above the superconducting transition temperature. Our main conclusion is that, in this regime, the SDW correlations are remarkably well described by a form similar to that predicted by Hertz-Millis theory, Eq.~(\ref{eq:hertz}) (although the temperature dependence of the SDW susceptibility deviates from the expected form). This holds both for the correlations of the bosonic SDW order parameter field, and for an SDW order parameter defined in terms of a fermion bilinear. In the same regime, we find evidence for strong scattering of quasiparticles near the hotspots, leading to a breakdown of Fermi liquid theory at these points on the Fermi surface. The scattering rate at the hotspots (extracted from the fermion self energy) is only weakly temperature and frequency dependent, down to $T\approx 2 T_c$, where we suspect that superconducting fluctuations begin to play a role; it is out of this unusual metallic state that the superconducting phase emerges.

In addition, we have studied the structure of the superconducting gap near the SDW transition.
Unlike the single-fermion Green's function in the normal state, it does not have a sharp feature at the hot spots; rather, it is found to vary smoothly across the Fermi surface.
Experimentally, a broad maximum of the superconducting gap near the hot spots was observed in a certain electron doped cuprate~\cite{Matsui2005}. Eliashberg theory predicts a peak of the gap function at the hot spots at weak coupling~\cite{Abanov2008} and it remains to be seen whether such behavior appears in our model at weaker coupling.

It is interesting to discuss our results in the context of the existing theories of metallic SDW transitions. First, the fact that Hertz-Millis theory successfully describes many features of our data is non-trivial, in view of the fact that it has no formal justification, even in the large $N$ limit~\cite{Lee2009,Metlitski2010}.
However, as we saw, an extension of the Hertz-Millis analysis to finite temperature predicts that at criticality, $\chi(T)\sim 1/T$, in apparent disagreement with our data. This may be due to the limited temperature window we can access without hitting the superconducting T$_c$, or to effects beyond the one-loop approximation.

An important conclusion of our study is that the SDW critical point is always masked by a superconducting phase~\footnote{Even for the smallest value of $\lambda$ that we have studied, where the maximal T$_c$ is smaller than the lowest temperature in our study, we see enhancement of diamagnetic fluctuations, indicating that we are not too far above T$_c$.}. As a result, it seems likely that the critical metallic regime is never parametrically broad, and one cannot sharply define scaling exponents within the metallic phase~\footnote{In the superconducting phase, we expect criticality of the $d=2+1$ XY universality class, since the Fermi surface is gapped.}. As mentioned above, the SDW correlations follow a Hertz-Millis form -- and hence it is tempting to associate with them critical exponents, i.e. a mean-field value $\nu=1/2$ for the correlation length exponent, and a dynamical critical exponent $z=2$. The fermionic quasiparticles at the hot spots, however, do not exhibit this scaling behavior. In particular, the expected scaling law $\Sigma(\omega_n) \sim \sqrt{\omega_n}$ for the fermion self-energy at the hot spots is not seen within our accessible temperature range.

One can imagine trying to access the ``bare'' metallic quantum critical point by suppressing the superconducting transition. Presumably, this can be done by adding to the model a term that breaks time reversal and inversion symmetries (such a term would lift the degeneracy of fermionic states with opposite momenta, and hence remove the Cooper instability). Breaking time reversal symmetry, however, gives rise to a sign problem, so it is not clear whether the critical behavior can be accessed within the QMC technique.

Alternatively, one could try to understand the metallic regime above T$_c$ in our model as a crossover regime of an underlying ``nearby'' metallic critical point, where some correlators already exhibit their asymptotic behavior (such as the SDW order parameter correlations), while others do not (e.g., the single-fermion Green's function). Interestingly, a simple, non-self consistent one-loop calculation of the fermionic self energy in our model does show a broad range of temperature and frequency where the self energy at the hot spot is nearly constant, before eventually settling into the expected $\sqrt{\omega_n}$ behavior (see Appendix \ref{sec:perturbation_theory}). This calls for a more detailed comparison between our numerically exact DQMC results and a detailed self-consistent one-loop analysis. Preliminary results show that this approximation is surprisingly successful in capturing at least some of the physics of our model~\cite{Wang2016}.

To what extent  such a crossover behavior, characterized by a nearly-constant fermionic lifetime at the hotspots, is ubiquitous across different models, as well as in real materials, remains to be seen. It is interesting to note, however, that a similar behavior has been observed in a study of a \emph{nematic} transition in a metal~\cite{Lederer2016}. It would be interesting to systematically look for such behavior in angle-resolved photoemission spectroscopy in the electron-doped cuprates, where anomalously large broadening of the quasiparticle peaks is seen near the hot spots~\cite{Armitage2001}.

Another important aspect of the metallic state in the critical regime, which we have not addressed in this work, is the electrical conductivity. The optical conductivity  may be strongly affected by the presence of an SDW critical point, even without quenched disorder~\cite{Hartnoll2011,Maslov2016}. Extracting the conductivity from quantum Monte Carlo simulations requires an analytic continuation, and is therefore intrinsically more difficult (and involves more uncertainties) than calculating thermodynamic and imaginary-time quantities. Nevertheless, we have obtained preliminary results showing strong effects of the critical fluctuations on the low-frequency optical conductivity~\cite{Schattner2016b}. A full analysis of the conductivity is deferred to future work.

\acknowledgments
We thank A. Chubukov and C. Varma for useful discussions on this work. E. B. and Y. S. also thank R. Fernandes, S.A. Kivelson, S. Lederer and X. Wang for numerous discussions and collaborations on related topics. E.B. thanks the hospitality of the Aspen Center for Physics, where part of this work was done.
The numerical simulations were performed on the CHEOPS cluster at RRZK Cologne, the JUROPA/JURECA clusters at the Forschungszentrum J\"ulich, and the ATLAS cluster at the Weizmann Institute. Y.S. and E.B. were supported by the Israel Science Foundation under Grant No. 1291/12, by the US-Israel BSF under Grant No. 2014209, and by a Marie Curie reintegration grant. E. B. was supported by an Alon fellowship. M. G. thanks the Bonn-Cologne Graduate School of Physics and Astronomy (BCGS) for support.

\appendix

\section{DQMC simulations}
\label{sec:DQMCAppendix}
In this Appendix we elaborate on a number of specific technical
aspects of our numerical implementation of the determinantal quantum
Monte Carlo (DQMC) approach.  We refer readers looking for a more
comprehensive discussion of the general DQMC setup to our previous
paper~\cite{Schattner2015a} and in particular its supplementary online
material.

We study the lattice model described by the action~(\ref{eq:action}) at finite
temperature in the grand canonical ensemble. After discretizing imaginary time and
integrating out the fermionic degrees of freedom, the partition function reads
\begin{equation}
  Z = \int\! D\vec{\varphi}\, e^{-\Delta \tau \sum_\tau  L_\varphi (\tau)} \det G_\varphi^{-1} \,,
  \label{eq:partition_function}
\end{equation}
where $G_\varphi$ is the equal-time Green's function matrix for a
fixed configuration of the bosonic order parameter field
$\vec{\varphi}$.  We use an imaginary time step of $\Delta \tau = 0.1$
in all calculations.  The DQMC method samples configurations of
$\vec{\varphi}$ according to their weight
$\exp\left(-\Delta \tau \sum_\tau L_\varphi (\tau)\right) \cdot \det
G_\varphi^{-1}$.  For efficient Monte Carlo sampling, is it highly
advantageous to consider models in which the determinant
in~\eqref{eq:partition_function} is guaranteed to be positive, thereby
avoiding the notorious fermion sign problem.  For the two-band model
that has been proposed in Ref. \cite{Berg2012} this is ensured by an
antiunitary symmetry of the action~\cite{Wu2005,Wei2016,Li2016signproblem}, written in
first quantization as
\begin{align}
  \mathcal{U} = is_y \tau_z K
  && \text{with} &&
  \mathcal{U}^2 = -\Id \,.
  \label{eq:antiunitary}
\end{align}
Here $K$ is the complex conjugation operator and $s_y$ ($\tau_z$) are
Pauli matrices acting in spin (flavor) space, respectively.

In this manuscript we have modified the model of Ref.~\cite{Berg2012} in two regards.
First, as in our previous paper~\cite{Schattner2015a}, we consider an easy-plane SDW order parameter, rather than
an $O(3)$ symmetric order parameter used in \cite{Berg2012}.
Second, we couple the system to a fictitious, spin and band dependent orbital
``magnetic'' field, whose flux through the system is given by
$\vec{\Phi}_{\alpha,s} = \Phi^{(x)}_{\alpha,s} \vec{e\,}{}^x + \Phi^{(y)}_{\alpha,s} \vec{e\,}{}^y + \Phi^{(z)}_{\alpha,s} \vec{e\,}{}^z$.
Here, we place the two-dimensional lattice ($x,y$ directions) on a torus.
The $x$ and $y$ components of the flux twist the boundary conditions along the $y$ and $x$ directions, respectively,
while the $z$ component acts as an orbital magnetic field with a uniform flux per plaquette through the torus.
As in the definition of the action \eqref{eq:action}, $\alpha=x,y$  is a fermion flavor index and
$s={\uparrow},{\downarrow}$ is a spin index. The flux $\Phi^{z}$ is restricted to be an integer multiple of the flux quantum $\Phi_0$.

In order to preserve the symmetry~\eqref{eq:antiunitary}, fermions of different
spin and flavor are coupled to this fictitious flux as
\begin{equation}
  \begin{alignedat}{3}
    \vec{\Phi}_{\alpha,\uparrow} &= -\vec{\Phi}_{\alpha,\downarrow} && \quad\text{ and}\\
    \vec{\Phi}_{x,s} &= -\vec{\Phi}_{y,s}, &&
  \end{alignedat}
  \label{eq:flux}
\end{equation}
with $\vec{\Phi}_{x,\uparrow} = \vec{\Phi}$.
Note that the inter-band sign change is not strictly necessary to avoid the fermion sign problem.
Importantly, in the thermodynamic limit $L \to \infty$
the fictitious field vanishes.
In our previous paper~\cite{Schattner2015a} and for some of the
results in the present one, we have chosen $\vec{\Phi}=(0,0,\Phi_0)$.
Although such a flux is useful in reducing
spurious finite-size effects at low temperatures
\cite{Assaad2002a,Schattner2015}, it also breaks lattice
translational symmetry, hampering the analysis of non-local, fermionic
correlation functions such as the momentum-resolved Green's function.
In order to measure such quantities, for the present paper, we have run
additional simulations without applying a perpendicular flux $\Phi^{(z)}$, but instead
with {\em in-plane} fields, such that $\vec{\Phi} = (n_x,n_y,0) \frac{\Phi_0}{4}$,
where $n_i=0,1,2,3$.
This procedure is equivalent to having twisted boundary
conditions, such that the allowed momenta for the $\psi_{x,\uparrow}$ fermions are
\begin{equation}
\mathbf{k}=\frac{2\pi}{4L}(4 j_x - n_y, 4 j_y + n_x),
\end{equation}
where $j_x$, $j_y$ are integers, thereby enhancing the momentum-space resolution
of fermionic observables fourfold.

\begin{figure}[t!]
  \newlength{\fh}
  \setlength{\fh}{0.63\linewidth}
  \raggedright
  \textbf{(a)} $\lambda=1.5$\\
  \includegraphics[height=\fh]{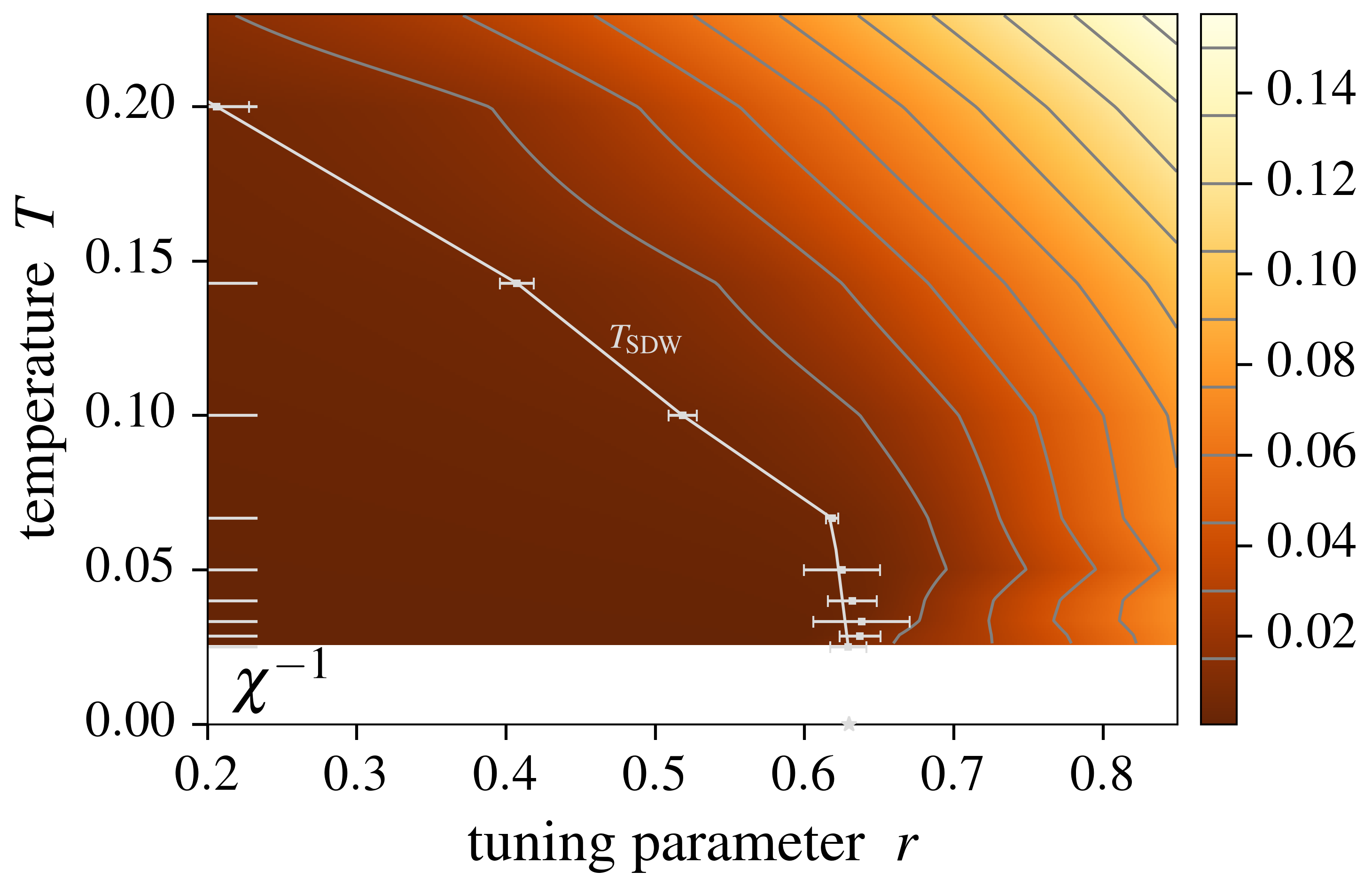}
  \\[5mm]
  \textbf{(b)} $\lambda=2$\\
  \includegraphics[height=\fh]{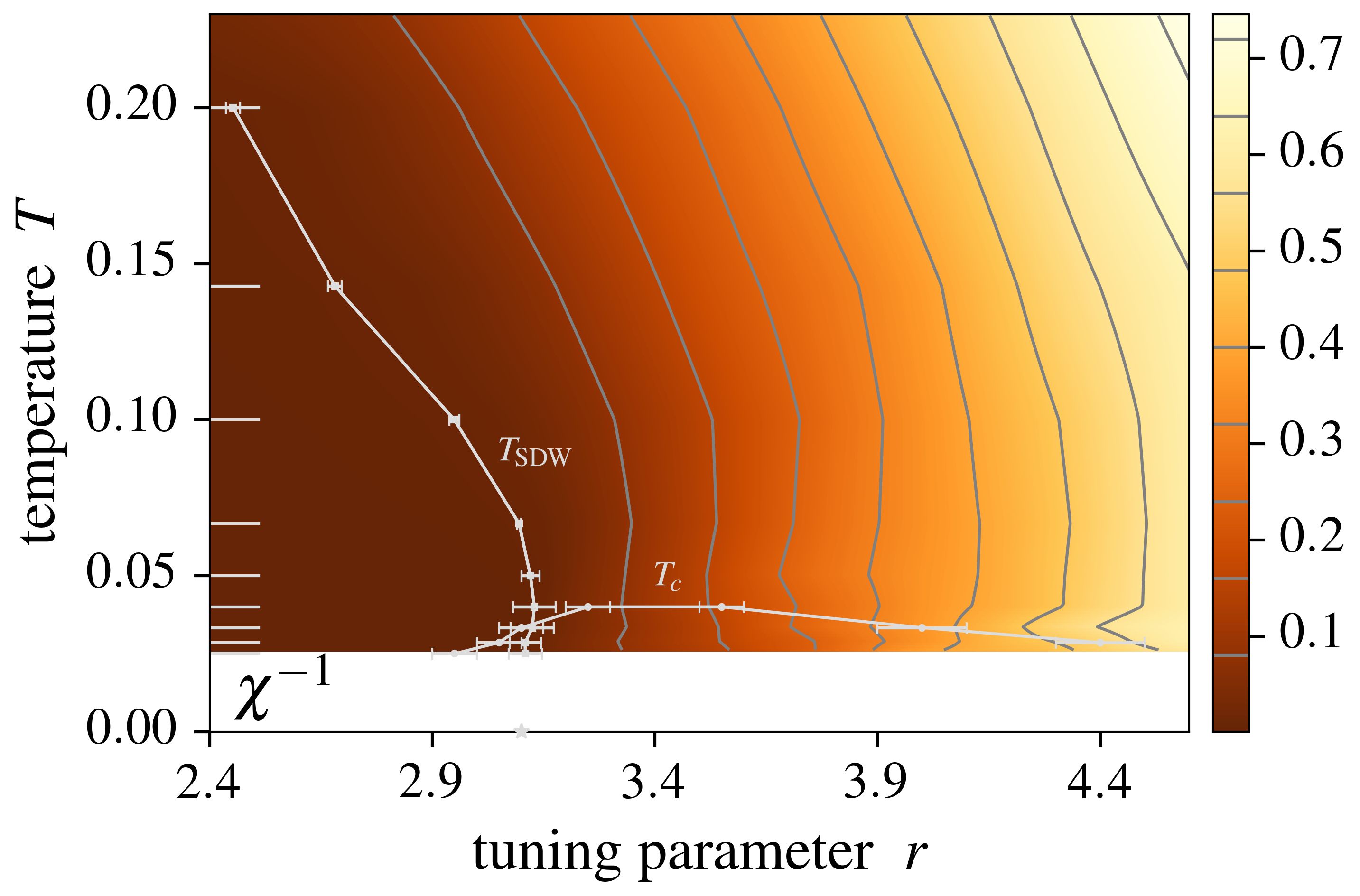}
  \caption{Companion figure to Fig.~\ref{fig:chi-inv-color-l1} for
    (a) $\lambda=1.5$ and (b) $\lambda=2$.}
  \label{fig:chi-inv-color-l1.5-l2}
\end{figure}

\section{Magnetic correlations for $\lambda = 1$ and $\lambda = 2$, and
  inside the superconducting phase}
\label{sec:magn-corr-at}

In this appendix we present additional data for the bosonic and
fermionic SDW susceptibilities $\chi(\mathbf{q}, i \omega_n, r, T)$
and $S_{xx}(\mathbf{q}, i \omega_n, r, T)$.

We begin with Fig.~\ref{fig:chi-inv-color-l1.5-l2}, which shows
$\chi^{-1}(\mathbf{q} = \mathbf{Q}, i\omega_n=0)$ across the phase
diagrams for $\lambda=1.5$ and $\lambda=2$. It illustrates how the
bending of the SDW finite-temperature transition line is also
visible in $\chi^{-1}$ for $r > r_{c0}$, similarly to the behavior for
$\lambda=1$ in Fig.~\ref{fig:chi-inv-color-l1}, where, however, no
superconducting phase has been observed within our temperature
resolution.

\begin{figure*}[h!]
  \centering
  \includegraphics[width=.45\linewidth]{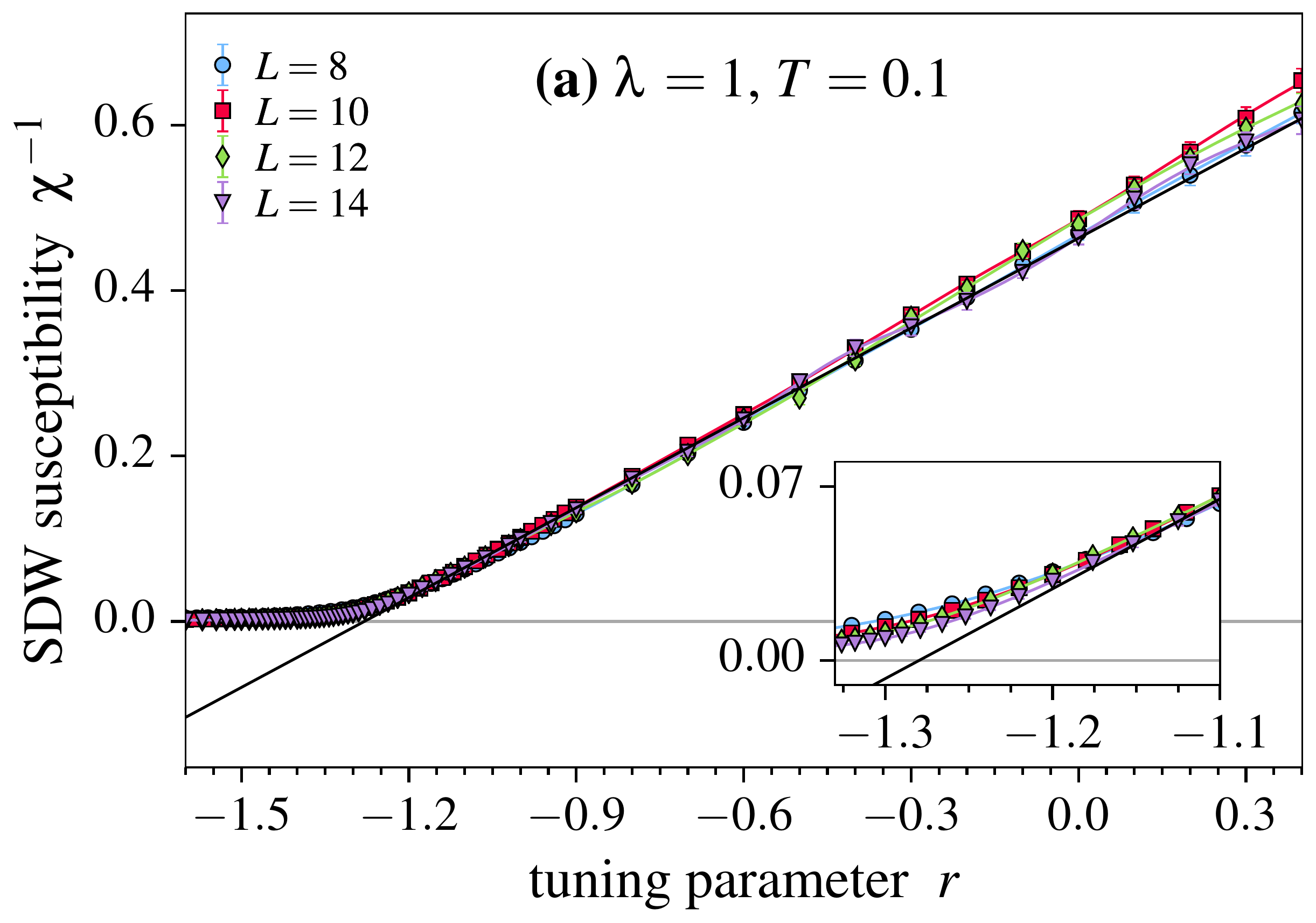}  \hspace{0.05\linewidth}
  \includegraphics[width=.45\linewidth]{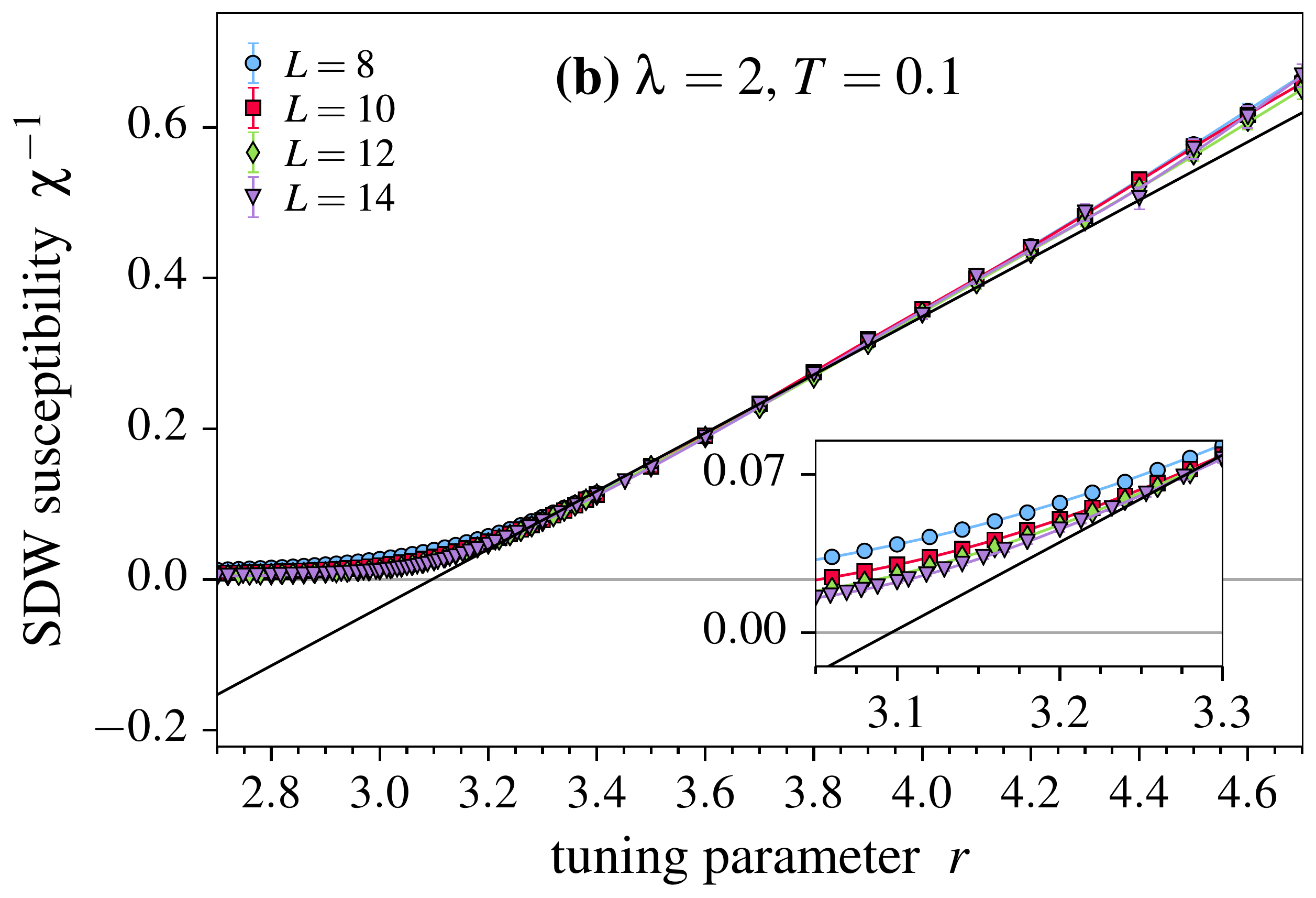}
  \caption{
    Companion figure of Fig.~\ref{fig:chi_r_l1.5} for Yukawa couplings (a) $\lambda=1$
    and (b) $\lambda=2$. The black lines are linear fits of the $L=14$ data for
    (a) $r > -1.2$ and (b) $r > 3.2$.
    }
\label{fig:chi_r_l1_l2}
\end{figure*}

\begin{figure*}[h!]
  \centering
  \includegraphics[width=.45\linewidth]{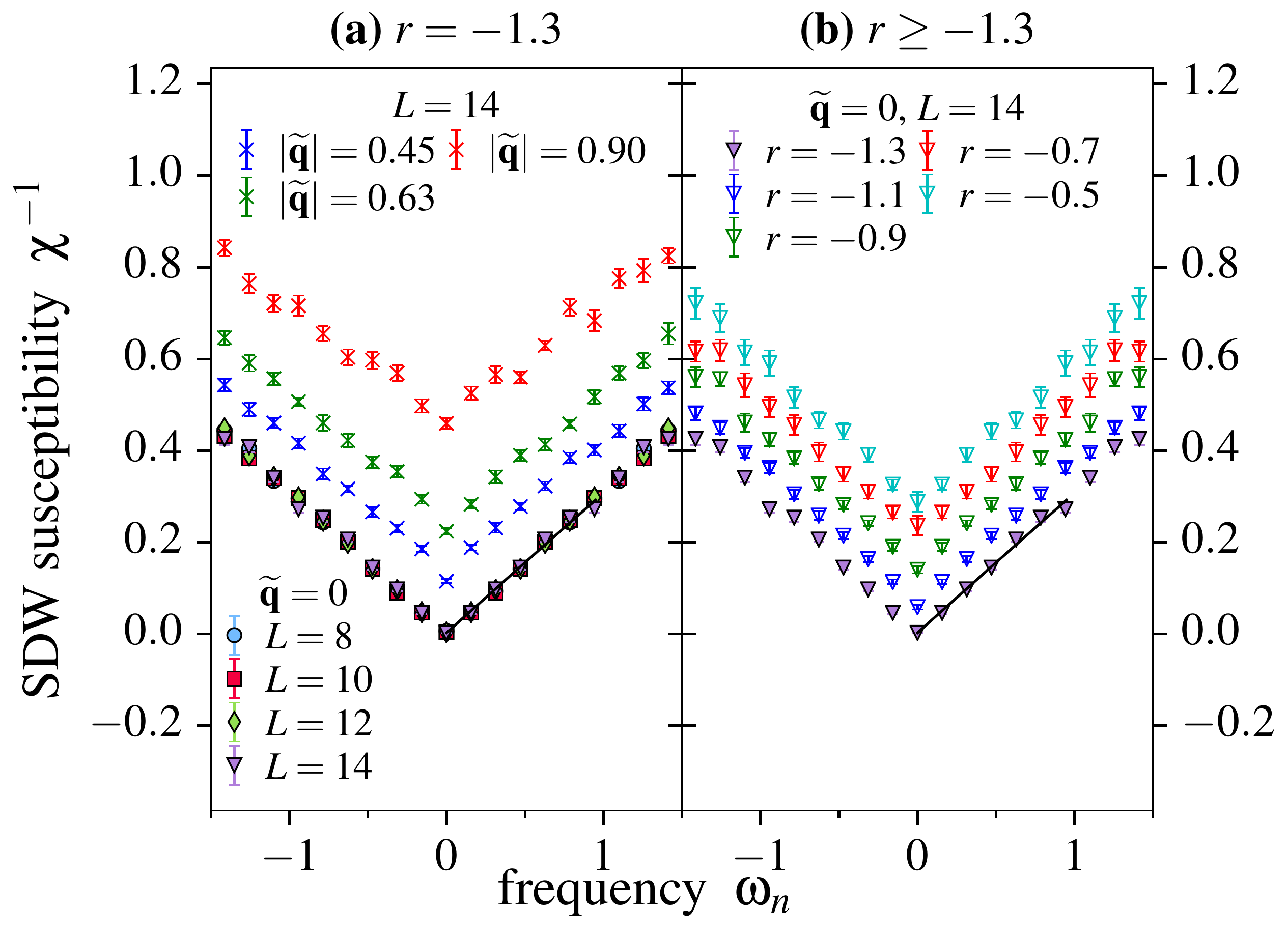}  \hspace{0.05\linewidth} 
  \includegraphics[width=.45\linewidth]{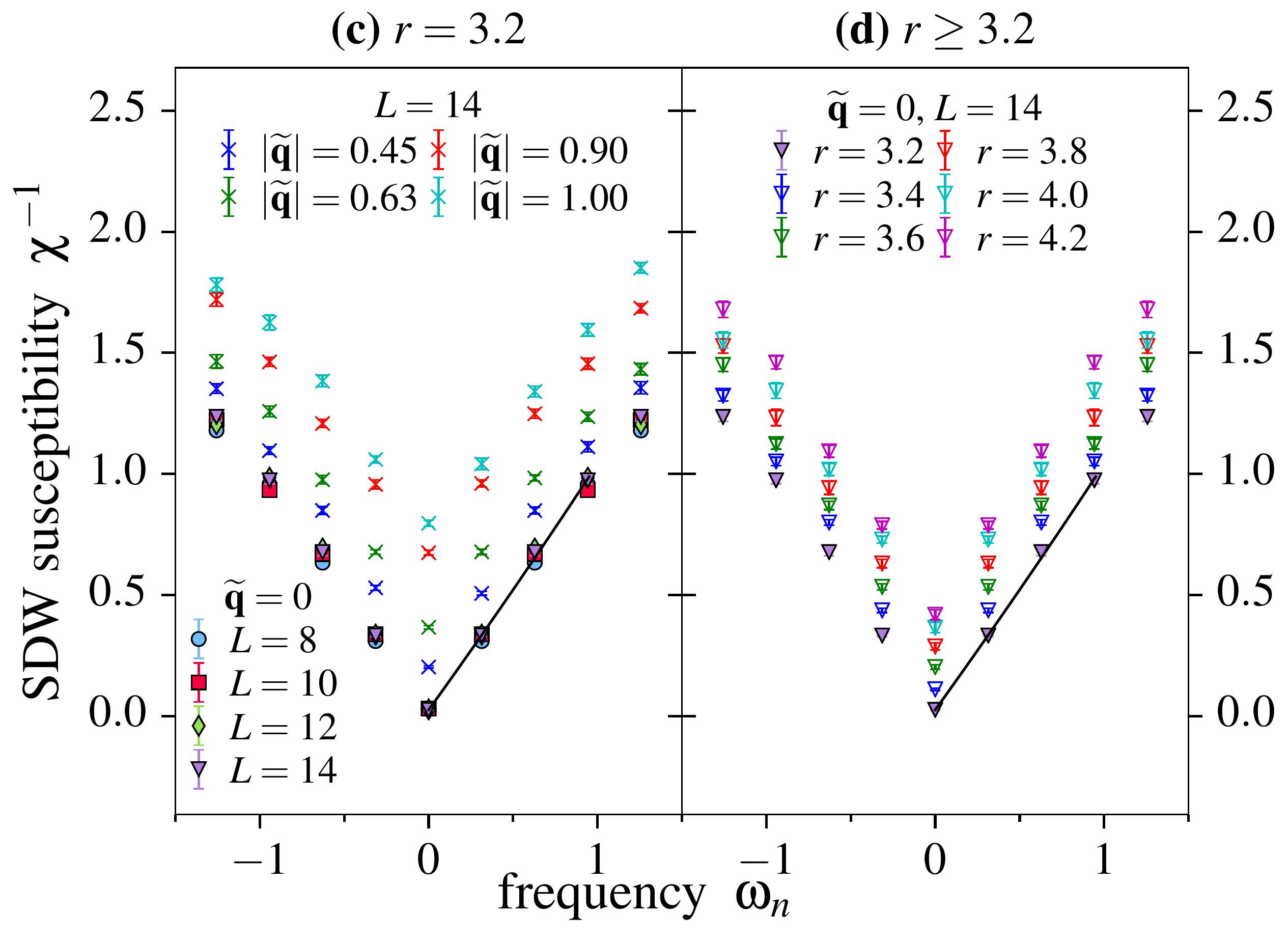} 
  \caption{
    Companion figure of Fig.~\ref{fig:chi_freq_multir_l1.5} for Yukawa
    couplings $\lambda=1$ at $T=1/40$ (left panel)
    and $\lambda=2$ at $T=1/20$ (right panel).
    }
  \label{fig:chi_freq_multir_l1_l2}
\end{figure*}

\begin{figure*}[h!]
  \centering
  \includegraphics[width=.45\linewidth]{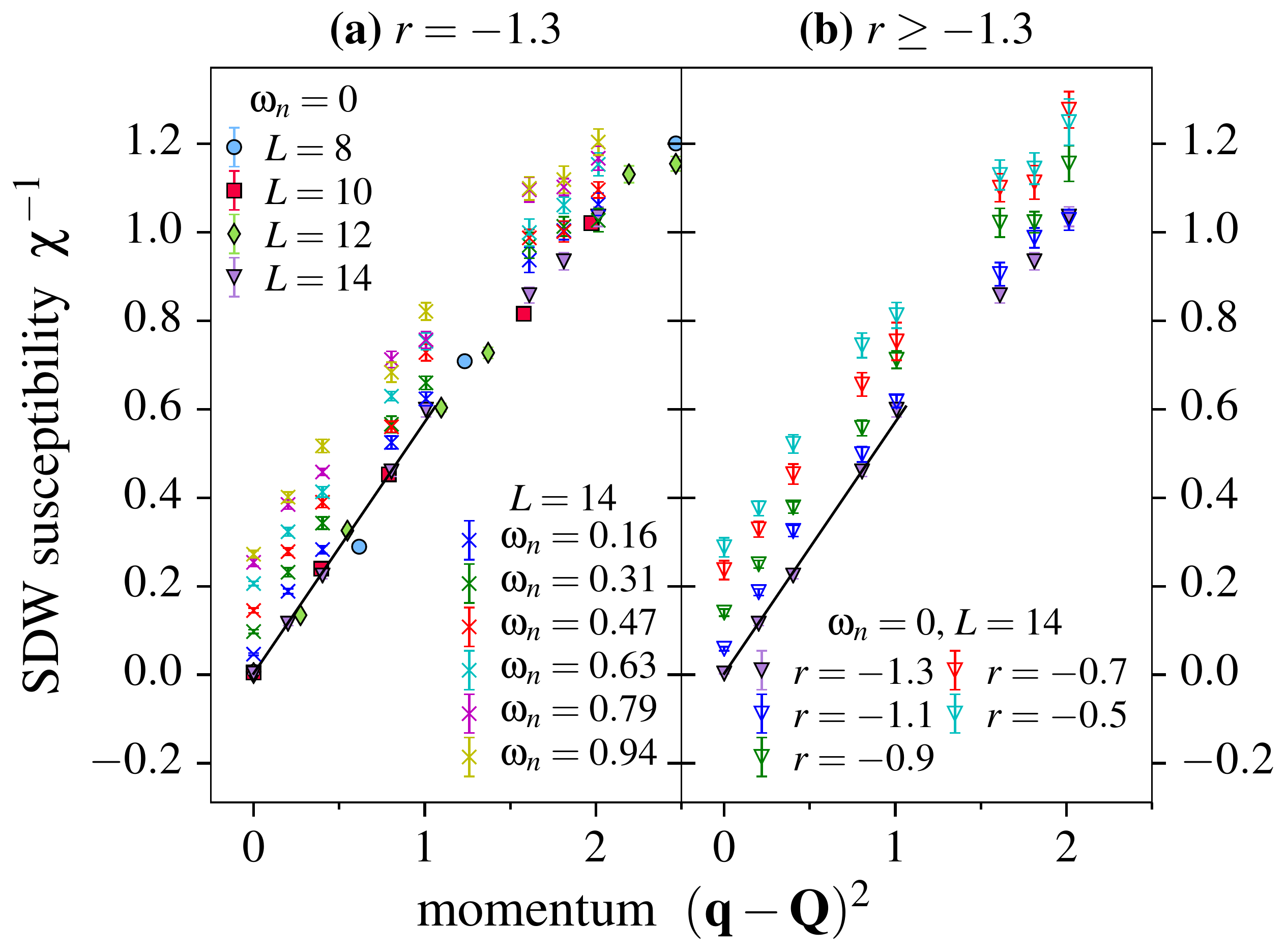}  \hspace{0.05\linewidth} 
  \includegraphics[width=.45\linewidth]{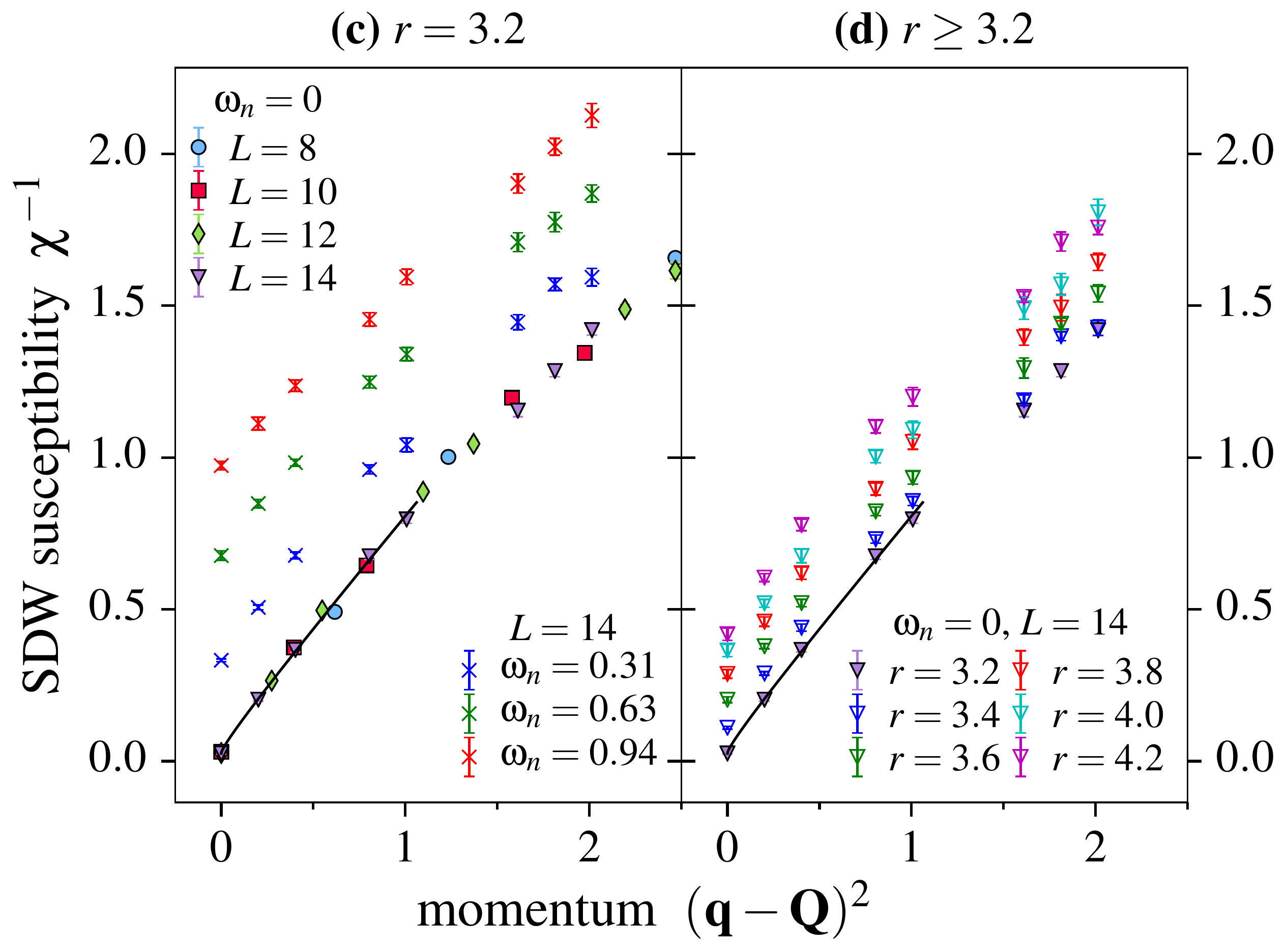}
  \caption{
    Companion figure of Fig.~\ref{fig:chi_mom_multir_l1.5} for Yukawa
    couplings $\lambda=1$ at $T=1/40$ (left panel)
    and $\lambda=2$ at $T=1/20$ (right panel).
	}
  \label{fig:chi_mom_multir_l1_l2}
\end{figure*}

\begin{figure*}[h!]
  \centering
  \includegraphics[width=.45\linewidth]{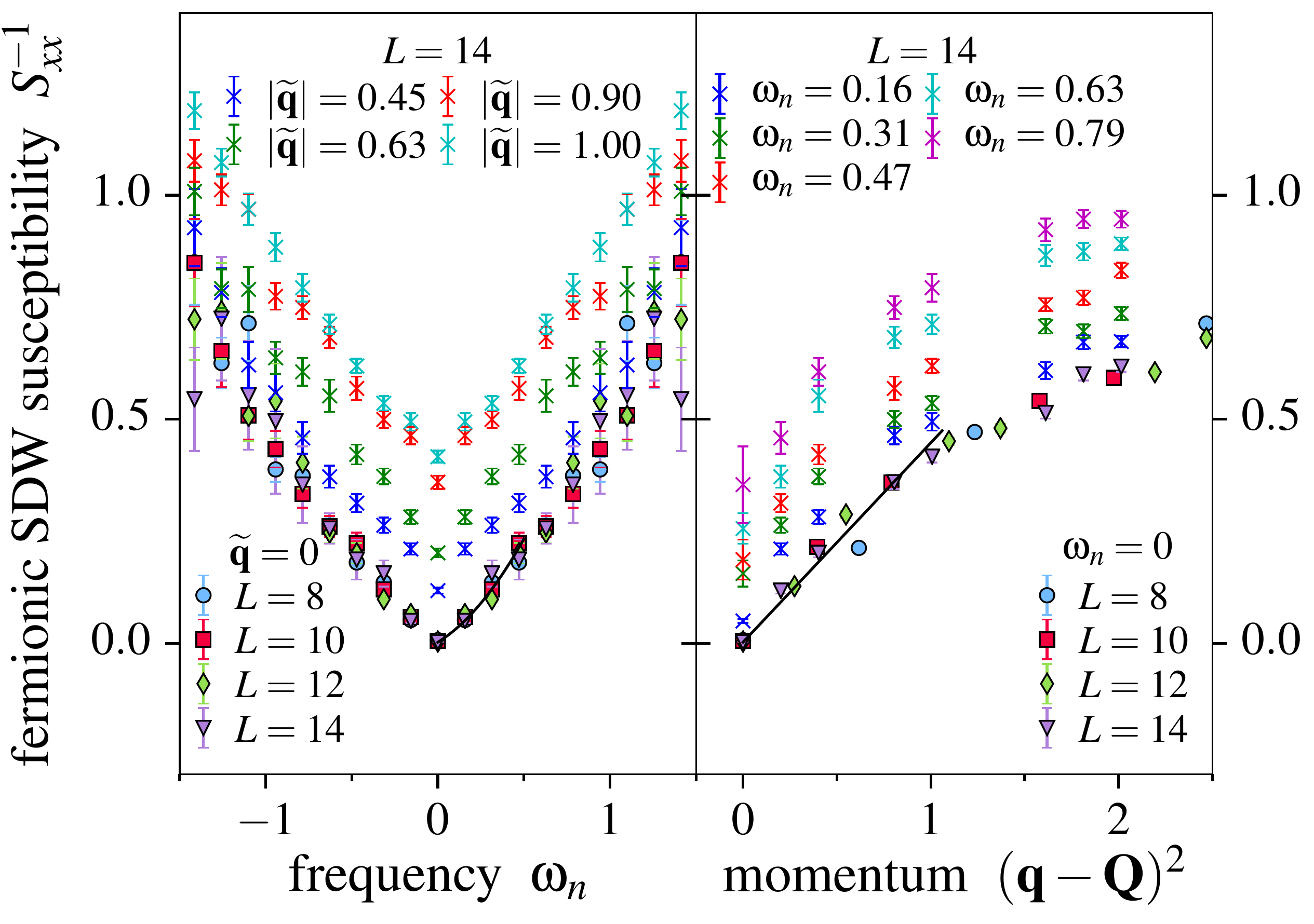}  \hspace{0.05\linewidth} 
  \includegraphics[width=.45\linewidth]{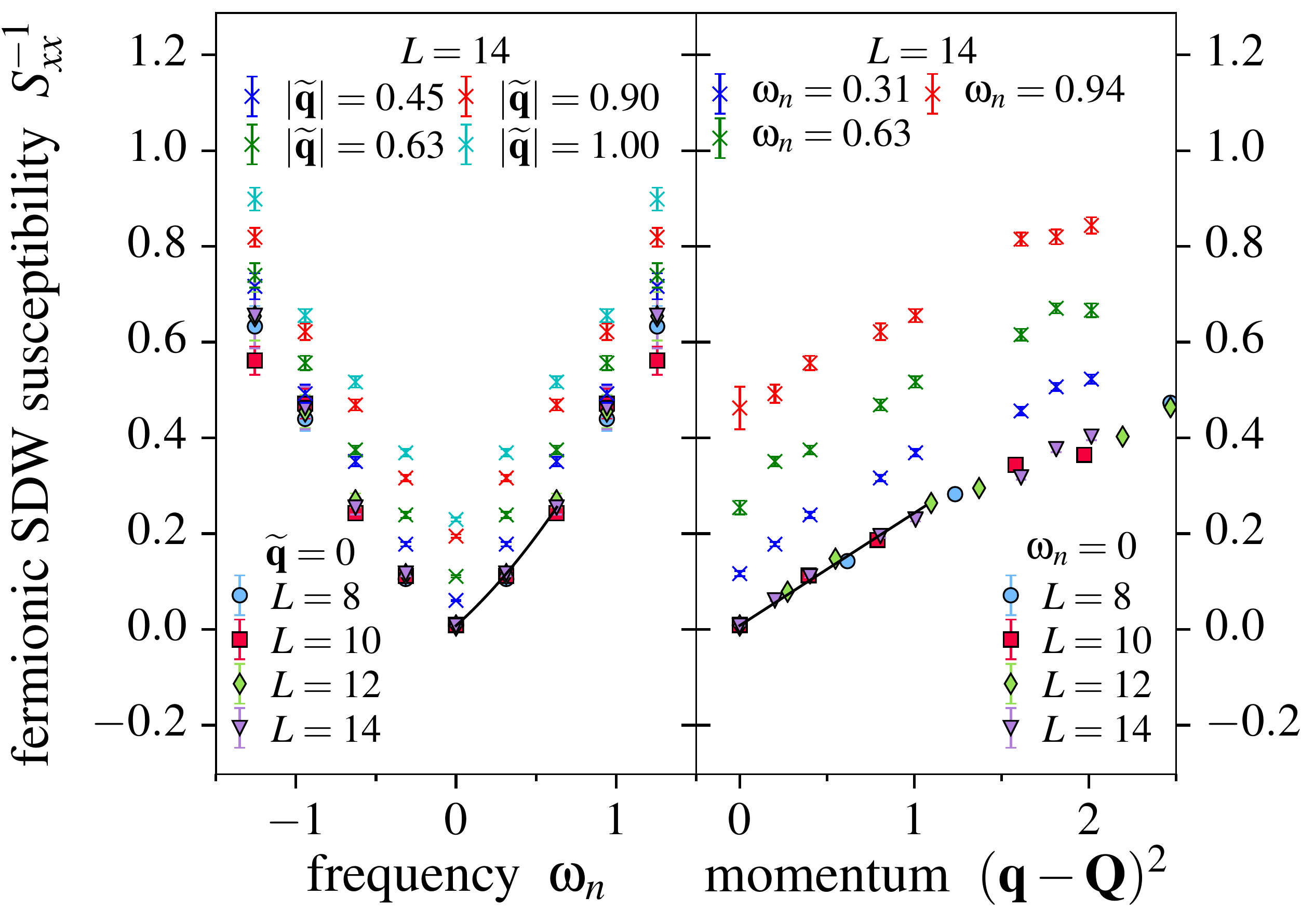}
  \caption{
    Companion figure of Fig.~\ref{fig:chi_freq_mom_f_l1.5} for Yukawa
    couplings $\lambda=1$ at $T=1/40$ (left panel)
    and $\lambda=2$ at $T=1/20$ (right panel).
    }
  \label{fig:chi_freq_mom_f_l1_l2}
\end{figure*}

Next, to complement the discussion in Sec.~\ref{sec:SDW_corr}, where we have
focused on the Yukawa coupling $\lambda = 1.5$ and a bosonic velocity
of $c = 3$, we here show detailed data for values $\lambda = 1$ and
$\lambda = 2$ at the same velocity $c$.  In Fig.~\ref{fig:chi_r_l1_l2} we
show $\chi^{-1}(\mathbf{q}=\mathbf{Q}, i \omega_n = 0)$ as a function
of $r$ for constant $T = 0.1$, where the same linear dependence as for
$\lambda=1$ (Fig.~\ref{fig:chi_r_l1.5}) is apparent.  As we show in
Fig.~\ref{fig:chi_freq_multir_l1_l2} both for $r \approx r_{c0}$ and a
range of $r > r_{c0}$ the leading frequency dependence of $\chi^{-1}$ is
clearly linear, similarly to $\lambda=1.5$
(Fig.~\ref{fig:chi_freq_multir_l1.5}), whereas the leading momentum
dependence shown in Fig.~\ref{fig:chi_mom_multir_l1_l2} is quadratic
in $\mathbf{q} - \mathbf{Q}$, which is again comparable to
$\lambda=1.5$ (Fig.~\ref{fig:chi_mom_multir_l1.5}).  Note that for
$\lambda=2$ we show data at a higher temperature $T=1/20$ rather than
at $T=1/40$, because otherwise the system would be in the
superconducting phase, where the frequency dependence is significantly
altered.  For small frequencies and momenta the fermionic
susceptibility $S_{xx}^{-1}$ in Fig.~\ref{fig:chi_freq_mom_f_l1_l2}
behaves similarly to the bosonic $\chi^{-1}$ (see
Fig.~\ref{fig:chi_freq_mom_f_l1.5} for $\lambda=1.5$).

\begin{figure*}[h!]
  \centering
  \includegraphics[width=0.45\linewidth]{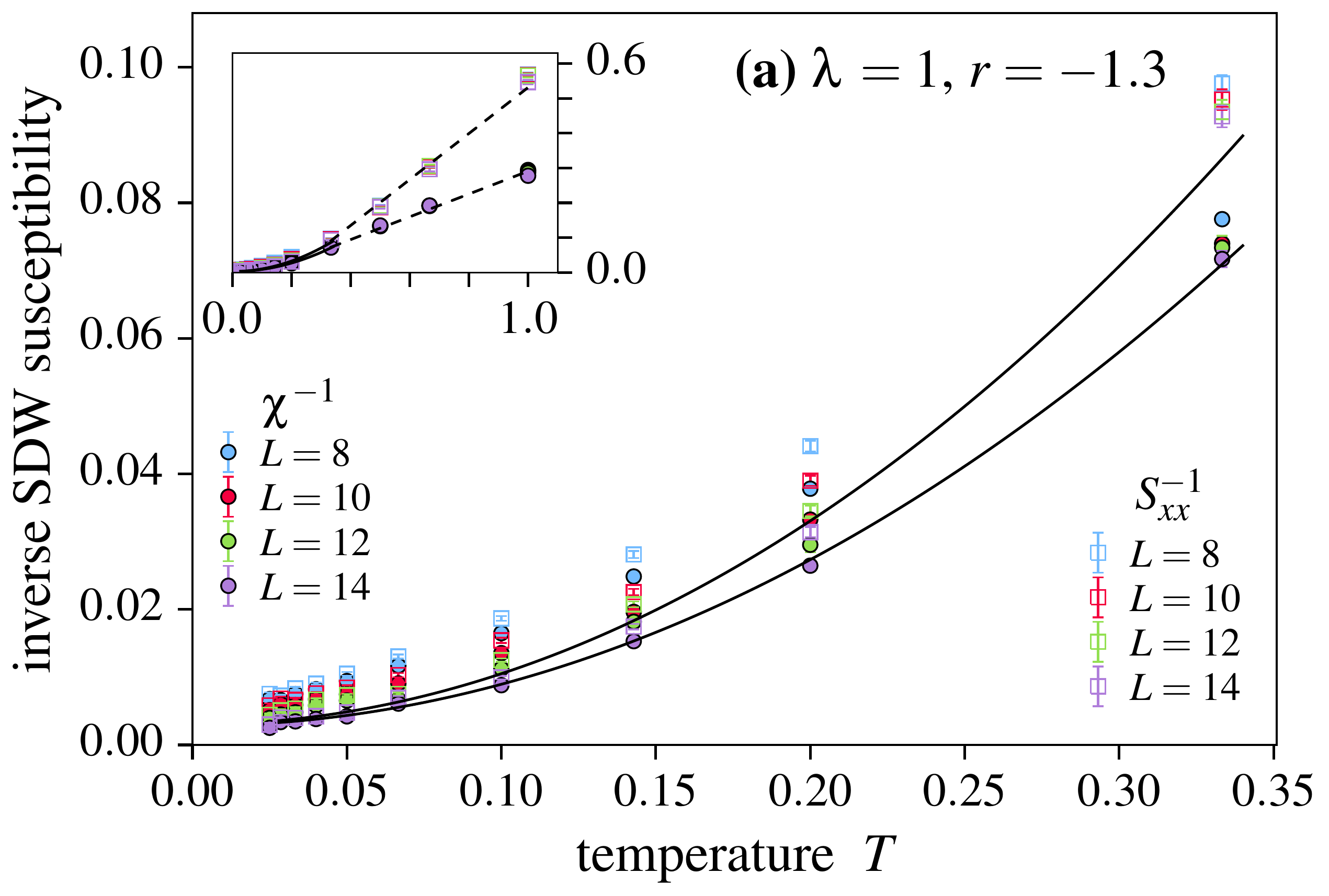} \hspace{0.05\linewidth}
  \includegraphics[width=0.45\linewidth]{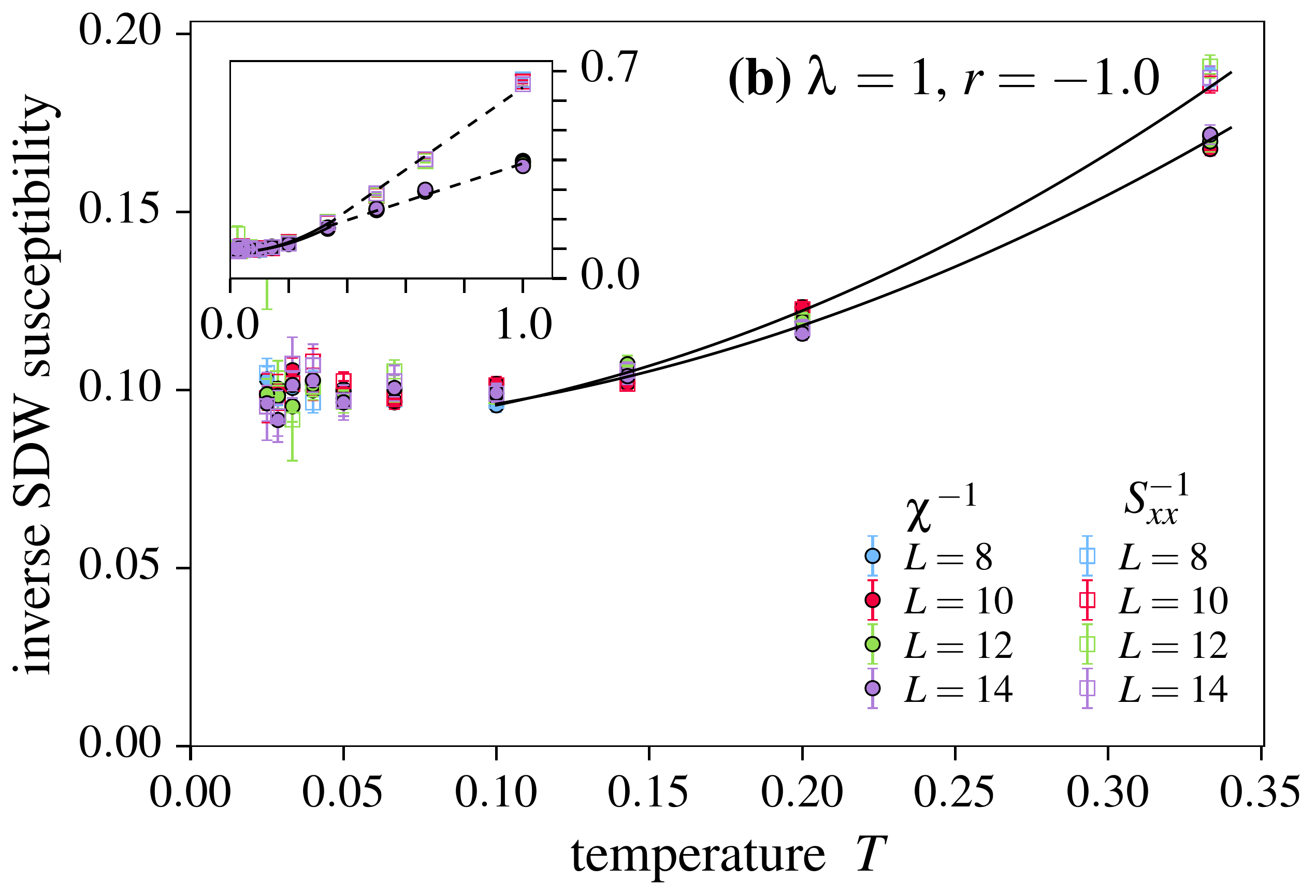}\\[5mm]
  \includegraphics[width=0.45\linewidth]{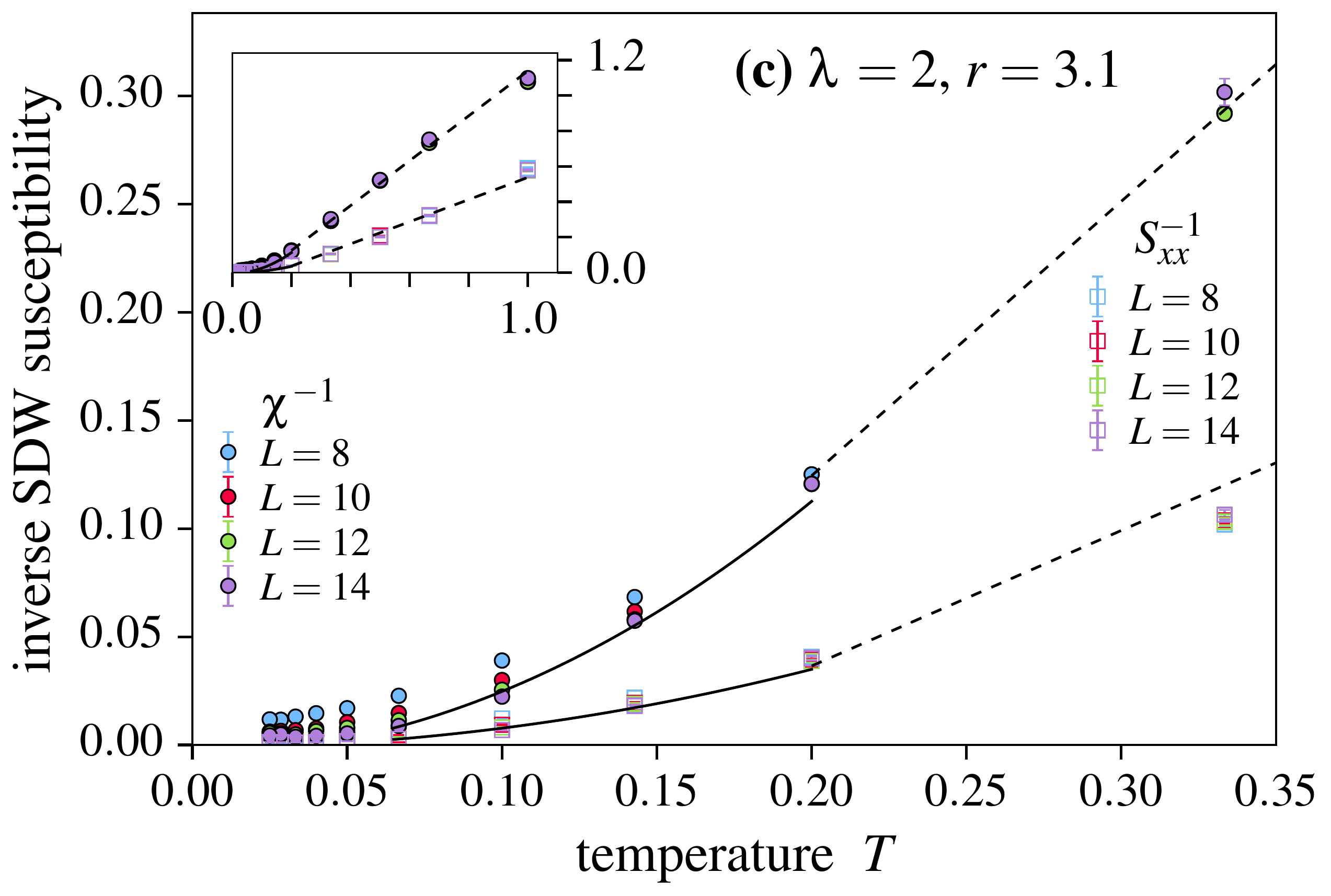} \hspace{0.05\linewidth}
  \includegraphics[width=0.45\linewidth]{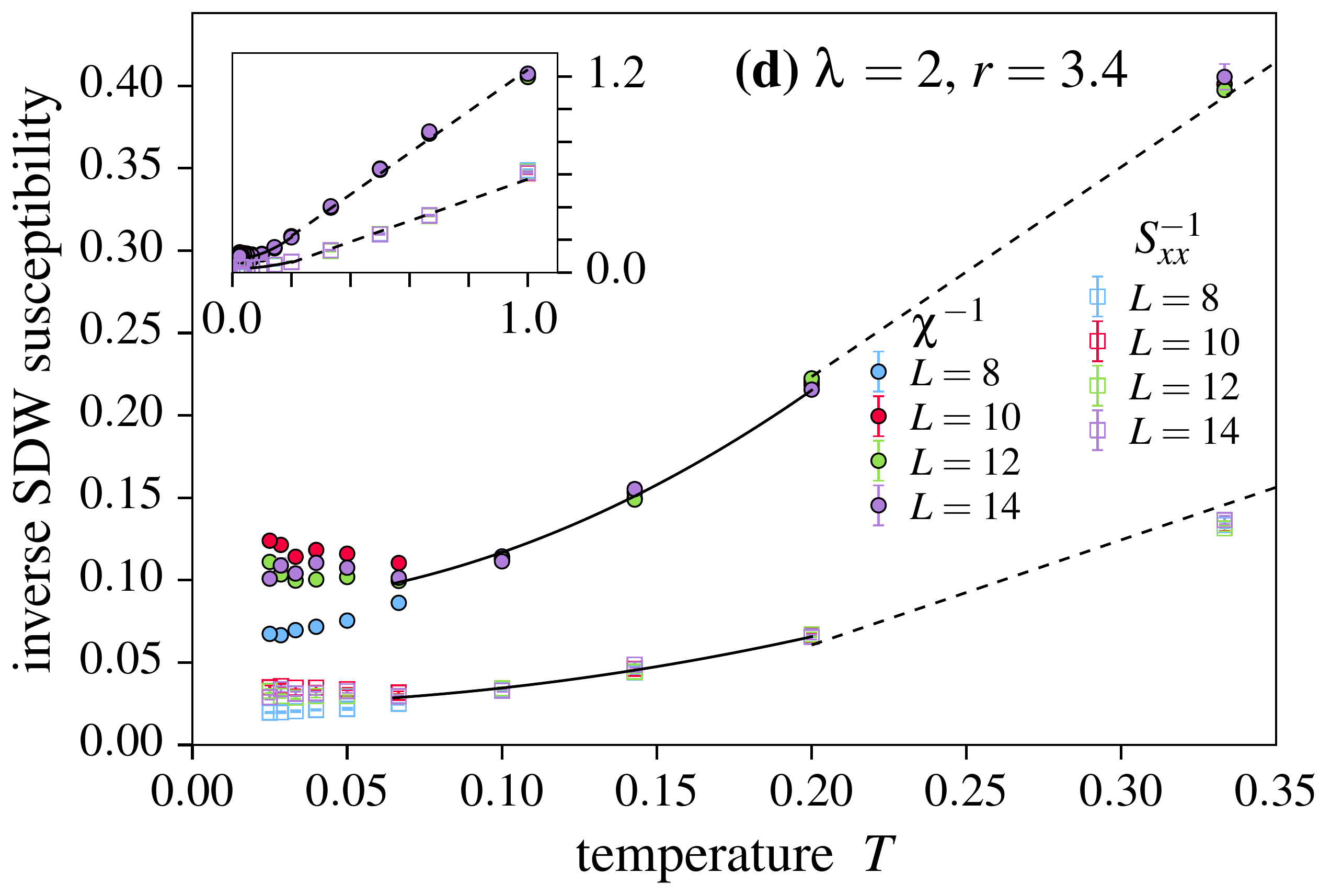}
  \caption{Companion figure of Fig.~\ref{fig:chi_temp_l1.5} for
    for (top row) $\lambda = 1$ at (a)
    $r = -1.3 \approx r_{c0}$ and at (b) $r = -1.0 > r_{c0}$, and for
    (bottom row) $\lambda = 2$ at
    (c) $r = 3.1 \approx r_{c0}$ and at (d)
    $r = 3.4 > r_{c0}$.
  }
\label{fig:chi_temp_l1_l2}
\end{figure*}

The temperature dependence of both $\chi^{-1}$ and $S_{xx}^{-1}$ is
demonstrated in Fig.~\ref{fig:chi_temp_l1_l2}.  As in the case of
$\lambda = 1.5$ (Fig.~\ref{fig:chi_temp_l1.5}), we observe a linear
regime at high temperatures, shown in the insets, and a crossover
region, where we can fit a quadratic law.  For $\lambda=1$ this second
region extends down to lower temperatures than for $\lambda = 1.5$,
where $T_c$ is higher.  At $\lambda = 2$ the data for $T < 0.05$ is
from within the superconducting phase.  Moreover, at
$r = 3.1 \approx r_{c0}$ the system is partially inside the magnetic
quasi-long range order phase (cf. the phase diagram in
Fig.~\ref{fig:phasediagrams}).

\begin{figure}[h!]
  \centering
  \includegraphics[width=0.675\linewidth]{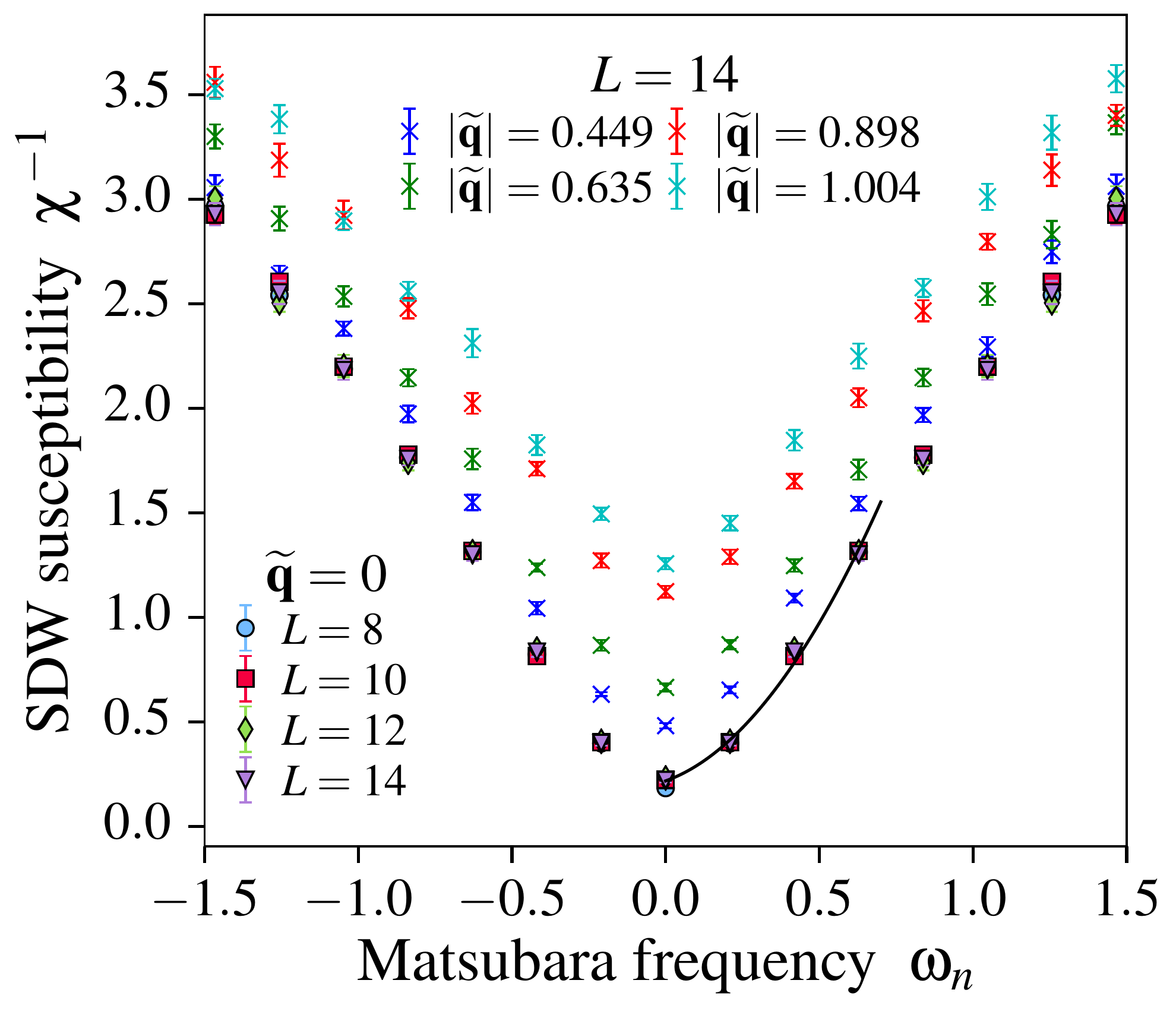}
  \caption{Frequency dependence of the SDW susceptibility $\chi^{-1}$
    for $\lambda=3$, $c=2$ at $T=1/30 < T_c$ and
    $r=10.37 \approx r_{\mathrm{opt}}$, close to where $T_c$ is
    highest for this set of parameters, shown for various momenta
    $\mathbf{q} = \mathbf{Q}+\mathbf{\widetilde{q}}$.  The black line
    is the best fit of a second degree polynomial
    $b_0 + b_1 |\omega_n| + b_2 \omega_n^2$ to the
    $\mathbf{q}=\mathbf{Q}$, $L=14$ low-frequency data.}
  \label{fig:chi_freq_l3_c2}
\end{figure}

Finally, to illustrate the influence of superconductivity on the
frequency dependence of the SDW susceptibility, we show data from deep
within the superconducting phase in Fig.~\ref{fig:chi_freq_l3_c2}.
Here we have chosen a data set with different values of the Yukawa
coupling $\lambda=3$ and the bosonic velocity $c=2$ (as in
Ref.~\cite{Schattner2015a}) since $T_c^{\mathrm{max}} \approx 0.08$ is
about twice as high for these parameters as for $\lambda = 2$, $c=3$.  In contrast to the
data at $T > T_c$ shown in Figs.~\ref{fig:chi_freq_multir_l1.5} and
\ref{fig:chi_freq_multir_l1_l2}, the low-frequency behavior is clearly
no longer  purely linear -- indicative of a suppression of Landau damping
in the superconducting phase.

\section{Comparison with a one loop approximation for the fermion self energy}
\label{sec:perturbation_theory}
In this Appendix, we consider the fermionic self energy in a one-loop approximation. To this order, the self energy is given by
\begin{equation}
\Sigma_{\mathbf{k},\alpha=y}(\omega_{n})=\frac{\lambda^{2}}{\beta L^{2}}\sum_{\mathbf{q},m}\chi_{\mathbf{q}}(\Omega_{m})G_{\mathbf{k}+\mathbf{q},\alpha=x}^{0}(\omega_{n}+\Omega_{m}),
\label{eq:perturbation_Sigma}
\end{equation}
where $G^0$ is the non-interacting Green's function and $\Omega_m=2\pi m T$ is a bosonic Matsubara frequency. Deferring more systematic calculations for future work, here we do not attempt a full, self-consistent solution of the coupled Eliashberg equations~\cite{Abanov2003} for the SDW correlations and the fermionic Green's function. Instead, we use the non-interacting Green's function and $\chi$ taken from a lattice, discretized imaginary time version of Eq.~\eqref{eq:chi_functional}
\begin{equation}
\begin{split}
\chi_{\mathbf{q}}^{-1}(\Omega_{m})&=a_{r}(r-r_{c})\\
&+4a_{q}\left[\sin^{2}\left(\frac{q_{x}-Q_{x}}{2}\right)+\sin^{2}\left(\frac{q_{y}-Q_{y}}{2}\right)\right]\\
&+\frac{2a_{\omega}}{\Delta\tau}\left|\sin\left(\frac{\Delta\tau\Omega_{m}}{2}\right)\right|,
\end{split}
\label{eq:chi_perturbative}
\end{equation}
where the parameters $a_r, a_q, a_\omega$ are taken from a fit to the DQMC data, see Section \ref{sec:boson-sdw-susc}.
Strictly speaking, this procedure is not justified. However, since within a self-consistent Eliashberg theory, $\chi$ has the form (\ref{eq:chi_functional},\ref{eq:chi_perturbative}), we expect our simplified approximation to capture the general behavior of the self-consistent theory.

\begin{figure}[th]
  \centering
  \includegraphics[width=\columnwidth]{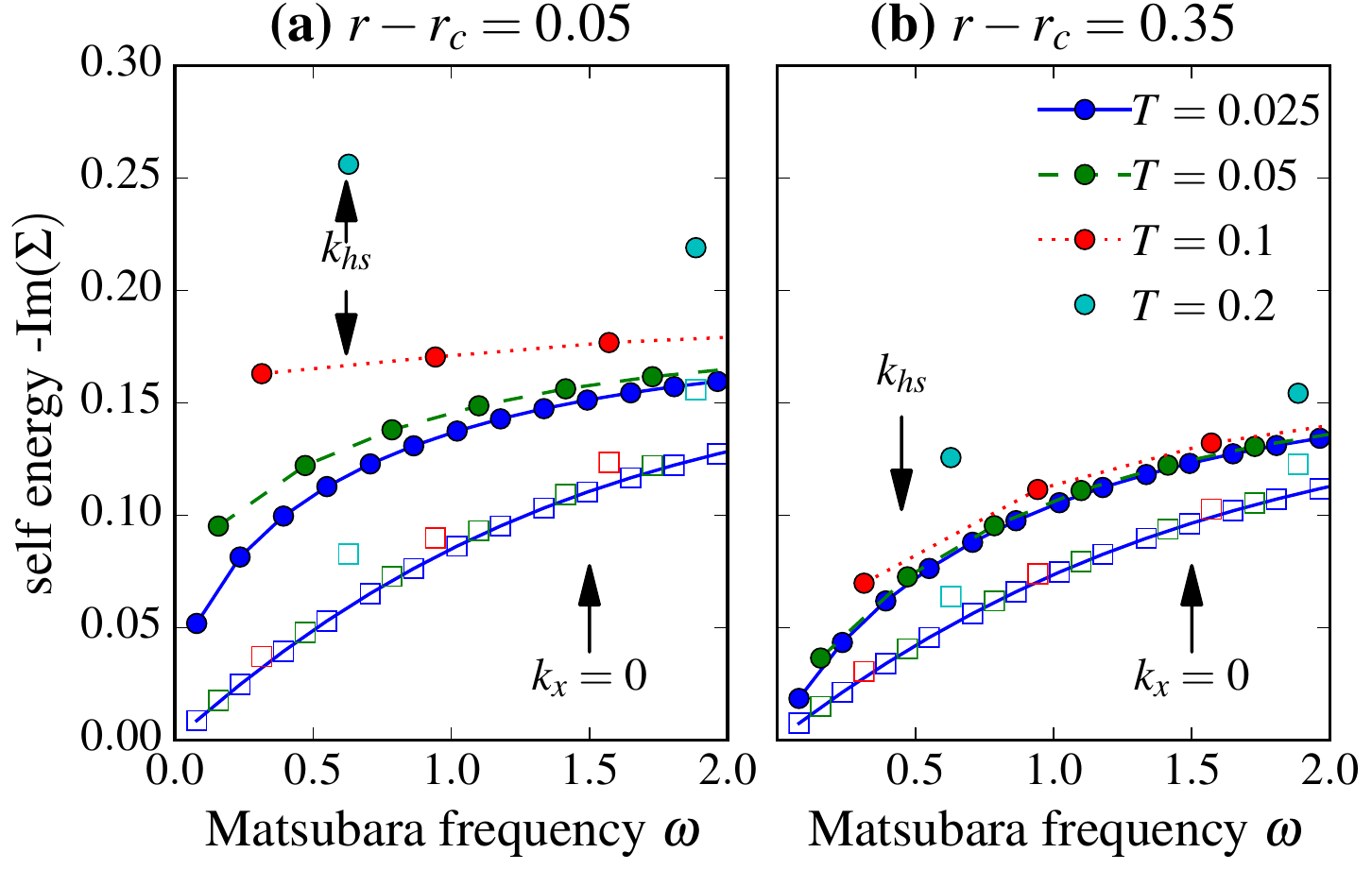}
  \caption{Imaginary part of the self energy \eqref{eq:perturbation_Sigma}, calculated within a one-loop approximation for several temperatures for two momenta on the Fermi surface and
  (a) close to the QCP at $r=r_c+0.05$, (b) further away from the QCP.
  An SDW correlator of the form \eqref{eq:chi_perturbative} was used, with the parameters taken from the fit to the DQMC data for $\lambda=1.5$, $c=3$, with $\Delta\tau=0.1$, and $L=200$, as shown in Fig.~\ref{fig:chi_collapses}(b).
  The data for $\mathbf{k}=\mathbf{k}_{hs}$ is indicated by full circles, the momentum away from the hotspot is indicated by empty squares.
  }
  \label{fig:perturbative_self_energy}
\end{figure}
In Fig.~\ref{fig:perturbative_self_energy} we show the imaginary part of the self energy. The results bear some similarities to the DQMC data, shown in Fig.~\ref{fig:self_energy}.
Whereas at moderate $r-r_c$ or away from the hotspots the self energy is rapidly diminished as the frequency $\omega_n$ is lowered, the behavior at the hotspots as $r$ approaches $r_c$ is different.
There, as a function of temperature, a change of slope occurs in the frequency dependence of the self energy, where at intermediate temperatures $T\approx 0.1$ the self energy is nearly frequency independent.
Only at lower temperatures, $\Sigma_{\mathbf{k}_{hs}}(\omega_n)$ starts resembling the expected $~\sqrt(\omega_n)$ form.
In comparing with the DQMC results in Fig.\ref{fig:self_energy}, we note the similar magnitude of the self energy. However, the DQMC results show a far weaker temperature dependence at the hotspots for $r$ close to $r_c$.

\section{Extracting the superconducting gap}
\label{sec:Delta}
\begin{figure}[th]
  \centering
  \includegraphics[width=\columnwidth]{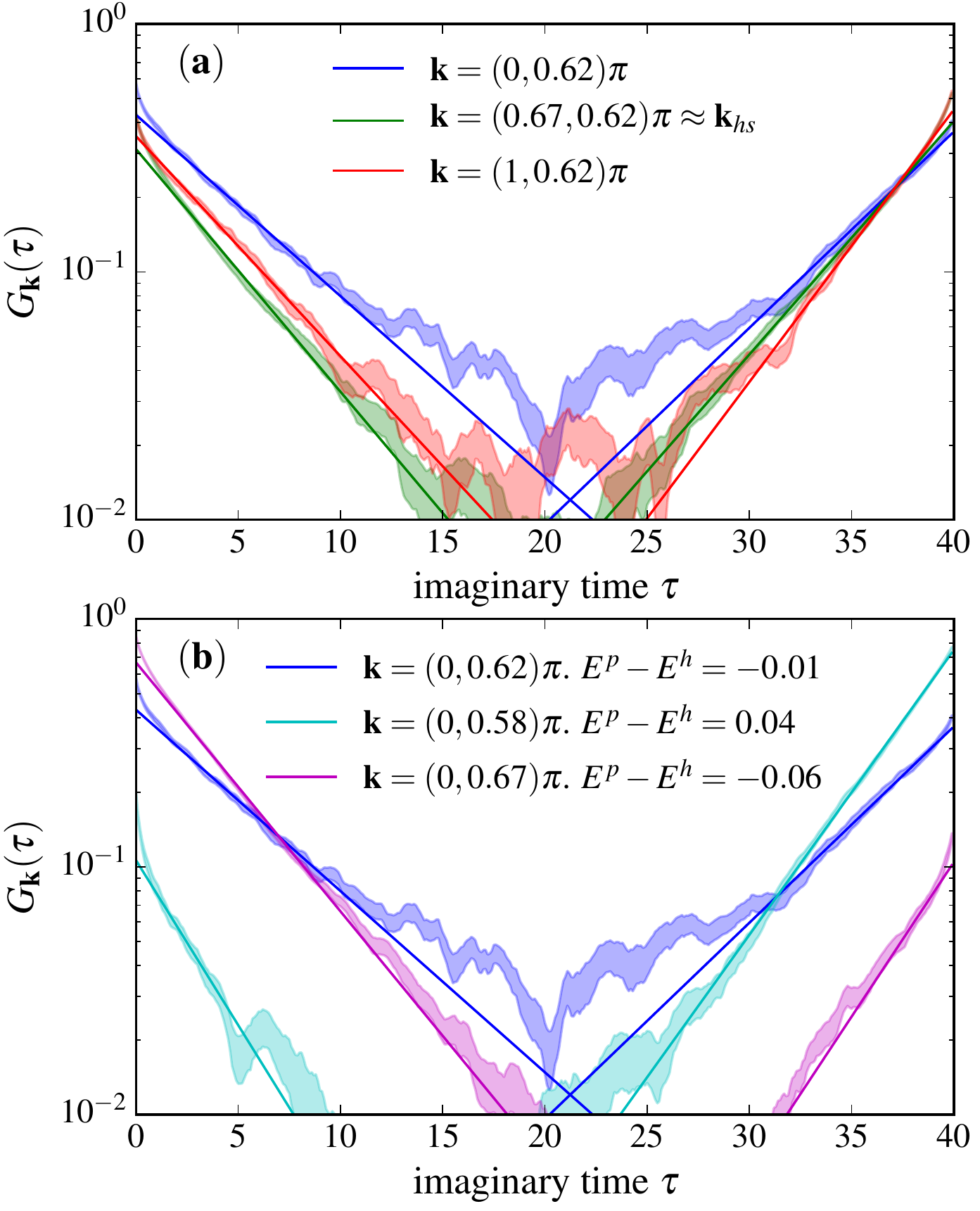}
  \caption{Imaginary time evolution of the single-particle Green's
    function $G_{\mathbf{k}}(\tau)$ for (a) several momenta along the
    noninteracting Fermi surface and (b) several momenta along a
    cut perpendicular to the Fermi surface.
    Here, $L=12$, $T=1/40$, $\lambda=3$, and $c=2$. Shaded
    regions indicate the statistical uncertainty. The solid lines are
    exponential fits.}
  \label{fig:Ek_from_Gtau_fits}
\end{figure}

In this Appendix we provide a detailed description of the procedure
by which we extract the single-particle excitation energy $E_{\mathbf{k}}$
in the superconducting state, which we discuss in Sec.~\ref{sec:SC} of the main text.
The single-particle Green's function $G_{\mathbf{k}}(\tau)$ is found to exhibit
a characteristic imaginary-time evolution as shown in Fig.~\ref{fig:Ek_from_Gtau_fits}.
At intermediate times, $\tau_0 < \tau < \beta/2$, where $\tau_0\sim 1$ is some microscopic time scale,
the single-particle Green's function decays exponentially as
\begin{equation}
	G_{\mathbf{k}}(\tau) \propto e^{-E^p_{\mathbf{k}}\tau} \,,
\label{eq:G_particle}
\end{equation}
and
similarly, for times $\tau_0 < \beta -\tau <\beta/2$,
\begin{equation}
	G_{\mathbf{k}}(\tau) \propto e^{-E^h_{\mathbf{k}}(\beta-\tau)} \,.
\label{eq:G_hole}
\end{equation}
At long times, $\tau\approx \beta/2$ the Green's function is
substantially suppressed and statistical errors dominate the signal.  We
extract the decay constants $E^p_{\mathbf{k}}$ and $E^h_{\mathbf{k}}$
from appropriate exponential fits and define the single-particle
excitation energy as their minimum
$E_{\mathbf{k}} = \min \left\{E^p_{\mathbf{k}},
  E^h_{\mathbf{k}}\right\}$.

For a qualitative understanding of these results, we consider the behavior of the Green's
function in a Fermi liquid and in a Bardeen-Cooper-Schrieffer (BCS) superconductor.
The fact that  $G_{\mathbf{k}}(\tau)$ exhibits exponential behavior can be interpreted as arising from a peak in the
spectral function $A_{\mathbf{k}} (\omega)$, occuring at a non-zero frequency, as can be seen from \eqref{eq:G_tau}.
In a Fermi liquid, the spectral function at a given momentum has a single, sharp peak at
the energy of the quasiparticle. It then follows that $G_{\mathbf{k}}(\tau)$ has the form of a single exponential,
with hole-like quasiparticles obeying \eqref{eq:G_hole} and
particle-like quasiparticles obeying \eqref{eq:G_particle}.
Indeed, in our simulations in the normal state we find monotonic behavior of $G_{\mathbf{k}} (\tau)$ (not shown).
It is illuminating to contrast this behavior with the BCS
state, where a superposition of hole-like and particle-like excitations is allowed.
In this case, the spectral function consists of two delta-function peaks at $\omega=\pm E_\mathbf{k}$, such that the Green's function takes the form
\begin{equation}
G_{\mathbf{k}}(\tau)=\frac{1}{1+e^{-\beta
    E_{\mathbf{k}}}}\left(u_{\mathbf{k}}^{2}e^{-E_{\mathbf{k}}\tau}+v_{\mathbf{k}}^{2}e^{-E_{\mathbf{k}}(\beta-\tau)}\right) \,.
\label{eq:BCS}
\end{equation}
Here
$E_{\mathbf{k}} = \sqrt{\Delta_{\mathbf{k}}^2 +
  \epsilon_{\mathbf{k}}^2}$ with the quasiparticle dispersion
$\epsilon_{\mathbf{k}}$ and the gap $\Delta_{\mathbf{k}}$, and
$u_{\mathbf{k}}$, $v_{\mathbf{k}}$ are particle and hole
amplitudes, respectively. The resulting Green's function is non-monotonic,
showing a minimum at a finite imaginary time.

While our numerical data below $T_c$ shares some similarities with the BCS form \eqref{eq:BCS},
it differs in two notable ways.  First, clear exponential behavior is not seen at
short  times $\tau \lesssim \tau_0 \sim 1$.  Second, the particle and hole excitation energies
differ, i.e. $E^p _{\mathbf{k}} \neq E^h _{\mathbf{k}}$, with the
difference more pronounced away from the Fermi surface, as illustrated
in Fig.~\ref{fig:Ek_from_Gtau_fits}(b).

\clearpage

\end{document}